\documentclass[10pt]{elsart}

\usepackage[dvips]{graphicx}
\graphicspath{{./img/},{./img.old/}}
\usepackage{amsmath}
\usepackage{amssymb}
\usepackage{cite}

\usepackage{hyperref}
\begin{document}
%
\renewcommand{\d}{\mathrm{d}}
\renewcommand{\i}{\mathrm{i}}
\newcommand{\Tr}{\mathop\mathrm{Tr}}
\newcommand{\tr}{\mathop\mathrm{tr}}
\renewcommand{\Re}{\mathop\mathrm{Re}}
\renewcommand{\Im}{\mathop\mathrm{Im}}
\newcommand{\drangle}{\rangle\!\rangle}
\newcommand{\dlangle}{\langle\!\langle}

%
\newcommand{\T}{\mathcal{T}}
\newcommand{\kBT}{k_\mathrm{B}T}

\sloppy

\begin{frontmatter}

\title{Driven quantum transport on the nanoscale}

\author{Sigmund Kohler\corauthref{cor}}, 
\corauth[cor]{Corresponding author.} 
\ead{Sigmund.Kohler@Physik.Uni-Augsburg.DE} 
\author{J\"org Lehmann\thanksref{now}}, and
\thanks[now]{Present address: Departement f\"ur Physik und Astronomie,
Universit\"at Basel, Klingelbergstrasse 82, CH-4056 Basel, Switzerland}
\author{Peter H\"anggi}
\address{Institut f\"ur Physik, Universit\"at Augsburg,
Universit\"atsstra\ss e~1, D-86135 Augsburg, Germany}

\begin{abstract}
We explore the prospects to control by use of time-dependent fields quantum
transport phenomena in nanoscale systems.  In particular, we study for
driven conductors the electron current and its noise properties.  We review
recent corresponding theoretical descriptions which are
based on Floquet theory.  Alternative approaches, as well as various
limiting approximation schemes are investigated and compared.  The general
theory is subsequently applied to different representative nanoscale
devices, like the non-adiabatic pumps, molecular gates, molecular quantum
ratchets, and molecular transistors.  Potential applications range from
molecular wires under the influence of strong laser fields to
microwave-irradiated quantum dots.
\end{abstract}

\end{frontmatter}

\tableofcontents
\vfill
\section*{Notation}
\begin{tabular}{lp{.8\textwidth}}
$n$                             & wire site index, $n=1,\ldots,N$ \\
$\ell$                          & $=\mathrm{L},\mathrm{R}$, lead index \\
$n_\ell$                        & wire site attached to lead $\ell$:
                                  $n_\mathrm{L}=1$, $n_\mathrm{R}=N$ \\
$\alpha,\beta$                  & Floquet state indices \\
$k$                             & side-band/Fourier index \\
$\epsilon_\alpha+\i\hbar\gamma_\alpha$
                                & complex quasienergy \\
$\Omega$                        & driving (angular) frequency \\
$\mathcal{T}$                   & $=2\pi/\Omega$, driving period \\
$\kBT$                          & Boltzmann constant times temperature \\
$\Gamma_\ell(\epsilon)$         & spectral density of lead $\ell$ \\
$\Sigma$                        & self energy \\
$|n\rangle$                     & wire site, $n=1,\ldots,N$ \\
$|u_{\alpha}(t)\rangle$         & $=\sum_k\exp(-\i k\Omega t)
                                  |u_{\alpha,k}\rangle$, Floquet state
                                  for finite self energy \\
$|u_{\alpha,k}\rangle$          & $k$th Fourier coefficient of Floquet state
                                  $|u_{\alpha}(t)\rangle$ \\
$|\phi_{\alpha}(t)\rangle$      & Floquet state for self energy $\Sigma=0$\\
$E_\alpha$, $|\alpha\rangle$    & eigenenergy and eigenstate of a static Hamiltonian \\
$P_{\alpha\beta}(t)$            & $=\langle c_\beta^\dagger c_\alpha\rangle$,
                                  single particle density matrix in
                                  Floquet basis \\
$f(x)$                          & $=[\exp(x/\kBT)+1]^{-1}$, Fermi function \\
\end{tabular}
\clearpage
\section{Introduction}
\label{sec:introduction}

As anticipated by Richard Feynman in his visionary lecture ``There's plenty
of room at the bottom'' \cite{Feynman1960a}, we witness an ongoing progress
in the study of physical phenomena on ever smaller scales. Partly, this has
been made possible by the continuous technical achievements in fabrication
and miniaturization of electronic devices. However, it was the invention of
scanning probe microscopes~\cite{Binnig1984a}, which brought about the
realization of Feynman's dream, namely the selective manipulation of matter
on the nanoscale.
Since then, much progress has been made in nano sciences.  In particular,
the field of molecular electronics has emerged, which deals with the
realization of electronic devices based on the properties of a single or a
few molecules.  The theoretical proposal of a molecular rectifier by Aviram
and Ratner \cite{Aviram1974a} has been trend-setting for investigating the
distinct features of electrical transport on the nanoscale.  On the
experimental side, an ancestor of molecular electronics was the pioneering
work by Mann and Kuhn \cite{Mann1971a} on transport through hybrid
acid-salt surface adlayers.
The ongoing advance in contacting single molecules by nano-electrodes
allows one to perform transport measurements~\cite{Reed1997a, Cui2001a,
Reichert2002a, Nitzan2003a, Heath2003a}.  In these experiments, the quantum
nature of the electrons and the quantum coherence across the wire, which is
connected to adjacent macroscopic lead electrodes, influence various
physical properties such as the conductance and the corresponding current
noise statistics.  The rapid evolution of molecular conduction is
documented by recent monographs and article collections
\cite{Hanggi2002elsevier, Balzani2003a, Goser2004a, Cuniberti2005a}.

For the corresponding theoretical investigations, two lines of
research are presently pursued.  A first one starts out from the
\textit{ab-initio} computation of the orbitals relevant for the motion
of excess charges through the molecular
wire~\cite{DiVentra2000a,DiVentra2002a,Xue2002a,Damle2002a,Heurich2002a}
At present, however, the results of such computations generally differ
by more than one order of magnitude from experimental data, possibly
due to the equilibrium treatment of exchange correlations
\cite{Evers2004a}.
The second line employs corresponding phenomenological models in order to
gain a qualitative understanding of the transport mechanisms involved
\cite{Mujica1994a, Segal2000a, Boese2001a, Petrov2001a, Nitzan2001a,
Hettler2003a}.  Two particular problems addressed within model calculations
are the conduction mechanism in the presence of electron-phonon coupling
\cite{Olson1998a, Yu1999a, Emberly2000a, Mikrajuddin2000a, Segal2000a,
Ness2001a, Boese2001a, Petrov2001a, Petrov2002a, May2002a, Petrov2004a} and
the length dependence of the current-voltage characteristics
\cite{Mujica1994a, Nitzan2001a}.  The present work also employs rather
universal models: We describe the molecules by a linear arrangement of
tight-binding levels with the terminating sites attached to leads.
Still it is possible to suitably parametrize such tight-binding models
in order to obtain qualitative results for real
systems~\cite{Fagas2001a, Cuniberti2002a, Gutierrez2002a}.
Furthermore, these models also capture the physics of the so-called artificial
molecules, i.e.\ coupled quantum dots and quantum dot arrays
\cite{Blick1996a, vanderWiel2003a}.

One particular question that arises in this context is the influence of
excitations by electromagnetic fields and gate voltages on the electron
transport.  Such excitations bear intriguing phenomena like photon-assisted
tunneling \cite{Fujisawa1997a, Oosterkamp1998a, vanderWiel2003a,
Platero2004a} and the adiabatic \cite{Thouless1983a, Altshuler1999a,
Switkes1999a} and non-adiabatic pumping \cite{Wagner1999a, Levinson2000a}
of electrons.  From a fundamental point of view, these effects are of
interest because the external fields enable selective electron excitations
and allow one to study their interplay with the underlying transport mechanism.
In practical applications, time-dependent effects can be used to control
and steer currents in coherent conductors.  However, such control schemes
can be valuable only if they operate at tolerable noise levels.  Thus, the
corresponding current noise is also of equal interest.

An intuitive description of the coherent electron transport through
time-independent mesoscopic systems is provided by the Landauer scattering
formula \cite{Landauer1957a} and its various generalizations.  Both the
average current \cite{Imry1986a, Datta1995a, Landauer1992a, Imry1999a} and
the transport noise characteristics \cite{Blanter2000a} can be expressed in
terms of the quantum transmission coefficients for the respective
scattering channels.  By contrast, the theory for driven quantum transport
is less developed.
Scattering of a single particle by arbitrary time-dependent potentials has
been considered \cite{Henseler2000a, Henseler2001a, Li1999a} without
relating the resulting transmission probabilities to a current between
electron reservoirs.  Such a relation is indeed non-trivial since the
driving opens inelastic transport channels and, therefore, in contrast to
the static case, an \textit{ad hoc} inclusion of the Pauli principle is no
longer unique.  This gave rise to a discussion about ``Pauli blocking
factors'' \cite{Datta1992a, Wagner2000a}.  In order to resolve such
conflicts, one should start out from a many-particle description.  In this
spirit, within a Green function approach, a formal solution for the current
through a time-dependent conductor has been presented \cite{Jauho1994a,
Stafford1996a} without taking advantage of the full Floquet theory for the
wire and without obtaining a ``scattering form'' for the current in the
general driven case.
The spectral density of the current fluctuations has been derived for the
low-frequency ac conductance \cite{Pretre1996a, Pedersen1998a} and the
scattering by a slowly time-dependent potential \cite{Lesovik1994a}.  For
arbitrary driving frequencies, the noise can be characterized by its
zero-frequency component.  A remarkable feature of the current noise in the
presence of time-dependent fields is its dependence on the phase of the
transmission \textit{amplitudes} \cite{Lesovik1994a, Camalet2003a,
Camalet2004a}.  By clear contrast, both the noise in the static case
\cite{Blanter2000a} and the current in the driven case \cite{Camalet2003a}
depend solely on transmission \textit{probabilities}.

In Section~\ref{sec:scattering}, we derive within a \textit{Floquet approach}
explicit expressions for both the current and the noise properties of the
electron transport through a driven nanoscale conductor under the influence
of time-dependent forces \cite{Camalet2003a, Camalet2004a}.  This approach
is applicable to arbitrary periodically driven tight-binding systems and,
in particular, is valid for arbitrary driving strength and extends beyond
the adiabatic regime.  The dynamics of the electrons is solved by
integrating the Heisenberg equations of motion for the electron creation
and annihilation operators in terms of the single-particle propagator.  For
this propagator, in turn, we provide a solution within a generalized
Floquet approach.  Such a treatment is valid only for effectively
non-interacting electrons, i.e., in the absence of strong correlations.
Moreover, this \textit{Floquet scattering approach} cannot be generalized
straightforwardly to the case with additional electron-vibrational
coupling.  Better suited for this situation is a quantum kinetic equation
formalism which, however, is perturbative in both the wire-lead coupling
and the electron-vibrational coupling \cite{Lehmann2003b, Lehmann2004a}.

An experimental starting point for the investigation of the influence of
electromagnetic fields on molecular conduction is the excitation of
electrons to higher orbitals of the contacted molecule.  In molecular
physics, specific excitations are usually performed with laser fields.  The
resulting changes of the current through a contacted molecule due to the
influence of a laser field are studied in Section~\ref{sec:resonances}.  In
particular, we focus on the modification of the length dependence of the
conductivity \cite{Kohler2002a, Kohler2004b}.

An intriguing phenomenon in strongly driven systems is the so-termed
ratchet or Brownian motor effect \cite{Hanggi1996a,
Astumian1997a, Julicher1997a, Reimann2002a,Reimann2002b, Astumian2002a},
originally discovered for overdamped classical Brownian motion in
asymmetric non-equilibrium systems.  Counter-intuitively to the second law
of thermodynamics, one then observes a directed transport although none of
the acting forces possesses any net bias.  This effect has been established
also within the regime of dissipative, incoherent quantum Brownian motion
\cite{Reimann1997a,Astumian2002a,Grifoni2002a}.  A mesoscopic device
related to ratchets is an electron pump \cite{Thouless1983a, Brouwer1998a,
Altshuler1999a, Switkes1999a, Wagner1999a, Levinson2000a, Wang2002a} which
indeed might be regarded as a localized ratchet.  Such systems have already
been realized in the quantum domain, but almost exclusively operating
in the regime of incoherent tunneling \cite{Linke1999a, Linke2002a,
Majer2003a, Haan2004a, Yasutomi2004a}.  In Section~\ref{sec:pump}, we study
the possibilities for molecular wires to act as coherent quantum ratchets
and explore the crossover from electron pumps to quantum ratchets.  This
requires to investigate thoroughly such quantum ratchet systems in the
coherent tunneling regime \cite{Lehmann2002b, Lehmann2003b}.

The tunneling dynamics of a particle in a bistable potential can be altered
significantly by ac fields.  In particular, it is possible to bring
tunneling to a standstill by the purely coherent influence of a
time-periodic driving \cite{Grossmann1991a,Grossmann1991b}.  This so-called
coherent destruction of tunneling has also been found in other systems
\cite{Grossmann1992a, Holthaus1992b, Creffield2002a}.  In
Section~\ref{sec:control}, we address the question whether a related effect
exists also for the electron transport through a driven conductor between
two leads.  Moreover, we study the noise properties of the resulting
transport process \cite{Lehmann2003a, Camalet2003a, Kohler2004a,
Camalet2004a}.

\subsection{Experimental motivation}

\subsubsection{Coupled quantum dots}

The experimental achievement of the coherent coupling of quantum dots
\cite{Blick1996a} enabled the measurement of intriguing phenomena in
mesoscopic transport \cite{vanderWiel2003a}.
A remarkable feature of coupled quantum dots---the so-called artificial
molecules with the single dots representing the atoms---is that the energy
levels of each ``atom'' can be controlled by an appropriate gate voltage.
In particular, the highest occupied levels of neighboring dots can be tuned
into resonance.  At such resonances, the conductance as a function of the
gate voltage exhibits a peak.
This behavior is modified by the influence of microwave radiation: With
increasing microwave intensity, the resonance peaks become smaller and
side-peaks emerge.  The distance between the central peak and the
side-peaks is determined by the frequency of the radiation field which
provides evidence for photon-assisted tunneling \cite{Fujisawa1997a,
Oosterkamp1998a, vanderWiel2003a, Platero2004a}.
Photon-assisted tunneling through quantum dots is, in comparison to its
counterpart in superconductor-insulator-superconductor junctions
\cite{Dayem1962a}, a potentially richer phenomenon.  The reason for this is
that quantum dots form a multi-barrier structure which permits real
occupation and resonant tunneling.  Therefore, a theoretical description
requires to also take into account the influence of the field on the
dynamics of the electrons localized in the central region between the
barriers.  The quantum dot setup used for the observation of
photon-assisted tunneling can also be employed as an implementation
\cite{vanderWiel1999a} of the theoretically suggested non-adiabatic pump
\cite{Stafford1996a, Stoof1996a, Brune1997a}.

Related experiments have been performed also with single quantum dots
exposed to laser pulses which resonantly couple the highest occupied
orbital and the lowest unoccupied orbital of the quantum dot
\cite{Zrenner2002a}.  Such a pulse can create an electron-hole pair which
in turn is transformed by a transport voltage into a current pulse.
Depending on their duration, pulses may not only excite an electron but
also coherently de-excite the electron and thereby reduce the resulting
current \cite{Rabi1937a}.  In the ideal case, the electron-hole pair is
excited with probability unity and finally yields a dc current consisting
of exactly one electron per pulse.  This effect might be employed for the
realization of a current standard.  At present, however, the deviations
from the ideal value of the current are still of the order of a few
percent.

\subsubsection{Molecular wires}

During the last years, it became possible to chemisorb organic
molecules via thiol groups to a metallic gold surface.  Thereby a
stable contact between the molecule and the gold is established.  This
enables reproducible measurements of the current not only through
artificial but also through real molecules.  Single molecule
conductance can be achieved in essentially two ways:
One possible setup is an open break junction bridged by a molecule
\cite{Reed1997a, Kergueris1999a, Reichert2002a}.  This setup can be kept
stable for several hours.  Moreover, it provides evidence for
\textit{single} molecule conductance because asymmetries in the
current-voltage characteristics reflect asymmetries of the molecule
\cite{Reichert2002a, Weber2002a}.
Alternatively, one can use a gold substrate as a contact and grow a
self-assembled monolayer of molecules on it.  The other contact is provided
by a gold cluster on top of a scanning tunneling microscope tip which
contacts one or a few molecules on the substrate \cite{Datta1997a,
Cui2001a}.  Yet another interesting device is based on the setup of a
single-molecule chemical field effect transistor in which the current
through a hybrid-molecular diode is controlled by nanometer-sized charge
transfer complex which is covalently linked to a molecule in a scanning
tunneling microscope junction \cite{Jackel2004a}.  Therein, the effect is
due to an interface dipole which shifts the substrate work function.
Naturally, the experimental effort with such molecular wires is accompanied
by vivid theoretical interest \cite{Nitzan2001a, Hanggi2002elsevier,
Nitzan2003a}.

Typical energy scales of molecules lie in the infrared regime where most of
today's lasers work.  Hence, lasers represent a natural possibility to
excite the electrons of the molecular wire and, thus, to study the
corresponding changes of the conduction properties.  At present, such
experiments are attempted, but still no clear-cut effect has been reported.
The molecule-lead contacts seem stable even against relatively intense
laser fields, but a main problem is the exclusion of side effects like,
e.g.\ heating of the break junction which might distort the molecule-tip
setup and, thus, be responsible for the observed enhancement of the
conductance \cite{Wurfel}.

In a recent experiment, Yasutomi \etal~measured the photocurrent
induced in a self-assembled monolayer of asymmetric molecules
\cite{Yasutomi2004a}.  They have found that even the current direction
depends on the wavelength of the irradiating light.  Albeit not a
single-molecules experiment, this measurement represents a first
experimental demonstration of a ratchet-like effect in molecular wires.

\section{Basic concepts}

Before going \textit{in medias res} and addressing specific quantum
transport situations, we introduce the reader to our archetypal working
model and the main theoretical methods and tools.  

\subsection{Model for driven conductor coupled to leads}
\label{sec:model}

The entire setup of our nanoscale system is described by the
time-dependent Hamiltonian
\begin{equation}
\label{H.wirelead}
H(t) = H_\mathrm{wire}(t) + H_{\rm leads} +
H_\mathrm{contacts},
\end{equation}
where the different terms correspond to the wire, the leads, and the
wire-lead couplings, respectively.  We focus on the regime of
coherent quantum transport where the main physics at work occurs on the
wire itself.  In doing so, we neglect other possible influences originating
from driving-induced hot electrons in the leads, dissipation on the wire
and, as well, electron-electron interaction effects.  Then, the wire
Hamiltonian reads in a tight-binding approximation with $N$ orbitals
$|n\rangle$
\begin{equation}
H_{\rm wire}(t)= \sum_{n,n'} H_{nn'}(t) c^{\dag}_n c^{\phantom{\dag}}_{n'}\;.
\label{Hw}
\end{equation}
For a molecular wire, this constitutes the so-called H\"uckel description
where each site corresponds to one atom.  The fermion operators $c_n$,
$c_n^{\dag}$ annihilate and create, respectively, an electron in the
orbital $|n\rangle$.  Note that in the absence of driving a diagonalization of
the system Hamiltonian would yield the stationary eigenvalues of the
wire levels.  The influence of an externally applied ac field
with frequency $\Omega=2\pi/\T$ results in a periodic time-dependence of
the wire Hamiltonian: $H_{nn'}(t+\T)=H_{nn'}(t)$.
In an experiment, the driving is switched on at a specific time and, thus,
the Hamiltonian is, strictly speaking, not time-periodic.  This can be
modeled by a slowly time-dependent driving amplitude that assumes its
ultimate value after a transient stage in the ``infinite past''.  Within
this work, however, we focus on the transport properties at asymptotically
long times where the amplitude has already settled \cite{Grossmann1991b,
Bavli1993a} and, thus, the driving can be assumed periodic.  This provides
the basis for the applicability of a Floquet transport theory.

\begin{figure}[tb]
\centerline{\includegraphics{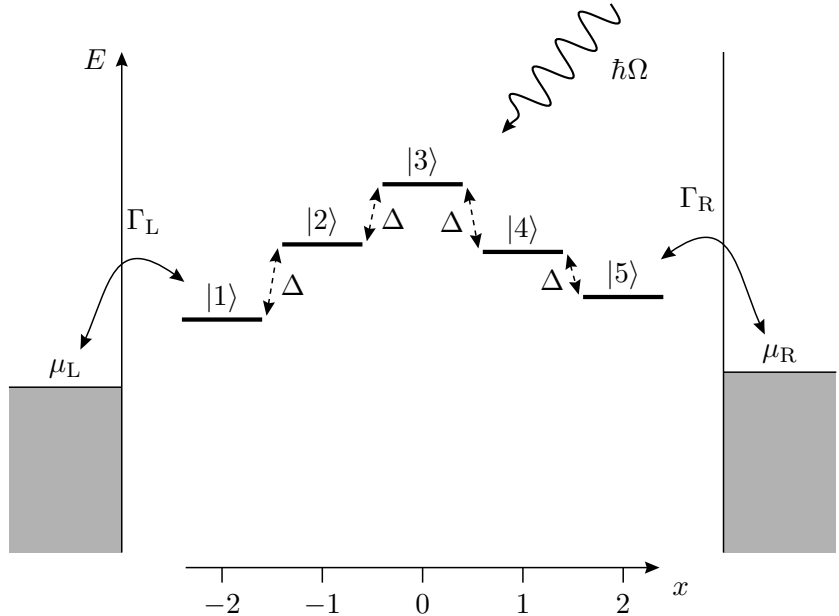}}
\caption{\label{fig:wire3} Level structure of a nano-conductor with
$N=5$ orbitals.  The end sites are coupled to two leads with chemical
potentials $\mu_\mathrm{L}$ and $\mu_\mathrm{R}=\mu_\mathrm{L}+eV$.}
\end{figure}

The leads are modeled by ideal electron gases,
\begin{equation}
H_\mathrm{leads}=\sum_q \epsilon_{q} (c^{\dag}_{\mathrm{L}q}
c^{\phantom{\dag}}_{\mathrm{L}q} + c^{\dag}_{\mathrm{R}q} c^{\phantom{\dag}}_{\mathrm{R}q}),
\label{Hl}
\end{equation}
where $c_{\mathrm{L}q}^{\dag}$ ($c_{\mathrm{R}q}^{\dag}$) creates an electron in the
state $|\mathrm{L}q \rangle$ ($|\mathrm{R}q \rangle$) in the left (right) lead.
The tunneling Hamiltonian
\begin{equation}
H_{\rm contacts} = \sum_{q} \left( V_{\mathrm{L}q} c^{\dag}_{\mathrm{L}q} c^{\phantom{\dag}}_1
+ V_{\mathrm{R}q} c^{\dag}_{\mathrm{R}q} c^{\phantom{\dag}}_N
\right) + \mathrm{h.c.}
\label{Hc}
\end{equation}
establishes the contact between the sites $|1\rangle$, $|N\rangle$
and the respective lead, as depicted with Fig.~\ref{fig:wire3}.
This tunneling coupling is described by the spectral density
\begin{equation}
\label{Gamma}
\Gamma_\ell (\epsilon) = 2\pi \sum_q |V_{\ell q}|^2
\delta(\epsilon-{\epsilon}_q)
\end{equation}
of lead $\ell = \mathrm{L},\mathrm{R}$ which becomes a smooth function if the lead
modes are dense.
If the leads are modeled by a tight-binding lattice, the
$\Gamma_\ell(\epsilon)$ assume a semi-elliptic shape, the so-called
Newns-Anderson density of states \cite{Newns1969a}, which is sometimes
employed in the context of molecular conduction \cite{Mujica1994a,
Mujica1996a, Hall2000a}.
Within the present context, however, we are mainly interested in the
influence of the driving field on the conductor and not in the details of
the coupling to the leads.  Therefore, we later on often choose for
$\Gamma_\ell(\epsilon)$ a rather generic form by assuming that in the
relevant regime, it is practically energy-independent,
\begin{equation}
\label{wbl}
\Gamma_\ell(\epsilon) \longrightarrow \Gamma_\ell .
\end{equation}

To fully specify the dynamics, we choose as an initial condition
for the left (right) lead a grand-canonical electron ensemble at
temperature $T$ and electro-chemical potential $\mu_{\mathrm{L}(\mathrm{R})}$.
Thus, the initial density matrix reads
\begin{equation}
\rho_0 \propto e^{-(H_\mathrm{leads} -\mu_\mathrm{L} N_\mathrm{L} -\mu_\mathrm{R} N_\mathrm{R})/\kBT} ,
\label{ic}
\end{equation}
where $N_{\ell}=\sum_q c^{\dag}_{\ell q} c^{\phantom{\dag}}_{\ell q}$
is the number of electrons in lead $\ell$
and $\kBT$ denotes the Boltzmann constant multiplied by the temperature.
An applied voltage $V$ maps to a chemical potential
difference $\mu_\mathrm{R}-\mu_\mathrm{L}=eV$ with $-e$ being the electron charge.
Then, at initial time $t_0$, the only nontrivial expectation values of the
lead operators read
$
\langle c_{\ell'q'}^\dagger c_{\ell q}\rangle
= f_\ell(\epsilon_q) \delta_{\ell\ell'}\delta_{qq'}
$
where $f_\ell(\epsilon)=(1+\exp[(\epsilon-\mu_\ell)/\kBT])^{-1}$ denotes
the Fermi function.

Below, we specify the wire Hamiltonian as a tight-binding model composed of
$N$ sites as sketched in Fig.~\ref{fig:wire3}.  Each orbital is coupled to
its nearest neighbor by a hopping matrix element $\Delta$, thus, the
single-particle wire Hamiltonian reads
\begin{equation}
\label{Hdipole}
\mathcal{H}_\mathrm{wire}(t)
=  -\Delta\sum_{n=1}^{N-1}\big( |n\rangle\langle n{+}1|
   + |n{+}1\rangle\langle n| \big)
   + \sum_n [E_n + x_n\, a(t)]\,|n\rangle\langle n| ,
\end{equation}
where $E_n$ stands for the on-site energies of the tight-binding levels.
Although the theoretical approach derived below is valid for an arbitrary
periodically driven wire Hamiltonian, we always assume that the
time dependence results from the coupling to an oscillating dipole field
that causes the time-dependent level shifts $x_n a(t)$, where
$x_n=(N+1-2n)/2$ denotes the scaled position of site $|n\rangle$.  The
energy $a(t)=a(t+\T)$ is determined by the electrical field strength
multiplied by the electron charge and the distance between two neighboring
sites.

An applied transport voltage $V$ is
mapped to a symmetric shift of the leads' chemical potentials, $\mu_\mathrm{R} =
-\mu_\mathrm{L} = eV/2$.  Moreover, for the evaluation of the dc current and the
zero-frequency noise, we restrict ourselves to zero temperature.  The
zero-temperature limit is physically well justified for molecular wires at
room temperature and for quantum dots at helium temperature since in both
cases, thermal electron excitations do not play a significant role.

In a realistic wire molecule, the hopping matrix element $\Delta$ is of the
order $0.1\,\mathrm{eV}$.  Thus, a typical wire-lead hopping rate
$\Gamma=0.1 \Delta$ yields a current $e\Gamma/\hbar=2.56\times10^{-5}$\,Amp\`ere and
$\Omega\approx10\Delta/\hbar$ corresponds to a laser frequency in the near
infrared, i.e., to wavelengths of the order $1\,\mu\mathrm{m}$.  For a typical
distance of {$5$\AA} between two neighboring sites, a driving amplitude
$A=\Delta$ is equivalent to an electrical field strength of
$2\times10^6\,\mathrm{V/cm}$. It has to be emphasized that the amplitude~$A$
is determined by the local electrical field between the contacts.  The
difference to the incident field can be huge: Model calculations
demonstrated that the presence of metallic tips enhances the local field by
several orders of magnitude \cite{Demming1998a, Otto2001a}.  This explains
the observation that the Raman scattering intensity increases drastically
once the molecules are adsorbed to a metallic surface
\cite{Fleischmann1974a, Jeanmaire1977a}.
Coupled quantum dots typically \cite{Blick1996a, Oosterkamp1998a,
vanderWiel2003a} have a distance of less than $1\,\mu\mathrm{m}$ while the
coupling matrix element $\Delta$ is of the order of $30\,\mu\mathrm{eV}$
which corresponds to a wavelength of roughly $1\,\mathrm{cm}$.
The dipole approximation inherent to the time-dependent part of the
Hamiltonian~\eqref{Hdipole} neglects the propagation of the electromagnetic
field and, thus, is valid only for wavelengths that are much larger than
the size of the sample \cite{Pellegrini1993a}.  This condition is indeed
fulfilled for both applications we have in mind.

\subsection{AC transport voltage}

Within this work, we focus on models presented in the previous subsection,
i.e., models where the driving enters solely by means of time-dependent
matrix elements of the wire Hamiltonian while the leads and the wire-lead
couplings remain time-independent.  However, it is worthwhile to
demonstrate that a setup with an oscillating external voltage can be mapped
by a gauge transformation to the model introduced above.  Consequently, it
is possible to apply the formalism derived below also to situations with an
oscillating voltage.

We restrict the discussion to a situation where the electron energies
of the left lead are modified by an external $\mathcal{T}$-periodic voltage
$V_\mathrm{ac}(t)$ with zero time-average, thus
\begin{equation}
\label{Vac}
\epsilon_q \to \epsilon_q - eV_\mathrm{ac}(t) .
\end{equation}
The generalization to a situation where also the levels in the right lead
are $\mathcal{T}$-periodically time-dependent, is straightforward.
Since an externally applied voltage causes a potential drop along the wire
\cite{Nitzan2002a,Pleutin2003a,Liang2004a}, we have to assume for
consistency that for an ac voltage, the wire Hamiltonian also obeys a
time-dependence.  Ignoring such a time-dependent potential profile enables
a treatment of the transport problem within the approach of
Refs.~\cite{Tucker1979a,Tucker1985a}.  In the general case, however, we
have to resort to the approach put forward with this work.

We start out by a gauge transformation of the Hamiltonian
\eqref{H.wirelead} with the unitary operator
\begin{equation}
\label{Uac}
U_\mathrm{ac}(t)
= \exp\Big\{-\i\phi(t)\Big(c_1^\dagger c_1^{\phantom{\dagger}}
  + \sum_q c_{\mathrm{L}q}^\dagger c_{\mathrm{L}q}^{\phantom{\dagger}} \Big)\Big\}
\end{equation}
where
\begin{equation}
\phi(t) = -\frac{e}{\hbar}\int^t \d t'\,V_\mathrm{ac}(t')
\end{equation}
describes the phase accumulated from the oscillating voltage.  The
transformation \eqref{Uac} has been constructed such that the new
Hamiltonian $\widetilde H(t)=U_\mathrm{ac}^\dagger H(t) U_\mathrm{ac}
-\i\hbar U_\mathrm{ac}^\dagger \dot U_\mathrm{ac}^{\phantom{\dagger}}$
possesses a time-independent tunnel coupling.  Since, the operator $c_1$
transforms as $c_1 \to c_1\exp(-\i\phi(t))$, the matrix elements
$H_{nn'}(t)$ of the wire Hamiltonian acquire an additional time-dependence,
\begin{equation}
\label{Hw,mod}
H_{nn'}(t) \to \widetilde H_{nn'}(t)
= H_{nn'}(t)e^{-\i\phi(t)(\delta_{n'1}-\delta_{n1})}
   + eV_\mathrm{ac}(t) \delta_{n1}\delta_{n'1} .
\end{equation}
The second term in the Hamiltonian \eqref{Hw,mod} stems from $-\i\hbar
U_\mathrm{ac}^\dagger \dot U_\mathrm{ac}^{\phantom{\dagger}}$.  Owing to
the zero time-average of the voltage $V_\mathrm{ac}(t)$, the phase
$\phi(t)$ is $\mathcal{T}$-periodic.  Therefore, the transformed wire
Hamiltonian is also $\mathcal{T}$-periodic while the
contact and the lead contributions are time-independent, thus,
$\widetilde H(t)$ is of the same form as the original Hamiltonian
\eqref{H.wirelead}.

\subsection{Tien-Gordon theory}
\label{sec:tien-gordon}

In order to explain the steps in the current-voltage characteristics of
microwave-irradiated superconductor-insulator-superconductor junctions
\cite{Dayem1962a}, Tien and Gordon \cite{Tien1963a} proposed a heuristical
theoretical treatment which is of appealing simplicity but nevertheless
captures some essential features of driven transport.  The central idea of
this approach is to model the influence of the driving fields by a periodic
shift of the energies in the, e.g.\ left lead according
$\tilde\epsilon_{\mathrm{L}q}(t) = \epsilon_{\mathrm{L}q} + A\cos(\Omega t)$, cf.\
Eq.~\eqref{Vac}.  Then the corresponding lead eigenstates evolve as
\begin{align}
|\mathrm{L}q\rangle_t
= & \exp\Big( -\frac{\i}{\hbar}\epsilon_{\mathrm{L}q}t -
\i\frac{A}{\hbar\Omega}\sin(\Omega t) \Big) |\mathrm{L}q\rangle
\\
= & \sum_{k=-\infty}^\infty J_k(A/\hbar\Omega)
    \exp\Big( -\frac{\i}{\hbar}(\epsilon_{\mathrm{L}q}+k\hbar\Omega)t \Big)
|\mathrm{L}q\rangle ,
\label{tien.gordon.lead}
\end{align}
where $J_k$ denotes the $k$th order Bessel function of the first kind.  The
interpretation of the Fourier decomposition \eqref{tien.gordon.lead} is
that each state consists of sidebands whose energies are shifted by
multiples of $\hbar\Omega$.  For the evaluation of the dc current, this is
equivalent to replacing the Fermi function of the left lead by
\begin{equation}
\label{tien.gordon}
f_\mathrm{L}(E) \longrightarrow \sum_k J_k^2(A/\hbar\Omega) f_\mathrm{L}(E+k\hbar\Omega)
\end{equation}
and formally treating the system as time-independent \cite{Tien1963a}.
While this effective static treatment indeed captures the
photon-assisted
dc current, it naturally fails to describe any time-dependent response.

For time-dependent wire-lead models where the driving shifts all wire
levels simultaneously, it is possible to map the driving field by a gauge
transformation to oscillating chemical potentials.  Then, the average
current can be evaluated from an effective electron distribution like the
one in Eq.~\eqref{tien.gordon} \cite{Wingreen1990a, Kislov1991a,
Aguado1996a}.  However, generally the time-dependent field also influences
the dynamics of the electrons on the wire.  In particular, this is the case
for the dipole driving \eqref{Hdipole}.  Then, a treatment beyond
Tien-Gordon theory becomes necessary.  Deriving an approach which is valid
in the general case is the objective of Section~\ref{sec:scattering}.

\subsection{Scattering approach for static conductors}
\label{sec:intro.scattering}

In the absence of a driving field, the computation of the coherent
transport through mesoscopic structures has become a standard procedure
\cite{Imry1986a, Datta1995a, Landauer1992a, Imry1999a}.
The crucial idea goes back to Landauer who postulated already in 1957
\cite{Landauer1957a} that in the absence of both inelastic effects and
electron-electron interaction, conduction can be described as a
coherent scattering process of independent electrons.  Then, an
infinitesimal voltage $V$ causes the current $I=GV$ with the (linear)
conductance
\begin{equation}
\label{landauer}
G = \frac{e^2}{h} T,
\end{equation}
of a one-dimensional conductor, where $T$ is the transmission probability
of an electron at the Fermi surface.  Since conductors may have
non-vanishing reflection probability $1-T$, the transmission probability does not
necessarily assume an integer value.  The prefactor $e^2/h =
(25.8\,\mathrm{k\Omega})^{-1}$ is the so-called conductance quantum.

Originally \cite{Landauer1957a}, the conductance \eqref{landauer} has been
proposed with $T$ replaced by $T/(1-T)$.  In the beginning of the 1980's,
there has been a theoretical debate \cite{Economou1981a, Langreth1981a,
Stone1988a} whether or not, the reflection coefficient $1-T$
has to be included.  The controversy was resolved by considering
four-terminal devices where two terminals act as voltage probes and are
considered as a part of the mesoscopic conductor \cite{Engquist1981a,
Buttiker1986a}.  Then, $V$ represents the probed voltage and the factor
$1/(1-T)$ indeed is justified.  In a two-terminal device, however, $V$
denotes the externally applied voltage and the conductance includes a
contact resistance and is given by Eq.~\eqref{landauer}.

With the same ideas, Landauer theory can be generalized to the
case of a finite voltage for which the current reads
\begin{equation}
\label{current.static}
I = \frac{e}{h}\int\d E\, \big[f_\mathrm{R}(E)-f_\mathrm{L}(E)\big]T(E) ,
\end{equation}
with $T(E)$ being the electron transmission probability at energy $E$.  The electron
distribution in the left (right) lead is given by the Fermi function
$f_{\mathrm{L}(\mathrm{R})}$ with the chemical potential $\mu_{\mathrm{L}(\mathrm{R})}$ whose difference
$\mu_\mathrm{R}-\mu_\mathrm{L} = eV$ is determined by the applied voltage.  The linearization for
small voltages yields the conductance \eqref{landauer}.
The current formula \eqref{current.static} and the conductance
\eqref{landauer} have been derived from Kubo formula \cite{Engquist1981a,
Langreth1981a, Fisher1981a, Stone1988a, Sols1991a} and by means of non-equilibrium
Green function methods~\cite{Caroli1971a, Fisher1981a, Meir1992a, Wingreen1993a}
for various microscopic models.  In doing so, one usually starts by
defining a current operator, e.g.\ as the change of the electron charge
$eN_\mathrm{L}$ in the left lead, i.e.\ $I = \i e[H,N_\mathrm{L}]/\hbar$.  Finally, one
obtains the expected expression for the current together with the relation
\begin{equation}
\label{fisher-lee}
T(E) = \tr [ G^\dagger(E)\, \Sigma_\mathrm{R}(E)\, G(E)\, \Sigma_\mathrm{L}(E)\, ]
\end{equation}
between the transmission probability $T(E)$ and the Green function of the electrons.
The trace sums over all single-particle states of the wire and
$\Sigma_{\ell} = |n_\ell \rangle\frac{\Gamma_\ell}{2}\langle n_\ell |$
denotes the imaginary part of the self-energy of the terminating wire sites
which results from the coupling to the respective leads.

In order to obtain an expression for the related current noise, one
considers the symmetrized correlation function
\begin{equation}
\label{current.correlation}
S (t,t')
= \frac{1}{2} \big\langle [\Delta I(t),\Delta I(t')]_+
\big\rangle
\end{equation}
of the current fluctuation operator $\Delta I(t) = I(t)-\langle
I(t)\rangle$, where the anticommutator $[A,B]_+=AB+BA$ ensures hermiticity.
For a stationary process, the correlation function $S(t,t') = S(t-t')$ is a
function of only the time difference.  Then, the noise strength can be
characterized by the zero-frequency component
\begin{equation}
\label{barS.static}
S = \int_{-\infty}^\infty \d\tau\, S (\tau) ,
\end{equation}
which obeys $S\geq 0$ according to the Wiener-Khinchine theorem.
In terms of the transmission function~$T(E)$, the noise strength reads
\cite{Blanter2000a}
\begin{equation}
\label{noise.static}
\begin{split}
S = \frac{e^2}{h}\int\d E\,\Big\{ &
T(E) \big[f_\mathrm{L}(E)[1-f_\mathrm{L}(E)] + f_\mathrm{R}(E)[1-f_\mathrm{R}(E)] \big] \Big. \\
&+ T(E)\big[1-T(E)\big] \big[f_\mathrm{R}(E)-f_\mathrm{L}(E)\big]^2 \Big\} .
\end{split}
\end{equation}
A dimensionless measure for the \textit{relative} noise strength, is the
so-called Fano factor \cite{Fano1947a}
\begin{equation}
\label{Fano}
F = \frac{S}{e|I|} \,.
\end{equation}
Note that in a two-terminal device, both the absolute value of the average
current and the noise strength are independent of the contact $\ell$.
Historically, the zero-frequency noise \eqref{barS.static} contains a
factor $2$, i.e., one considers $S'=2S$, resulting from a different
definition of the Fourier transform.  Then, the Fano factor is defined as
$F= S'/2e|I|$.  The definition \eqref{Fano} is such that a Poisson process
corresponds to $F=1$.

The generalization of the noise expression \eqref{noise.static} to driven
systems must also account for absorption and emission.  Owing to this energy
non-conserving processes, the zero-frequency noise is no longer given
solely in terms of transmission \textit{probabilities} but also depends on
the phases of the transmission \textit{amplitudes} \cite{Lesovik1994a,
Camalet2003a, Camalet2004a}; cf.\ Eq.~\eqref{barS}, below.

\subsection{Master equation}
\label{sec:intro.master}

A different strategy for the computation of stationary currents relies on
the derivation of a master equation for the dynamics of the wire electrons.
There, the central idea is to consider the contact Hamiltonian \eqref{Hc}
as a perturbation, while the dynamics of the leads and the wire, including
the external driving, is treated exactly.  From the Liouville-von Neumann
equation $\i\hbar \dot \varrho(t)=[H(t),\varrho(t)]$ for the total density
operator $\varrho(t)$ one obtains by standard techniques
\cite{Nakajima1958a, Zwanzig1960a} the approximate equation of motion
\begin{equation}
\label{mastereq}
\begin{split}
\dot\varrho(t)
= & -\frac{\i}{\hbar}[H_{\rm wire}(t)+H_\mathrm{leads},\varrho(t)] \\
  & -\frac{1}{\hbar^2}\int_0^\infty \d\tau [H_\mathrm{contacts},
     [\widetilde H_\mathrm{contacts}(t-\tau,t),\varrho(t)]] .
\end{split}
\end{equation}
The tilde denotes operators in the interaction picture with respect to the
molecule and the lead Hamiltonian without the molecule-lead coupling,
$\widetilde X(t,t')=U_0^\dagger(t,t')\,X\,U_0(t,t')$, where $U_0$ is the
propagator without the coupling.
For the evaluation of Eq.~\eqref{mastereq} it is essential to use an exact
expression for the zeroth-order time evolution operator $U_0(t,t')$.  The
use of any approximation bears the danger of generating artifacts, which,
for instance, may lead to a violation of fundamental equilibrium
properties~\cite{Novotny2002a, May2004a}.

In order to make practical use of equation \eqref{mastereq}, one has to
trace over the lead degrees of freedom and thereby obtains a master
equation for the reduced density operator of the wire electrons.
Subsequently, the reduced density operator is decomposed into the
eigenstates of the wire Hamiltonian $H_\mathrm{wire}$---or the
corresponding Floquet states if the system is driven.
As a further simplification, one might neglect off-diagonal matrix elements
and, thus, obtain a master equation of the Pauli type, i.e., a closed
equation for the occupation \textit{probabilities} of the eigenstates
\cite{Hanggi1982a, Bruder1994a, Brune1997a}.  For driven systems close to
degeneracies of the quasienergies, however, such a Pauli master equation is
not reliable as has been exemplified in Ref.~\cite{Lehmann2004a}.

\section{Floquet approach to the driven transport problem}
\label{sec:scattering}

In the following, we present the Floquet approach for our working model of
Section~\ref{sec:model}.  This derivation is rigorous and exact: It is
equivalent to an exact treatment in terms of a Keldysh Green
function calculation \cite{Jauho1994a}.  However, the chosen Floquet
derivation is here more direct and technically less cumbersome.

We start out from the Heisenberg equations of motion for the annihilation
operators in lead $\ell$, i.e.,
\begin{equation}
\label{Heisenberg:lead}
\dot c_{\ell q}
= -\frac{\i}{\hbar} \epsilon_q c_{\ell q} - \frac{\i}{\hbar}V_{\ell q}\,c_{n_\ell} ,
\end{equation}
where $n_\ell$ denotes the conductor site attached to lead $\ell$,
i.e., $n_\mathrm{L}=1$ and $n_\mathrm{R}=N$. These equations 
are straightforwardly integrated to read
\begin{equation}
\label{c:lead}
c_{\ell q}(t)
=  c_{\ell q}(t_0)e^{-\i\epsilon_q(t-t_0)/\hbar}
   -\frac{\i}{\hbar} V_{\ell q} \int_0^{t-t_0}\! \d\tau\,
   e^{-\i\epsilon_q\tau/\hbar} c_{n_\ell}(t-\tau) .
\end{equation}
  Inserting \eqref{c:lead} into the
Heisenberg equations for the wire operators yields in the asymptotic
limit $t_0\to-\infty$
\begin{align}
\label{c:xi:1N}
\dot c_{n_\ell}(t)
=& -\frac{\i}{\hbar} \sum_{n'} H_{n_\ell,n'}(t)\, c_{n'}(t)
   -\frac{1}{\hbar}\int_0^\infty \d\tau\, \Gamma_{\ell}(\tau)\,
    c_{n_\ell}(t-\tau) +\xi_{\ell}(t) ,\\
\dot c_n(t)
=& -\frac{\i}{\hbar} \sum_{n'} H_{nn'}(t)\, c_{n'}(t)\,, \quad n=2,\ldots,N-1,
\label{c:xi:n}
\end{align}
where the lead response function $\Gamma_\ell(t)$ results from the
Fourier transformation of the spectral density \eqref{Gamma},
\begin{equation}
\label{Gammat}
\Gamma_\ell(t) = \int \frac{\d\epsilon}{2\pi\hbar}
 \e^{-\i\epsilon t/\hbar} \Gamma_\ell(\epsilon) .
\end{equation}
In the wide-band limit \eqref{wbl}, one obtains $\Gamma_\ell(t) =
\Gamma_\ell\, \delta(t)$ and, thus, the equations of motion for the
wire operators are memory-free.
The influence of the operator-valued Gaussian noise
\begin{equation}
\label{xi}
\xi_{\ell}(t)
= -\frac{\i}{\hbar} \sum_q V^*_{\ell q}\,
  e^{-\i\epsilon_q(t-t_0)/\hbar} \, c_{\ell q}(t_0)
\end{equation}
is fully specified by the expectation values
\begin{align}
\label{xi1}
\langle\xi_\ell(t)\rangle =  {}& 0,
\\
\label{xi2}
\langle\xi^\dagger_{\ell'}(t')\, \xi_{\ell}(t)\rangle
 = {}& \delta_{\ell\ell'} \int \frac{\d\epsilon}{2\pi\hbar^2}\,
   \e^{-\i\epsilon(t-t')/\hbar}\,\Gamma_\ell(\epsilon)f_\ell(\epsilon) \, ,
\end{align}
which for the uncorrelated initial state \eqref{ic} follow
from the definition \eqref{xi}.
It is convenient to define the Fourier representation of the noise operator,
$\xi_\ell(\epsilon) = \int\d t\exp(\i\epsilon t/\hbar) \xi_\ell(t)$ whose
correlation function
\begin{equation}
\label{xi2.epsilon}
\langle\xi_\ell^\dagger(\epsilon) \xi_{\ell'}(\epsilon')\rangle
= 2\pi \Gamma_\ell(\epsilon) f_\ell(\epsilon)\,
  \delta(\epsilon-\epsilon')\, \delta_{\ell\ell'}
\end{equation}
is obtained directly from Eq.\ \eqref{xi2}.

\subsection{Retarded Green function}

The equations of motion \eqref{c:xi:1N} and \eqref{c:xi:n} represent a set
of linear inhomogeneous equations and, thus, can be solved with the help of
the retarded Green function $G(t,t')=U(t,t')\Theta(t-t')$ which obeys
\begin{equation}
\label{G.def}
\Big(\i\hbar\frac{\d}{\d t} - \mathcal{H}(t)\Big) G(t,t')
+ \i\int_0^\infty \d\tau\, \Gamma(\tau)\, G(t-\tau,t') = \delta(t-t') ,
\end{equation}
where $\Gamma(t) = |1\rangle\Gamma_\mathrm{L}(t)\langle 1| +
|N\rangle\Gamma_\mathrm{R}(t)\langle N|$.
At this stage, it is important to note that in the asymptotic limit $t_0
\to -\infty$, the l.h.s.\ of this equation is periodic in $t$.  As
demonstrated in the Appendix, this has the consequence that for the
propagator of the homogeneous equations obeys $U(t,t')=U(t+\T,t'+\T)$ and,
accordingly, the retarded Green function
\begin{equation}
\label{G(t,epsilon)}
G(t,\epsilon)
=  -\frac{\i}{\hbar} \int_0^\infty \! \d\tau\, \e^{\i\epsilon\tau/\hbar}
    U(t,t-\tau)
= G(t+\mathcal{T},\epsilon)
\end{equation}
is also $\T$-periodic in the time argument.  Thus, we can employ the
Fourier decomposition $G(t,\epsilon) = \sum_k \e^{-\i k\Omega t}
G^{(k)}(\epsilon)$, with the coefficients
\begin{align}
\label{G(k,epsilon)}
G^{(k)}(\epsilon) = & \frac{1}{\mathcal{T}} \int_0^\mathcal{T} \d t\,
\e^{\i k\Omega t} G(t,\epsilon) .
\end{align}
Physically, $G^{(k)}(\epsilon)$ describes the propagation of an electron
with initial energy $\epsilon$ under the absorption (emission) of $|k|$
photons for $k>0$ ($k<0$).
In the limiting case of a time-independent situation, $G(t,\epsilon)$
becomes independent of $t$ and, consequently, identical to
$G^{(0)}(\epsilon)$ while all sideband contributions with $k\neq 0$ vanish.

From the definition of the Green function, it can be shown that the
solution of the Heisenberg equation \eqref{c:xi:1N}, \eqref{c:xi:n}
reads
\begin{equation}
\label{cn.tau}
c_n(t)
= \i\hbar \sum_\ell \int_0^\infty \d\tau\, G_{n,n_\ell}(t,t-\tau)\, \xi_\ell(t-\tau) .
\end{equation}
Inserting for $G_{n,n_\ell}(t,t') = \langle n|G(t,t')|n_\ell\rangle$
the Fourier representation \eqref{G(t,epsilon)}, one obtains the form
\begin{equation}
\label{cn.epsilon}
c_n(t) = \frac{\i}{2\pi} \sum_\ell\int\d\epsilon \e^{-\i\epsilon t/\hbar}
G_{n,n_\ell}(t,\epsilon)\, \xi_\ell(\epsilon),
\end{equation}
which proves more convenient.

Below, we need for the elimination of back-scattering terms the relation
\begin{equation}
\label{datta}
\begin{split}
G^\dagger(t,\epsilon')-G(t,\epsilon)
={} &  \big( \i\hbar \frac{\d}{\d t} - \epsilon' + \epsilon \big)
       G^\dagger(t,\epsilon') G(t,\epsilon)
\\
&  + \i\int_0^\infty\d\tau \e^{\i\epsilon\tau/\hbar}
     G^\dagger(t,\epsilon')\Gamma(\tau) G(t-\tau,\epsilon)
\\
&  + \i\int_0^\infty\d\tau \e^{-\i\epsilon'\tau/\hbar}
     G^\dagger(t-\tau,\epsilon')\Gamma^\dagger(\tau) G(t,\epsilon) .
\end{split}
\end{equation}
A proof of this relation starts from the definition of the Green function,
Eq.\ \eqref{G.def}.  By Fourier transformation with respect to $t'$, we
obtain
\begin{equation}
\label{G:epsilon:def}
\Big(\i\hbar\frac{\d}{\d t} + \epsilon - \mathcal{H}(t)\Big) G(t,\epsilon)
+ \i\int_0^\infty \d\tau\, \e^{\i\epsilon\tau/\hbar}
  \Gamma(\tau)\, G(t-\tau,\epsilon) = \mathbf{1}
\end{equation}
which we multiply by $G^\dagger(t,\epsilon)$ from the left.  The difference
between the resulting expression and its hermitian adjoint with $\epsilon$
and $\epsilon'$ interchanged is relation~\eqref{datta}.

\subsection{Current through the driven nano-system}
\label{sec:IS}

The (net) current flowing across the contact of lead $\ell$ into the
conductor is determined by the negative change of the electron number
in lead $\ell$ multiplied by the electron charge $-e$.  Thus, the
current operator reads $I_\ell = \i e[H(t),N_\ell]/\hbar$, where
$N_\ell = \sum_q c_{\ell q}^\dagger c_{\ell q}$ denotes the
corresponding electron number.  By using Eqs.\ \eqref{c:lead} and
\eqref{xi}, we obtain
\begin{equation}
\label{I.operator}
\begin{split}
I_\ell(t)
={} & \frac{e}{\hbar} \int_0^\infty \d\tau\,\big\{ \Gamma_\ell(\tau)
   c_1^\dagger(t)c_1(t-\tau)
   +\Gamma_\ell^*(\tau) c_1^\dagger(t-\tau) c_1(t) \big\}
\\
 &     -e\big\{c_1^\dagger(t)\xi_\ell(t)+\xi_\ell^\dagger(t)c_1(t)\big\} .
\end{split}
\end{equation}
This operator-valued expression for the time-dependent current is a convenient
starting point for the evaluation of expectation values like dc current, ac
current, and current noise.

\subsubsection{Average current}

In order to evaluate the current $\langle I_\mathrm{L}(t)\rangle$, we insert the
solution \eqref{cn.epsilon} of the Heisenberg equation into the current
operator \eqref{I.operator} and use the expectation values
\eqref{xi2.epsilon}.  The resulting expression
\begin{equation}
\label{I.operator.xi}
\begin{split}
\langle I_\mathrm{L}(t)\rangle
={} &  \frac{e}{h} \sum_\ell \int\d\epsilon \int_0^\infty \d\tau \big(
     \e^{\i\epsilon\tau/\hbar}
      G_{1\ell}^*(t,\epsilon)\, \Gamma_\mathrm{L}(\tau)\, G_{1\ell}(t-\tau,\epsilon)
      \Gamma_\ell(\epsilon) f_\ell(\epsilon)
\\
& \phantom{\frac{e}{\hbar} \sum_\ell \int\d\epsilon\int_0^\infty \d\tau}
      + \e^{-\i\epsilon\tau/\hbar}
      G_{1\ell}^*(t-\tau,\epsilon)\, \Gamma_\mathrm{L}^*(\tau)\, G_{1\ell}(t,\epsilon)
      \Gamma_\ell(\epsilon) f_\ell(\epsilon)
\big)
\\
& + \i e\int\d\epsilon\, \big(G_{11}^*(t,\epsilon) - G_{11}(t,\epsilon) \big)
     \Gamma_\mathrm{L}(\epsilon) f_\ell(\epsilon)
\end{split}
\end{equation}
still contains back-scattering terms $G_{11}$ and, thus, is not of a
``scattering form''.  Indeed, bringing \eqref{I.operator.xi} into a form that
resembles the static current formula \eqref{current.static} requires some
tedious algebra.  Such a derivation has been presented for the linear
conductance of time-independent systems \cite{Fisher1981a}, for finite voltage
in the static case for tunneling barriers \cite{Caroli1971a} and mesoscopic
conductors \cite{Meir1992a}, a wire consisting of levels that couple equally to
both leads \cite{Jauho1994a}, and for weak wire-lead coupling
\cite{Datta1992a}.  For the general time-dependent case in the absence of
electron-electron interactions, such an expression has been \textit{derived}
only recently \cite{Camalet2003a,Camalet2004a}.

Inserting the matrix element $\langle 1|\ldots|1\rangle$ of equation
\eqref{datta}, eliminates the back-scattering terms and we obtain for the
time-dependent current the expression
\begin{equation}
\label{I(t)}
\langle I_\mathrm{L}(t)\rangle
=  \frac{e}{h}  \int \d\epsilon\,
   \big\{ T_{\mathrm{L}\mathrm{R}} (t,\epsilon) f_\mathrm{R} (\epsilon)
        - T_{\mathrm{R}\mathrm{L}} (t,\epsilon) f_\mathrm{L} (\epsilon) \big\}
   - \frac{\d}{\d t} q_\mathrm{L}(t)
\end{equation}
where
\begin{equation}
\begin{split}
q_{\mathrm{L}}(t)
= & \frac{e}{2\pi} \int\d\epsilon \, \Gamma_{\mathrm{L}} (\epsilon)
  \sum_n \left| G_{n 1}(t,\epsilon) \right|^2 f_\mathrm{L} (\epsilon)
\end{split}
\end{equation}
denotes the charge oscillating between the left lead and the wire.
Obviously, since $q_\mathrm{L}(t)$ is time-periodic and bounded, its time derivative
cannot contribute to the average current.  The corresponding charge arising
from the right lead, $q_\mathrm{R}(t)$, is \textit{a priori} unrelated to $q_\mathrm{L}(t)$;
the actual charge on the wire reads $q_\mathrm{L}(t)+q_\mathrm{R}(t)$.  The time-dependent
current is determined by the time-dependent transmission probability
\begin{equation}
\label{T.LR(t)}
T_{\mathrm{L}\mathrm{R}}(t,\epsilon)
= 2 \Re \int_0^\infty \d\tau \e^{\i\epsilon\tau/\hbar}
  \Gamma_\mathrm{L}(\tau)\, G_{1N}^*(t,\epsilon)\, G_{1N}(t-\tau,\epsilon)\,
  \Gamma_\mathrm{R}(\epsilon) .
\end{equation}
The corresponding expression for $T_{\mathrm{R}\mathrm{L}}(t,\epsilon)$ follows from the
replacement $(\mathrm{L},1)\leftrightarrow(\mathrm{R},N)$.
We emphasize that \eqref{I(t)}
obeys the form of the current formula obtained for a \textit{static}
conductor within a scattering formalism.  In particular, consistent with
Refs.~\cite{Datta1992a, Datta1995a}, no ``Pauli blocking factors''
$(1-f_\ell)$ appear in our derivation.  In contrast to a static situation,
this is in the present context relevant since for a driven system generally
\begin{equation}
T_{\mathrm{R}\mathrm{L}}(t,\epsilon) \neq T_{\mathrm{L}\mathrm{R}}(t,\epsilon)
\end{equation}
such that a contribution proportional to $f_\mathrm{L}(\epsilon_{q'}) f_\mathrm{R}(\epsilon_q)$
would not cancel \cite{Datta1992a, Wagner2000a}.

In order to obtain an expression for the dc current, we insert for the
Green function the Fourier representation \eqref{G(k,epsilon)} followed by
performing the average over time $t$.  Then, the average current becomes
\begin{equation}
\label{barI}
\bar I
= \frac{e}{h}  \sum_{k=-\infty}^\infty
  \int\d\epsilon \left\{ T_{\mathrm{L}\mathrm{R}}^{(k)} (\epsilon)
  f_\mathrm{R} (\epsilon) - T_{\mathrm{R}\mathrm{L}}^{(k)} (\epsilon) f_\mathrm{L} (\epsilon) \right\} ,
\end{equation}
where
\begin{align}
\label{TLR}
T_{\mathrm{L}\mathrm{R}}^{(k)}(\epsilon)
=& \Gamma_\mathrm{L} (\epsilon+k\hbar\Omega) \Gamma_\mathrm{R} (\epsilon)
   \big|G_{1N}^{(k)}(\epsilon) \big|^2 , \\
\label{TRL}
T_{\mathrm{R}\mathrm{L}}^{(k)}(\epsilon)
=& \Gamma_\mathrm{R} (\epsilon+k\hbar\Omega) \Gamma_\mathrm{L} (\epsilon)
   \big|G_{N1}^{(k)}(\epsilon) \big|^2 ,
\end{align}
denote the transmission probabilities for electrons from the right lead,
respectively from the left lead, with initial energy $\epsilon$ and final
energy $\epsilon+k\hbar\Omega$, i.e., the probability for an scattering
event under the absorption (emission) of $|k|$ photons if $k>0$ ($k<0$).

For a static situation, the transmission probabilities $T_{\mathrm{L}\mathrm{R}}^{(k)}(\epsilon)$ and
$T_{\mathrm{R}\mathrm{L}}^{(k)}(\epsilon)$ are identical and contributions with $k\neq 0$
vanish.  Thus, it is possible to write the current \eqref{barI} in the form
\eqref{current.static} as a product of a \textit{single} transmission probability
$T(\epsilon)$, which is independent of the direction, and the difference of
the Fermi functions, $f_\mathrm{R}(\epsilon)-f_\mathrm{L}(\epsilon)$.  We emphasize that in
the driven case this is no longer true.

\subsubsection{Noise power}

Like in the static case, we characterize the noise power by the
zero-frequency component of the current-current correlation function
\eqref{current.correlation}.  However, in the driven case, $S_\ell(t,t') =
S_\ell(t+\mathcal{T},t'+\mathcal{T})$ is still time-dependent.  Since it
shares the time-periodicity of the driving, it is possible to characterize
the noise level by the zero-frequency component of $S_\ell(t,t-\tau)$
averaged over the driving period,
\begin{equation}
\label{barS.def}
\bar S_\ell = \frac{1}{\T} \int_0^{\T} \d t
\int_{-\infty}^{+\infty} \d\tau\, S_\ell (t,t-\tau) .
\end{equation}
It can be shown \cite{Camalet2004a} that for driven two-terminal devices,
$\bar S_\ell$ is independent of the contact $\ell$, i.e., $\bar S_\mathrm{L}=\bar S_\mathrm{R}
\equiv \bar S$.

We start by writing $S_\mathrm{L}(t,t-\tau)$ with the current operator
\eqref{I.operator} and insert the solution \eqref{cn.epsilon} of the
Heisenberg equations.  We again employ relation \eqref{datta} and
finally obtain
\begin{equation}
\label{barS}
\begin{split}
\lefteqn{
\bar S = \frac{e^2}{h} \sum_k \int \d\epsilon \Big\{
    \Gamma_\mathrm{R} (\epsilon^{(k)}) \Gamma_\mathrm{R} (\epsilon)
    \Big| \sum_{k'} \Gamma_\mathrm{L} ( \epsilon^{(k')})
    G^{(k'-k)}_{1N}(\epsilon^{(k)}) \big[G_{1N}^{(k')}(\epsilon)\big]^* \Big|^2
    f_\mathrm{R} (\epsilon) \bar f_\mathrm{R} (\epsilon^{(k)})
}
\\
&+  \Gamma_\mathrm{R} (\epsilon^{(k)}) \Gamma_\mathrm{L} (\epsilon)
    \Big| \sum^{(k')}
    \Gamma_\mathrm{L}(\epsilon_{k'}) G^{(k'-k)}_{1N}(\epsilon^{(k)})
    \big[ G_{11}^{(k')}(\epsilon) \big]^* - \i G^{(-k)}_{1N}(\epsilon^{(k)})
    \Big|^2 f_\mathrm{L}(\epsilon) \bar f_\mathrm{R}(\epsilon^{(k)})
    \Big\} \\
&+  \text{same terms with the replacement $(\mathrm{L},1) \leftrightarrow (\mathrm{R},N)$} .
\end{split}
\end{equation}
We have defined $\epsilon^{(k)}=\epsilon+k\hbar\Omega$ and $\bar f_\ell=1-f_\ell$.
It can be shown (cf.\ Section~\ref{sec:static}) that in the undriven limit, the
noise power \eqref{barS} depends solely on the transmission
\textit{probabilities} and is given by Eq.~\eqref{noise.static}.  In the
time-dependent case, however, the noise expression~\eqref{barS} cannot be
brought into such a convenient form and, thus, generally depends on the
phase of the transmission amplitude.

\subsubsection{Floquet decomposition}

For energy-independent wire-lead coupling, i.e.\ in the so-called wide-band
limit $\Gamma_\ell(\epsilon)=\Gamma_\ell$ the lead response function
\eqref{Gammat} reads $\Gamma_\ell(t)=\Gamma_\ell \delta(t)$.
Consequently, the integro-differential equation \eqref{G.def} for the Green
function becomes a pure differential equation.  Then, determining the Green
function is equivalent to computing a complete set of solutions for the
equation
\begin{equation}
\label{preFloquet}
\i\hbar\frac{\d}{\d t} |\psi(t)\rangle
= \big(\mathcal{H}_\mathrm{wire}(t) - \i\Sigma \big) |\psi(t)\rangle ,
\end{equation}
where the self-energy
\begin{equation}
\label{sigma}
\Sigma = |1\rangle\frac{\Gamma_\mathrm{L}}{2}\langle 1 |
       + |N\rangle\frac{\Gamma_\mathrm{R}}{2}\langle N |
\end{equation}
results from the coupling to the leads.
Equation \eqref{preFloquet} is linear and possesses time-dependent,
$\mathcal{T}$-periodic coefficients.  Thus, following the reasoning of
Appendix \ref{app:floquet}, it is possible to construct
a complete solution with the Floquet ansatz
\begin{align}
|\psi_\alpha(t)\rangle
= & \exp[(-\i\epsilon_\alpha/\hbar-\gamma_\alpha)t] |u_\alpha(t)\rangle
,
\\
|u_{\alpha}(t)\rangle
= & \sum_k |u_{\alpha,k}\rangle\exp(-\i k\Omega t) .
\end{align}
The so-called Floquet states $|u_{\alpha}(t)\rangle$ obey the
time-periodicity of $\mathcal{H}_\mathrm{wire}(t)$ and have been
decomposed
into a Fourier series.  In a Hilbert space that is extended by a
periodic
time coordinate, the so-called Sambe space \cite{Sambe1973a}, they
obey the
Floquet eigenvalue equation \cite{Grifoni1998a, Buchleitner2002a}
\begin{equation}
\Big(\mathcal{H}_\mathrm{wire}(t) - \i\Sigma
-\i\hbar\frac{\d}{\d t}\Big)|u_{\alpha}(t)\rangle
= (\epsilon_{\alpha} -  \i\hbar\gamma_{\alpha}) |u_{\alpha}(t)\rangle .
\label{Fs}
\end{equation}
Due to the Brillouin zone structure of the Floquet spectrum
\cite{Shirley1965a, Sambe1973a, Grifoni1998a}, it is sufficient to compute
all eigenvalues of the first Brillouin zone,
$-\hbar\Omega/2<\epsilon_\alpha \le \hbar\Omega/2$.  Since the operator on
the l.h.s.\ of Eq.~\eqref{Fs} is non-Hermitian, the eigenvalues
$\epsilon_{\alpha} - \i\hbar\gamma_{\alpha}$ are generally complex valued
and the (right) eigenvectors are not mutually orthogonal.  Thus, to
determine the propagator, we need to solve also the adjoint Floquet
equation yielding again the same eigenvalues but providing the adjoint
eigenvectors $|u_\alpha^+(t)\rangle$.  It can be shown that the Floquet
states $|u_\alpha(t)\rangle$ together with the adjoint states
$|u_\alpha^+(t)\rangle$ form at equal times a complete bi-orthogonal basis:
$\langle u^+_{\alpha}(t)|u_{\beta}(t)\rangle = \delta_{\alpha\beta}$ and
$\sum_{\alpha} |u_{\alpha}(t)\rangle \langle u^+_{\alpha} (t)|=
\mathbf{1}$.  A proof requires to account for the time-periodicity of the
Floquet states since the eigenvalue equation \eqref{Fs} holds in a Hilbert
space extended by a periodic time coordinate \cite{Jung1990a,
Grifoni1998a}.  For details, see Appendix~\ref{app:floquet}.

For the special case \cite{Stafford1996a} of a wire with $N=2$ sites which
couple equally strong to both leads, i.e., $\Gamma_\mathrm{L}=\Gamma_\mathrm{R}$, the
self-energy is proportional to the unity matrix.  Consequently, the Floquet
states $|u_\alpha^+(t)\rangle$ become independent of the self-energy.

Using the Floquet equation \eqref{Fs}, it is straightforward to show that
with the help of the Floquet states $|u_\alpha(t)\rangle$ the propagator
can be written as
\begin{equation}
\label{Gtt}
U(t,t')
= \sum_\alpha \e^{-\i(\epsilon_\alpha/\hbar-\i\gamma_\alpha)(t-t')}
  |u_\alpha(t)\rangle\langle u^+_\alpha(t')| ,
\end{equation}
where the sum runs over all Floquet states within one Brillouin zone.
Consequently, the Fourier coefficients of the Green function read
\begin{align}
G^{(k)}(\epsilon)
=& -\frac{\i}{\hbar}\int_0^\T \frac{\d t}{\T} e^{\i k\Omega t}
   \int_0^\infty\!\! \d\tau
   \e^{\i\epsilon\tau/\hbar} U(t,t-\tau)
\\
=& \sum_{\alpha,k'}
   \frac{|u_{\alpha,k'+k}\rangle\langle u_{\alpha,k'}^+|}
      {\epsilon-(\epsilon_\alpha+k'\hbar\Omega-\i\hbar\gamma_\alpha)} .
\label{G}
\end{align}

For the exact computation of current and noise, we solve numerically the
Floquet equation \eqref{Fs}.  With the resulting Floquet states and
quasienergies, we obtain the Green function \eqref{G(k,epsilon)}.  In the
zero temperature limit, the Fermi functions in the expressions for the
average current \eqref{barI} and the zero-frequency noise \eqref{barS}
become step functions.  Then, the remaining energy integrals can be
performed analytically since the integrands are rational functions.

\subsection{Symmetries}
\label{sec:symmetry}

A system obeys a discrete symmetry if its Hamiltonian is invariant under
a symmetry operation $\mathcal{S} = (\mathcal{S}^+)^{-1}$, i.e, if
$\mathcal{S}^{-1} H(t) \mathcal{S} = H(t)$.  Then the corresponding transition
amplitude in position representation fulfills the relation
\begin{equation}
\label{symmetry}
\langle x|\mathcal{S}^+ U(t',t) \mathcal{S}|x'\rangle
= \langle x|U(t,t')|x'\rangle^{(*)}
\end{equation}
such that the corresponding transmission \textit{probabilities} are identical.
The complex conjugation in Eq.\ \eqref{symmetry} holds if $\mathcal{S}$
includes time inversion \cite{Sakurai}; then the r.h.s.\ becomes $\langle x'|
U(t',t) |x\rangle$.  If $\mathcal{S}|x'\rangle \neq |x\rangle$, relation
\eqref{symmetry} means that two \textit{different} scattering processes occur
with the same probability.  Correspondingly, in a time-dependent transport
problem as defined by the Hamiltonian \eqref{H.wirelead}, the presence of a
symmetry implies that two different transport channels have equal
transmission probability.

Here, we identify the channel which is related to $T_{\mathrm{L}\mathrm{R}}^{(k)}(\epsilon)$
given a certain symmetry is present.  In particular, we consider systems that are
invariant under the transformations studied in the Appendix
\ref{app:symmetry} which are combinations of the transformations
\begin{align}
\mathcal{S}_\mathrm{P} &: x \to -x ,    \label{S:P} \\
\mathcal{S}_\mathrm{T} &: t \to -t ,    \label{S:T} \\
\mathcal{S}_\mathrm{G} &: t \to t+\T/2 .        \label{S:G}
\end{align}
For the tight-binding model sketched in Fig.~\ref{fig:wire3}, the parity
operation \eqref{S:P} maps the lead states and the wire sites according to
\begin{equation}
\label{S:P.tight-bind}
\mathcal{S}_\mathrm{P} : (\mathrm{L}q,n) \leftrightarrow (\mathrm{R}q, N+1-n),
\end{equation}
where $n=1,\ldots,N$ labels the wire sites and $\mathrm{L}q$ ($\mathrm{R}q$) the states in
the left (right) lead.  Both the parity $\mathcal{S}_\mathrm{P}$ and the
time inversion $\mathcal{S}_\mathrm{T}$ can be generalized by an additional
shift of position and time, respectively.  Alternatively, one can place the
origin of the corresponding axis properly.  For convenience, we choose the
latter option.

It should be mentioned that for the periodic driving considered in this work,
the system contains a further symmetry, namely the time-translation by a full
driving period.  This has already been taken into account when deriving a
Floquet transport theory and cannot be exploited further.
\begin{figure}[t]
\centerline{\includegraphics[scale=.5]{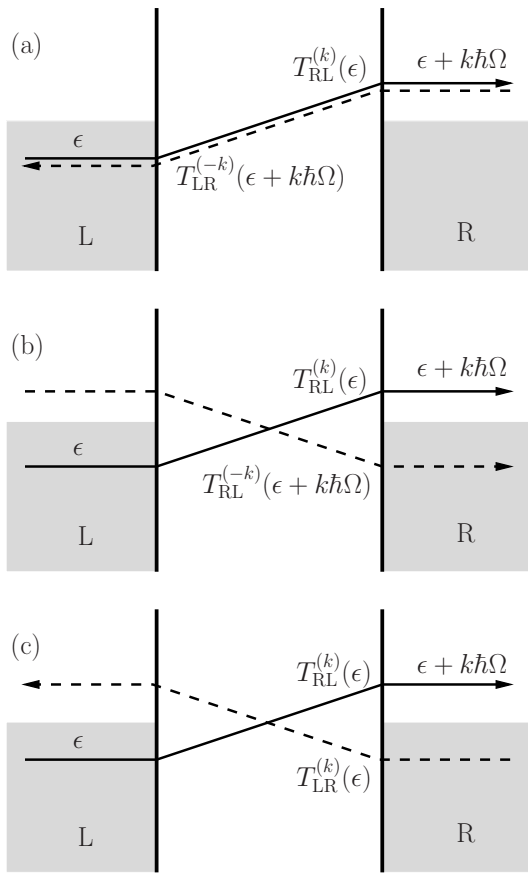}}
\caption{\label{fig:symmetry}
Transmission of an electron with energy $\epsilon$ under the absorption of
$k$ photons (solid line) and its symmetry related process (dashed) for (a)
time-reversal symmetry, (b) time-reversal parity, and (c) generalized
parity.  The sketched processes occur with equal probability.}
\end{figure}

\subsubsection{Time-reversal symmetry}

If the Hamiltonian obeys time-reversal symmetry $\mathcal{S}_\mathrm{T}$,
i.e., if $H(t)=H(-t)$, Eq.\ \eqref{symmetry} yields $\langle
1|U(t,t')|N\rangle=\langle N| U(-t',-t) |1\rangle$.  Inserting into the
definition of the Green function $G^{(k)}(\epsilon)$, Eqs.\
\eqref{G(t,epsilon)} and \eqref{G(k,epsilon)}, results in the relation
$G_{1N}^{(k)}(\epsilon) = G_{N1}^{(-k)}(\epsilon+k\hbar\Omega)$, where we
have shifted the limits of the $t$-integration using the relation
$G(t,\epsilon) = G(t+\T,\epsilon)$.  Thus, the transmission probabilities obey
\begin{equation}
\label{transmission:T}
T_{\mathrm{R}\mathrm{L}}^{(k)}(\epsilon) = T_{\mathrm{L}\mathrm{R}}^{(-k)}(\epsilon+k\hbar\Omega) ,
\end{equation}
i.e., the scattering processes sketched in Fig.~\ref{fig:symmetry}a occur with
equal probability.

A time-independent system in the absence of magnetic fields presents a
particular case of time-reversal symmetry since all transmissions probabilities with $k\neq
0$ vanish and, thus, $T_{\mathrm{R}\mathrm{L}}^{(0)}(\epsilon) = T_{\mathrm{L}\mathrm{R}}^{(0)}(\epsilon) =
T(\epsilon)$.

\subsubsection{Time-reversal parity}

Systems driven by a dipole force with purely harmonic time-dependence obey
the so-called time-reversal parity $\mathcal{S}_\mathrm{TP} \equiv
\mathcal{S}_\mathrm{T} \mathcal{S}_\mathrm{P}$, i.e., a combination of
time-reversal symmetry and parity.  This of course implies that the static
part of the Hamiltonian has to obey spatial parity which requires identical
wire-lead couplings, $\Gamma_\mathrm{L}(\epsilon) \equiv \Gamma_\mathrm{R}(\epsilon)$.  The
consequences for the Floquet states are discussed in the Appendix
\ref{app:symmetry} while here, we derive the consequences for the
transmission probabilities.

By the same reasoning as in the case of time-reversal symmetry discussed
above, but with additionally interchanging left and right, we find
$G_{1N}^{(k)}(\epsilon) = G_{1N}^{(-k)}(\epsilon+k\hbar\Omega)$ which yields
equal transmission probabilities for the scattering events sketched in Fig.\
\ref{fig:symmetry}b, i.e.
\begin{equation}
\label{transmission:TP}
T_{\mathrm{R}\mathrm{L}}^{(k)}(\epsilon) = T_{\mathrm{R}\mathrm{L}}^{(-k)}(\epsilon+k\hbar\Omega) .
\end{equation}
Interestingly, time-reversal parity relates two scattering events that both go
into the same direction.  Therefore, relation \eqref{transmission:TP} has no
obvious consequence for any dc current.  Still time-reversal parity entails an
intriguing and more hidden consequence for non-adiabatic electron pumping by
harmonic mixing as a function of the wire-lead coupling \cite{Lehmann2003b}.
We discuss this effect in the context of non-adiabatic electron pumping in
Section\ \ref{sec:mixing}.

\subsubsection{Generalized parity}

A further spatio-temporal symmetry that has an impact on the transmission
properties is the so-called generalized parity $\mathcal{S}_\mathrm{GP} =
\mathcal{S}_\mathrm{G} \mathcal{S}_\mathrm{P}$, i.e., a parity operation
combined with a time shift by half a driving period.  This symmetry also
explains qualitatively the quasienergy spectra found in the context of
driven quantum tunneling \cite{Grossmann1991a, Grossmann1991b, Peres1991a,
Holthaus1992a}.

If the wire-lead Hamiltonian is invariant under $\mathcal{S}_\mathrm{GP}$,
the time evolution operator obeys $\langle 1| U(t,t') |N\rangle = \langle
N| U(t+\T/2,t'+\T/2) |1\rangle$.  Inserting into Eq.~\eqref{G(t,epsilon)}
results in $G_{1N}^{(k)}(\epsilon) = G_{N1}^{(k)}(\epsilon)$ and, thus, the
scattering events sketched in Fig.~\ref{fig:symmetry}c obey
\begin{equation}
\label{transmission:GP}
T_{\mathrm{R}\mathrm{L}}^{(k)}(\epsilon) = T_{\mathrm{L}\mathrm{R}}^{(k)}(\epsilon) .
\end{equation}
Again, we have shifted the integration limits by using the time-periodicity of
the Green function $G(t,\epsilon)$.

\subsection{Approximations}

In Section~\ref{sec:IS}, expressions for the current and the noise power have
been derived for a periodic but otherwise arbitrary driving.
Within the wide-band limit, both quantities can be expressed in terms of
the solutions of the Floquet equation \eqref{Fs}, i.e., the solution of
a non-Hermitian eigenvalue problem in an extended Hilbert space.  Thus, for
large systems, the numerical computation of the Floquet states can be
rather costly.  Moreover, for finite temperatures, the energy integration
in the expressions \eqref{barI} and \eqref{barS} have to be performed
numerically.  Therefore, approximation schemes which allow a more efficient
computation are of much practical use.

\subsubsection{Weak-coupling limit}
\label{sec:approx.weakcoupling}

In the limit of a weak wire-lead coupling, i.e., for coupling constants
$\Gamma_\ell$ which are far lower than all other energy scales of the wire
Hamiltonian, it is possible to to derive within a master equation approach
a closed expression for the dc current \cite{Lehmann2003b}; cf.\
Section~\ref{sec:master}.  The corresponding approximation within the present
Floquet approach is based on treating the self-energy contribution
$-\i\Sigma$ in the non-Hermitian Floquet equation \eqref{Fs} as a
perturbation.  Then, the zeroth order of the Floquet equation
\begin{equation}
\label{Floquet0}
\Big(\mathcal{H}_\mathrm{wire}(t)-\i\hbar\frac{\d}{\d t}\Big)
|\phi_{\alpha}(t)\rangle
= \epsilon^0_{\alpha} |\phi_{\alpha}(t)\rangle ,
\end{equation}
describes the driven wire in the absence of the leads,
where $|\phi_\alpha(t)\rangle=\sum_k \exp(-\i k\Omega t) |\phi_{\alpha,k}
\rangle$ are the ``usual'' Floquet states with quasienergies
${\epsilon}_{\alpha}^0$.  In the absence of degeneracies the first
order correction to the quasienergies is $-\i\hbar\gamma_\alpha^1$ where
\begin{align}
\gamma_\alpha^1
={}& \frac{1}{\hbar}\int_0^{\mathcal{T}} \frac{\d t}{\mathcal{T}}
   \langle\phi_\alpha(t)| \Sigma |\phi_\alpha(t)\rangle
\\
={}& \frac{\Gamma_\mathrm{L}}{2\hbar} \sum_k |\langle 1|\phi_{\alpha,k}\rangle|^2
+  \frac{\Gamma_\mathrm{R}}{2\hbar} \sum_k |\langle N|\phi_{\alpha,k}\rangle|^2 .
\label{gamma_alpha}
\end{align}
Since the first order correction to the Floquet states
will contribute to neither the current nor the noise, the zeroth-order
contribution $|u_\alpha(t)\rangle = |u_\alpha^+(t)\rangle =
|\phi_\alpha(t)\rangle $ is already sufficient for the present purpose.
Consequently, the transmission probability \eqref{TLR} assumes the form
\cite{Camalet2004a}
\begin{equation}
\label{Tgamma0}
T^{(k)}_{\mathrm{L}\mathrm{R}}(\epsilon)
= \Gamma_\mathrm{L}\Gamma_\mathrm{R} \sum_{\alpha,\beta,k',k''}
\frac{\langle N|\phi_{\alpha,k'}\rangle\langle\phi_{\alpha,k'+k}|1\rangle
      \langle 1|\phi_{\beta,k''+k}\rangle\langle\phi_{\beta,k''}|N\rangle}
     {[\epsilon-(\epsilon_\alpha^0+k'\hbar\Omega+\i\hbar\gamma_\alpha^1)]
      [\epsilon-(\epsilon_\beta^0+k''\hbar\Omega-\i\hbar\gamma_\beta^1)]}
\end{equation}
and $T_{\mathrm{R}\mathrm{L}}^{(k)}(\epsilon)$ accordingly.  The transmission probability
\eqref{Tgamma0} exhibits for small values of $\Gamma_\ell$ sharp peaks at
energies $\epsilon^0_\alpha+k'\hbar\Omega$ and
$\epsilon^0_\beta+k''\hbar\Omega$ with widths $\hbar\gamma_\alpha^1$ and
$\hbar\gamma_\beta^1$.  Therefore, the relevant contributions to the sum
come from terms for which the peaks of both factors coincide and, in the
absence of degeneracies in the quasienergy spectrum, we keep only terms
with
\begin{equation}
\alpha=\beta,\quad k'=k'' .
\end{equation}
Then provided that $\gamma_\alpha^1$ is small, the fraction in
\eqref{Tweak} is a Lorentzian and can be approximated by
$\pi\delta(\epsilon-\epsilon_\alpha^0-k'\hbar\Omega)/\hbar\gamma_\alpha^1$
yielding the transmission probability
\begin{align}
\label{Tweak}
T^{(k)}_{\mathrm{L}\mathrm{R}}(\epsilon)
={} & \Gamma_\mathrm{L}\Gamma_\mathrm{R} \sum_{\alpha,k'}
  \frac{\pi}{\hbar\gamma_\alpha^1}
  |\langle 1|\phi_{\alpha,k'+k}\rangle\langle\phi_{\alpha,k'}|N\rangle|^2
  \delta(\epsilon-\epsilon_\alpha^0+k'\hbar\Omega)
\\
={} & T_{\mathrm{R}\mathrm{L}}^{(-k)}(\epsilon+k\hbar\Omega) .
\label{symmetry.weak}
\end{align}
The last line which follows by substituting $k'\to k'-k$, means that the
transmission probabilities in the weak-coupling limit obey the same
relation as in the case of time-reversal symmetry, cf. \
Eq.~\eqref{transmission:T} even in the absence of any symmetry.

The energy integration in \eqref{barI} can now be performed even for finite
temperature and we obtain for the dc current the expression
\begin{equation}
\label{I:RWA}
\bar I
= \frac{e}{\hbar}\sum_{\alpha,k,k'}
    \frac{\Gamma_{\mathrm{L}\alpha k} \Gamma_{\mathrm{R}\alpha k'}}
     {\Gamma_{\mathrm{L}\alpha}+\Gamma_{\mathrm{R}\alpha}}
   \big[ f_\mathrm{R}(\epsilon_{\alpha}^0+k'\hbar\Omega)-
                f_\mathrm{L}(\epsilon_{\alpha}^0+k\hbar\Omega)\big] .
\end{equation}
The coefficients
\begin{align}
\label{GammaLalpha}
\Gamma_{\mathrm{L}\alpha k} =& \Gamma_\mathrm{L}|\langle 1|\phi_{\alpha,k}\rangle|^2, &
 \Gamma_{\mathrm{L}\alpha} =& \sum_k \Gamma_{\mathrm{L}\alpha k} \,, \\
\label{GammaRalpha}
\Gamma_{\mathrm{R}\alpha k} =& \Gamma_\mathrm{R}|\langle N|\phi_{\alpha,k}\rangle|^2 , &
 \Gamma_{\mathrm{R}\alpha} =& \sum_k \Gamma_{\mathrm{R}\alpha k} \,,
\end{align}
denote the overlap of the $k$th sideband $|\phi_{\alpha,k}\rangle$ of the
Floquet state $|\phi_\alpha(t)\rangle$ with the first site and the last
site of the wire, respectively.  We have used $2\hbar\gamma_\alpha^1 =
\Gamma_{\mathrm{L}\alpha} + \Gamma_{\mathrm{R}\alpha}$ which follows from
\eqref{gamma_alpha}.  Expression \eqref{I:RWA} can been derived also within
a rotating-wave approximation of a Floquet master equation approach
\cite{Lehmann2003b}; cf. Sect.~\ref{sec:rwa}.

Within the same approximation, we expand the zero-frequency noise
\eqref{barS} to lowest-order in $\Gamma_{\ell}$: After inserting the
spectral representation \eqref{G} of the Green function, we again keep only
terms with identical Floquet index $\alpha$ and identical sideband index
$k$ to obtain
\begin{equation}
\label{S:RWA}
\begin{split}
\bar S
=  \frac{e^2}{\hbar} \sum_{\alpha,k,k'}
\frac{\Gamma_{\mathrm{R}\alpha k'}\bar f_\mathrm{R}(\epsilon_\alpha^0{+}k'\hbar\Omega)}
     {(\Gamma_{\mathrm{L}\alpha}+\Gamma_{\mathrm{R}\alpha})^3}
\big\{ &
  2\Gamma_{\mathrm{L}\alpha}^2 \Gamma_{\mathrm{R}\alpha k}f_\mathrm{R}(\epsilon_\alpha^0{+}k\hbar\Omega)
\\
+& (\Gamma_{\mathrm{L}\alpha}^2 + \Gamma_{\mathrm{R}\alpha}^2)
   \Gamma_{\mathrm{L}\alpha k} f_\mathrm{L}(\epsilon_\alpha^0 {+} k\hbar\Omega)
\big\}
\\
& \hspace{-20ex}+\text{same terms with the replacement $\mathrm{L} \leftrightarrow \mathrm{R}$} .
\end{split}
\end{equation}

Of particular interest for the comparison to the static situation is the
limit of a large applied voltage such that practically $f_\mathrm{R}=1$ and $f_\mathrm{L}=0$.
Then, in Eqs.\ \eqref{I:RWA} and \eqref{S:RWA}, the sums over the
sideband indices $k$ can be carried out such that
\begin{align}
\label{IinfG0}
\bar I_\infty
=& \frac{e}{\hbar} \sum_{\alpha}
\frac{\Gamma_{\mathrm{L}\alpha} \Gamma_{\mathrm{R}\alpha}}{\Gamma_{\mathrm{L}\alpha}+\Gamma_{\mathrm{R}\alpha}}
, \\
\label{SinfG0}
\bar S_\infty
=& \frac{e^2}{\hbar} \sum_{\alpha}
   \frac{\Gamma_{\mathrm{L}\alpha}\Gamma_{\mathrm{R}\alpha}(\Gamma_{\mathrm{L}\alpha}^2
   +\Gamma_{\mathrm{R}\alpha}^2)}{(\Gamma_{\mathrm{L}\alpha}+\Gamma_{\mathrm{R}\alpha})^3} .
\end{align}
These expressions resemble the corresponding expressions for the transport
across a \textit{static} double barrier \cite{Blanter2000a}.  If now
$\Gamma_{\mathrm{L}\alpha}=\Gamma_{\mathrm{R}\alpha}$ for all Floquet states
$|\phi_\alpha(t)\rangle$, we find $F=1/2$.  This is in particular the case
for systems obeying reflection symmetry.  In the presence of such
symmetries, however, the existence of exact crossings, i.e.\ degeneracies,
limits the applicability of the weak-coupling approximation and a
master equation approach (cf.\ Sect.~\ref{sec:master}) is more appropriate.

\subsubsection{High-frequency limit}
\label{sec:approx.hf}

Many effects occurring in driven quantum systems, such as coherent
destruction of tunneling \cite{Grossmann1991a} or current and noise control
\cite{Lehmann2003a, Camalet2003a}, are most pronounced for a large excitation
frequency $\Omega$.  Thus, it is particularly interesting to derive for the
present Floquet approach an expansion in terms of $1/\Omega$.  Thereby, the
driven system will be approximated by a static system with renormalized
parameters.  Such a perturbation scheme has been developed for two-level
systems in Ref.~\cite{Shirley1965a} and applied to driven tunneling
in bistable systems \cite{Grossmann1991b} and superlattices
\cite{Holthaus1992a}.  For open quantum system, the coupling to the
external degrees of freedom (e.g., the leads or a heat bath) bears
additional complications that have been solved heuristically in
Ref.~\cite{Kohler2004a} by replacing the Fermi functions by
effective electron distributions.  In the following, we present a rigorous
derivation of this approach based on a perturbation theory for the Floquet
equation~\eqref{Fs}.

We assume a driving that leaves all off-diagonal matrix elements of the
wire Hamiltonian time-independent while the tight-binding levels undergo a
position-dependent, time-periodic driving $f_n(t)=f_n(t+\T)$ with zero
time-average.  Then, the wire Hamiltonian is of the form
\begin{equation}
\label{hf:dipole}
\mathcal{H}_\mathrm{wire}(t)
=\mathcal{H}_0 + \sum_n f_n(t)\, |n \rangle \langle n| .
\end{equation}
If $\hbar\Omega$ represents the largest energy scale of the problem, we can
in the Floquet equation \eqref{Fs} treat the \textit{static} part of the
Hamiltonian as a perturbation.  Correspondingly, the eigenfunctions of the
operator $\sum_n f_n(t)|n\rangle\langle n| -\i\hbar \d/\d t$ determine the
zeroth order Floquet states
\begin{equation}
\e^{-\i F_n (t)}|n\rangle .
\label{zthorder}
\end{equation}
We have defined the accumulated phase
\begin{equation}
\label{Fn}
F_n(t) = \frac{1}{\hbar} \int_0^t \d t'\, f_n(t') = F_n(t+\T),
\end{equation}
which is $\T$-periodic due to the zero time-average of $f_n(t)$.  As a
consequence of this periodicity, to zeroth order the quasienergies are zero
($\mathrm{mod}\,\hbar\Omega$) and the Floquet spectrum is
given by multiples of the photon energy, $k\hbar\Omega$.  Each
$k=0,\pm1,\pm2,\ldots$ defines a degenerate subspace of the extended
Hilbert space.  If now $\hbar\Omega$ is larger than all other energy
scales, the first order correction to the Floquet states and the
quasienergies can be calculated by diagonalizing the perturbation in the
subspace defined by $k=0$.  Thus, we have to solve the time-independent
eigenvalue equation
\begin{equation}
(\mathcal{H}_\mathrm{eff}-\i\Sigma)|\alpha\rangle
= (\epsilon^{1}_{\alpha}-\i\hbar \gamma^{1}_{\alpha}) |\alpha\rangle .
\label{Hzthorder}
\end{equation}
The static effective Hamiltonian $\mathcal{H}_\mathrm{eff}$ is
defined by the matrix elements of the original static Hamiltonian
$\mathcal{H}_0$ with the zeroth order Floquet states \eqref{zthorder},
\begin{equation}
\label{hf:Heff}
(\mathcal{H}_\mathrm{eff})_{nn'}
= \int_0^\T \frac{\d t}{\T}
  \e^{\i F_n(t)} (\mathcal{H}_0)_{nn'} \e^{-\i F_{n'}(t)} .
\end{equation}
The $t$-integration constitutes the inner product in the Hilbert space
extended by a periodic time coordinate \cite{Sambe1973a} (for
details, see Appendix~\ref{sec:sambe}).
To first order in $1/\Omega$, the quasienergies $\epsilon^1_\alpha-\i\hbar
\gamma^1_\alpha$ are given by the eigenvalues of the static equation
\eqref{Hzthorder} and, consequently, the corresponding Floquet states read
\begin{equation}
\label{hf:states}
|u_\alpha(t)\rangle = \sum_n \e^{-\i F_n(t)}|n\rangle\langle n|\alpha\rangle .
\end{equation}
The fact that all $F_n(t)$ are $\T$-periodic, allows one to write in
\eqref{hf:states} the time-dependent phase factor as a Fourier series,
\begin{equation}
\label{a:hf}
\e^{-\i F_n(t)} = \sum_k a_{n,k}\,\e^{-\i k\Omega t} .
\end{equation}
Thus, $\langle n|u_{\alpha,k}\rangle = a_{n,k}\langle n|\alpha\rangle$ and
the Green function for the high-frequency driving reads
\begin{equation}
\label{UlO}
G_{nn'}^{(k)}(\epsilon) = \sum_{k'} a_{n,k'+k} a_{n',k'}^*
G_{nn'}^\mathrm{eff}(\epsilon-k'\hbar\Omega) ,
\end{equation}
where $G^\mathrm{eff}(\epsilon)$ denotes the Green function corresponding
to the static Hamiltonian $\mathcal{H}_\mathrm{eff}$ with the self-energy
$\Sigma$.  Finally, substituting $\epsilon\to\epsilon+k'\hbar\Omega$ and
using the sum rule $\sum_{k'} a_{n,k+k'} a_{n,k'}^* = \delta_{k,0}$, we obtain
\begin{equation}
\bar I = \frac{e}{h} \int \d\epsilon \; T_\mathrm{eff}(\epsilon)
\big\{ f_{\mathrm{R},\mathrm{eff}}(\epsilon) - f_{\mathrm{L},\mathrm{eff}}(\epsilon) \big\} .
\label{IlO}
\end{equation}
The effective transmission probability $T_\mathrm{eff}(\epsilon) = \Gamma_\mathrm{L} \Gamma_\mathrm{R}
|G_{1N}^\mathrm{eff}(\epsilon)|^2$ is computed from the effective
Hamiltonian \eqref{hf:Heff}; the electron distribution is given by
\begin{align}
f_{\mathrm{L},\mathrm{eff}}(\epsilon)
= \sum_k |a_{1,k}|^2 f_\mathrm{L}(\epsilon+k\hbar\Omega)
\label{hf:feff}
\end{align}
and $f_{\mathrm{R},\mathrm{eff}}$ follows from the replacement $(1,\mathrm{L})\to(N,\mathrm{R})$.

In order to derive a high-frequency approximation for the zero-frequency
noise $\bar S$, we insert \eqref{UlO} into \eqref{barS} and neglect
products of the type $G^\mathrm{eff}(\epsilon-k\hbar\Omega)\,
G^\mathrm{eff}(\epsilon-k'\hbar\Omega)$ for $k\neq k'$.  Employing the
above sum rule for the Fourier coefficients $a_{n,k}$, we obtain for the
noise the static expression \eqref{noise.static}, but with the transmission probability
$T(\epsilon)$ and the Fermi functions $f_{\mathrm{R},\mathrm{L}}(\epsilon)$ replaced by the
effective transmission probability $T_\mathrm{eff}(\epsilon)$ and the effective
distribution function \eqref{hf:feff}, respectively.

Note that in general, $a_{1,k}\neq a_{N,k}$ such that $f_{\mathrm{R},\mathrm{eff}}
\neq f_{\mathrm{L},\mathrm{eff}}$.  This means that the driving can create an
effective bias and thereby create a non-adiabatic pump current.
Moreover, if all $F_n$ are identical, the phase factors in \eqref{hf:Heff}
cancel each other and the effective Hamiltonian $\mathcal{H}_\mathrm{eff}$
equals the original static Hamiltonian.

\subsubsection{Linear-response limit}

For small driving amplitudes, it is often sufficient to treat the driving
in the linear-response limit \cite{Keller2002a}.  In doing so, we denote by
$g(t-t')$ the undriven limit of the Green function $G(t,t')$ and by
$\mathcal{H}_1(t)$ the time-dependent part of the Hamiltonian which is
considered as a perturbation.  Then, a formal solution of Eq.~\eqref{G.def}
is given by the Dyson equation
\begin{equation}
\label{lr:Dyson}
G(t,t-\tau) = g(\tau)
+ \int_{-\infty}^{+\infty} \d t'\, g(t-t')\, \mathcal{H}_1(t')\, G(t',t-\tau) ,
\end{equation}
as can be shown by inserting \eqref{lr:Dyson} into \eqref{G.def}.
A self-consistent solution of this equation has been presented by
Brandes \cite{Brandes1997a}.  Here, we restrict ourselves to the lowest
order in the driving and, thus, can replace in the integral $G(t',t-\tau)$ by
$g(t'-t+\tau)$.  Inserting moreover the Fourier representations
\begin{align}
\mathcal{H}_1(t)
&= \int\frac{\d\omega}{2\pi} \e^{-\i\omega t} \mathcal{H}_1(\omega) ,
\\
g(t) &= \int\frac{\d\epsilon}{2\pi\hbar} \e^{-\i\epsilon t/\hbar}
        g(\epsilon),
\end{align}
and Eq.~\eqref{G(t,epsilon)}, we obtain
\begin{equation}
G(t,\epsilon) = g(\epsilon) + \int\frac{\d\omega}{2\pi}\e^{-\i\omega t}
g(\epsilon+\hbar\omega)\, \mathcal{H}_1(\omega)\, g(\epsilon) .
\end{equation}

For purely harmonic driving, $\mathcal{H}_1(t) = \mathcal{H}_1 \cos(\Omega
t)$, one finds for the Fourier coefficients \eqref{G(k,epsilon)} of the
Green function the expressions
\begin{align}
G^{(0)}(\epsilon) &= g(\epsilon) ,
\\
G^{(\pm 1)}(\epsilon) &= \frac{1}{2} g(\epsilon\pm\hbar\Omega)\,
\mathcal{H}_1\, g(\epsilon) ,
\end{align}
while all Fourier components $G^{(k)}$ with $|k|>1$, vanish to linear order.
Consequently, the elastic transmission probability $T^{(0)}(\epsilon)$
is independent of the driving, i.e.\ it equals the result
in the absence of external driving.  The transmission probabilities under
emission/absorption of a single photon are, however, proportional to the
intensity of the driving field, i.e.\ $\propto |\mathcal{H}_1| ^2 $, and read
\begin{equation}
T_{\mathrm{L}\mathrm{R}}^{(\pm 1)}(\epsilon)
= \Gamma_\mathrm{L}(\epsilon \pm \hbar\Omega)\, \Gamma_\mathrm{R}(\epsilon)\,
  \big| \langle 1| g(\epsilon \pm \hbar\Omega)\, \mathcal{H}_1\, g(\epsilon)
  |N\rangle \big|^2 \;.
\end{equation}
$T_{\mathrm{R}\mathrm{L}}^{(\pm 1)}(\epsilon)$ follows from the replacement $(\mathrm{L},1)
\leftrightarrow (\mathrm{R},N)$.

\subsection{Special cases}

In some special cases, the results of our Floquet approach reduce to
simpler expressions.  In particular, this is the case for zero driving
amplitude, i.e.\ in the absence of driving, and for a driving that results
from a time-dependent gate voltage and, thus, is homogeneous along the
wire.

\subsubsection{Static conductor and adiabatic limit}
\label{sec:static}

For consistency, the expressions \eqref{barI} and \eqref{barI} for the
dc current and the zero-frequency noise, respectively, must coincide in the
undriven limit with the corresponding expressions of the time-independent
scattering theory, Eqs.~\eqref{current.static} and \eqref{noise.static},
respectively.  This is indeed the case because the
static situation is characterized by two relations: First, in the absence
of spin-dependent interactions, we have time-reversal symmetry and
therefore $T_{\mathrm{L}\mathrm{R}}(\epsilon) = T_{\mathrm{R}\mathrm{L}}(\epsilon)$.  Second, all sidebands with
$k\neq 0$ vanish, i.e., $T^{(k)}_{\mathrm{R}\mathrm{L}}(\epsilon) = T^{(k)}_{\mathrm{L}\mathrm{R}}(\epsilon) =
\delta_{k,0}T(\epsilon)$, where
\begin{equation}
\label{Tstat}
T(\epsilon) = \Gamma_\mathrm{L}(\epsilon)\, \Gamma_\mathrm{R}(\epsilon)\, |G_{1N}(\epsilon)|^2
\end{equation}
and $G(\epsilon)$ is the Green function in the absence of driving.  Then
the current assumes the known form \eqref{current.static}.
Moreover in a static situation, the matrix element $\langle
1|\ldots|1\rangle$ of Eq.~\eqref{datta} reads \cite{Datta1995a}
\begin{equation}
\label{datta.static}
| \Gamma_\mathrm{L}(\epsilon) G_{11}(\epsilon) + \i |^2 = 1-T(\epsilon) .
\end{equation}
This relation allows one to eliminate the backscattering terms in the second
line of Eq.~\eqref{barS} such that the zero-frequency noise becomes
\eqref{noise.static}.  Obviously if in a static situation both voltage and
temperature are zero, not only the current \eqref{current.static} but also
the noise \eqref{noise.static} vanishes.  In the presence of driving, this
is no longer the case.  This becomes particularly evident in the
high-frequency limit studied in Section~\ref{sec:approx.hf}.

It is known that in the adiabatic limit, i.e., for small driving
frequencies, the numerical solution of the Floquet equation \eqref{Fs}
becomes infeasible because a diverging number of sidebands has to be taken
into account.  In more mathematical terms, Floquet theory has no proper
limit as $\Omega\to 0$ \cite{Hone1997a}.  The practical consequence of this
is that for low driving frequencies, it is favorable to tackle the
transport problem with a different strategy: If $\hbar\Omega$ is the
smallest energy-scale of the Hamiltonian \eqref{H.wirelead}, one computes
for the ``frozen'' Hamiltonian at each instance of time the current and the
noise from the static expressions \eqref{barI} and \eqref{barS} being
followed up by time-averaging.

\subsubsection{Spatially homogeneous driving}
\label{sec:special.homogeneous}

In many experimental situations, the driving field acts as a time-dependent
gate voltage, i.e., it merely shifts all on-site energies of the wire
uniformly.  Thus, the wire Hamiltonian is of the form
\begin{equation}
\mathcal{H}_\mathrm{wire}(t)=\mathcal{H}_0 + f(t)\sum_n |n \rangle \langle n| ,
\end{equation}
where, without loss of generality, we restrict $f(t)$ to possess zero
time-average.  A particular case of such a homogeneous driving is realized
with a system that consists of only one level
\cite{Wingreen1990a,Kislov1991a,Aguado1996a}.  Then trivially, the time and
the position dependence of the Floquet states factorize and, therefore, the
dc current can be obtained
within the formalism introduced by Tien and Gordon \cite{Tien1963a}.  The
corresponding noise properties have been addressed by Tucker and Feldman
\cite{Tucker1979a,Tucker1985a}.  Here, we establish the relation between
such a treatment and the present Floquet approach.

Since the time-dependent part of the Hamiltonian is proportional to the
unity operator, the solution of the Floquet equation \eqref{Fs} is,
besides a phase factor, given by the eigenfunctions $|\alpha\rangle$ of
the time-independent operator $\mathcal{H}_0-\i\Sigma$,
\begin{equation}
\label{states:tucker}
|u_\alpha(t)\rangle = e^{-\i F(t)} |\alpha\rangle ,
\end{equation}
where $(\mathcal{H}_0-\i\Sigma)|\alpha\rangle =
(\epsilon_\alpha-\i\hbar\gamma_\alpha)|\alpha\rangle$ and
\begin{equation}
F(t) = \frac{1}{\hbar} \int_0^t \d t'\, f(t') .
\end{equation}
The quasienergies $(\epsilon_\alpha-\i\hbar\gamma_\alpha)$ coincide with the
eigenvalues of the static eigenvalue problem.  Note that $F(t)$ obeys the
$\T$-periodicity of the driving field since the time-average of $f(t)$
vanishes.  Thus, the phase factor in the Floquet states
\eqref{states:tucker} can be written as a Fourier series,
\begin{equation}
\e^{-\i F(t)} = \sum_k a_k\, \e^{-\i k\Omega t}
\end{equation}
and, consequently we find $|u_{\alpha,k}\rangle=a_k|\alpha\rangle$ and the
adjoint states accordingly.
Then, the Green function \eqref{G(k,epsilon)} becomes
\begin{equation}
\label{GTucker}
G^{(k)}(\epsilon)
= \sum_{k'}  a_{k'+k}\, a_{k'}^*\, G(\epsilon-k'\hbar\Omega) ,
\end{equation}
where $G(\epsilon)$ denotes the Green function in the absence of the
driving field.  Inserting \eqref{GTucker} into \eqref{barI} and
employing the sum rule $\sum_{k'} a_{k'}^* a_{k'+k}=\delta_{k,0}$, yields
\begin{equation}
\bar I = \sum_k |a_k|^2
\frac{e}{h} \int\! \d\epsilon \, T (\epsilon-k\hbar\Omega)
 [ f_{\mathrm{R}}(\epsilon) - f_{\mathrm{L}}(\epsilon) ] ,
\label{ITucker}
\end{equation}
where $T(\epsilon)$ is the transmission probability in the absence of the driving.
This expression allows the interpretation, that for homogeneous driving,
the Floquet channels contribute \textit{independently} to the current
${\bar I}$.  For the special case of a one-site conductor and a sinusoidal
driving, this relation to the static situation has been discussed in
Refs.~\cite{Wingreen1990a,Kislov1991a}.

Addressing the noise properties, we obtain by inserting the Green function
\eqref{GTucker} into \eqref{barS} the expression
\begin{equation}
\begin{split}
\label{STucker}
\lefteqn{
\bar S
=  \frac{e^2}{h} \sum_k \int\! \d\epsilon\, \Big\{
   \Big| \sum_{k'} a_{k'+k}^* a_{k'} T(\epsilon-k'\hbar\Omega) \Big|^2
   f_\mathrm{R}(\epsilon) \bar f_\mathrm{R}(\epsilon+k\hbar\Omega)
}
\\
&+  \Gamma_\mathrm{L} \Gamma_\mathrm{R}
   \Big| \sum_{k'} a_{k'+k}^* a_{k'} G_{1N}(\epsilon-k'\hbar\Omega)
   \big[ \Gamma_\mathrm{L} G_{11}^*(\epsilon-k'\hbar\Omega) - \i \big] \Big|^2
   f_\mathrm{L}(\epsilon) \bar f_\mathrm{R}(\epsilon+k\hbar\Omega)
\\
&+  \,\text{same terms with the replacement $(\mathrm{L},1) \leftrightarrow (\mathrm{R},N)$}
\Big\} .
\end{split}
\end{equation}
While the term in the first line contains only the static transmission probability at
energies shifted by multiples of the photon energies, the contribution in
the second line cannot be brought into such a convenient form.  The reason
for this is that the sum over $k'$ inhibits the application of relation
\eqref{datta}.  As a consequence, in clear contrast to the dc current,
the zero-frequency noise cannot be interpreted in terms of independent
Floquet channels.
Only in the limit of large driving frequencies (cf.\ Section
\ref{sec:approx.hf}), the channels become effectively independent and we end
up with an expression that depends only on the transmission probability in the
absence of the driving, and the Fourier coefficients $a_k$.

For large voltages where $f_\mathrm{L}=0$ and $f_\mathrm{R}=1$, the sums over the Fourier
coefficients in Eqs.\ \eqref{ITucker} and \eqref{STucker} can be
evaluated with the help of the sum rule $\sum_{k'} a_{k'}^*
a_{k'+k}=\delta_{k,0}$.  Then both the dc current and the zero-frequency
noise become identical to their value in the absence of the driving.  This
means that for a transport voltage which is sufficiently large, a
time-dependent gate voltage has no influence on the average current and the
zero-frequency noise.

\section{Master equation approach}
\label{sec:master}

An essential step in the derivation of the transmission within a
weak-coupling approximation, Eq.~\eqref{Tweak}, is the assumption that only
terms with $\alpha=\beta$ and $k=k'$ contribute significantly to
\eqref{Tgamma0}.  As discussed after Eq.~\eqref{Tweak}, this requires that
the separation of any pair of resonances is larger than their widths.  This
condition can be fulfilled only if the quasienergy spectrum does not
contain any degeneracies and if, in addition, the wire-lead coupling is
very weak.  Here, we refine the weak-coupling approximation scheme of
Section~\ref{sec:approx.weakcoupling} and derive a master equation approach
which yields reliable results also in the presence of degeneracies and for
intermediately strong wire-lead coupling \cite{Lehmann2002b, Lehmann2003b}.

\subsection{Current formula}

We start again from the asymmetric expression~\eqref{I.operator.xi} for the
time-dependent current through the left contact.  After averaging over the
driving period, we obtain the dc current
\begin{equation}
\label{Imeanunsym}
\begin{split}
\bar I
={} &  \frac{2e}{h\T} \sum_\ell \int\d\epsilon \int_0^\infty \!\!\d\tau
\int_0^\T \!\!\d t\,
\Gamma_\ell(\epsilon) f_\ell(\epsilon)
   \Im
   \e^{\i\epsilon\tau/\hbar}
   G_{1\ell}^*(t,\epsilon)\, \Gamma_\mathrm{L}(\tau)\, G_{1\ell}(t-\tau,\epsilon)
\\
& + 2 e\int\d\epsilon\,
     \Gamma_\mathrm{L}(\epsilon) f_\mathrm{L}(\epsilon)
     \Im G^{(0)}_{11}(\epsilon) ,
\end{split}
\end{equation}
for which we shall derive an approximation for small wire-lead coupling.

We start with the second term which is linear in the retarded Green
function $G^{(0)}_{11}(\epsilon)$.  For small values of $\Gamma$, we obtain
from \eqref{G} the approximation
\begin{equation}
\Im G^{(0)}(\epsilon) = 2\pi \sum_{\alpha,k} |\phi_{\alpha,k}\rangle
\langle\phi_{\alpha,k}|\, \delta(\epsilon-\epsilon_\alpha-k\hbar\Omega)
\end{equation}
which allows one to perform the energy integration in Eq.~\eqref{Imeanunsym}.
Then, we obtain the contribution
\begin{equation}
  -
  \frac{e}{\hbar}
  \sum_{\alpha,k}
  |\langle 1|\phi_{\alpha,k}\rangle|^2\,
  \Gamma_\mathrm{L}(\epsilon^0_{\alpha} + k \hbar\Omega)\,
 f(\epsilon^0_{\alpha} + k \hbar \Omega-\mu_\mathrm{L}) .
\end{equation}

The first term in Eq.~\eqref{Imeanunsym} is quadratic in the Green function
and, thus, requires a more elaborate treatment since otherwise, squares of
$\delta$-functions would emerge (cf.\ also the discussion in
Section~\ref{sec:approx.weakcoupling}).  For that purpose, it is
advantageous to go one step back and to use instead of the current
formula~\eqref{Imeanunsym} the current operator \eqref{I.operator} as a
starting point.  The time-average of the expectation value corresponding to
the first term of Eq.~\eqref{Imeanunsym} reads
\begin{equation}
  \frac{2e}{\hbar\T} \int_0^\infty \!\!\d\tau
\int_0^\T \!\!\d t\,
\Re \left[
  \Gamma_\mathrm{L}(\tau)\,
  \langle
  c^\dagger_1(t) c_1(t-\tau)
  \rangle
\right] .
\label{Ifirstterm}
\end{equation}
Assuming that $\Gamma_\mathrm{L}(\epsilon)$ is a slowly varying function
in the relevant energy range, we can replace the
time-evolution of $c_1$ from the time $t$ back to $t-\tau$ by the
Heisenberg operator $\tilde c_1(t-\tau,t) = U_0^\dagger(t-\tau,t) c_1
U_0(t-\tau,t)$ with $U_0$ being the propagator \eqref{Gtt} in the limit
$\Gamma_\mathrm{L/R}\to 0$.  This means that $\tilde c_1(t-\tau,t)$
represents the limit $\Gamma_\mathrm{L/R}\to 0$ of $c_1(t-\tau)$.
In order to include the coherent dynamics properly, it is convenient to
introduce the ``Floquet picture creation operators'' $c_\alpha(t)$ which
are defined by the time-dependent transformation~\cite{Lehmann2002b,
Lehmann2003b}
\begin{equation}
  c_\alpha(t) = \sum_n \langle\phi_\alpha(t)|n\rangle\, c_n\,.
  \label{c_alpha}
\end{equation}
Using the inverse transformation
$
c_n = \sum_{\alpha} \langle n|\phi_\alpha(t)\rangle\,c_\alpha(t)
$, which follows from the completeness of the Floquet states at equal times,
we obtain
\begin{equation}
    c_n(t-\tau,t) \approx
  \sum_{\alpha,k}
  \e^{-\i k\Omega t}\,
  \e^{\i(\epsilon^{0}_{\alpha} + k\hbar\Omega)\tau/\hbar} \langle n|\phi_{\alpha,k}\rangle\,
  c_{\alpha}(t) .
\label{cinter}
\end{equation}
Inserting \eqref{cinter} with $n=1$ into \eqref{Ifirstterm}, we arrive at an
expression that contains the time-dependent expectation values
$P_{\alpha\beta}(t) = \langle c^\dagger_\beta(t)\, c_\alpha(t)\rangle_t$
with both operators taken at time $t$.  The $P_{\alpha\beta}(t)$ at
asymptotic times, in turn, are determined from a kinetic equation which we
derive in the next subsection.  Before doing so, however, we simplify
Eq.~\eqref{Ifirstterm} further by using of the fact, that at asymptotically
long times, all $P_{\alpha\beta}(t)$ become $\T$-periodic functions and,
thus, can be decomposed into a Fourier series $P_{\alpha\beta}(t) = \sum_k
\exp(-\i k \Omega t) P_{\alpha\beta,k}$.  This brings
Eq.~\eqref{Ifirstterm} into the form
\begin{equation}
  \frac{2e}{\hbar}
  \sum_{\alpha,\beta,k,k'}
  \int_0^\infty \!\!\d\tau\,
  \Re \left[
  \Gamma_\mathrm{L}(\tau)\,
  \e^{\i(\epsilon^0_\alpha + k\hbar\Omega)\tau/\hbar}
  \langle \phi_{\beta,k+k'}|1\rangle\langle1|\phi_{\alpha,k}\rangle
  P_{\alpha\beta,k'}
 \right].
\end{equation}
By inserting for the lead response function $\Gamma(\tau)$ its
definition~\eqref{Gammat}, we finally find for the time-averaged current
through the wire the expression
\begin{equation}
  \begin{split}
    \bar I =
    \frac{e}{\hbar}
    \sum_{\alpha,k}
      \Gamma_\mathrm{L}(\epsilon^0_\alpha + k\hbar\Omega)
      \Big[
      &
      \sum_{\beta,k'}
    \Re
    \left\{
      \langle \phi_{\beta,k+k'}|1\rangle\langle1|\phi_{\alpha,k}\rangle
      P_{\alpha\beta,k'}
    \right\}\\
  &
  -
  |\langle 1|\phi_{\alpha,k}\rangle|^2\,
  f(\epsilon^0_{\alpha} + k \hbar \Omega-\mu_\mathrm{L})
  \Big]\,.
  \end{split}
  \label{Imastereq}
\end{equation}
Note that we have disregarded the principal value terms which correspond to
an energy-renormalization due to the wire-lead coupling.

\subsection{Floquet-Markov master equation}

Having expressed the current in terms of the wire expectation values
$P_{\alpha\beta}(t)$, we now derive for them an equation of motion valid in
the regime of weak to moderately strong wire-lead coupling.  We thus
consider the time-derivative $\dot P_{\alpha\beta}(t)$, which with the help
of the zeroth-order Floquet equation~\eqref{Floquet0}, can be written as
\begin{equation}
  \dot P_{\alpha\beta}(t) =
  - \frac{\i}{\hbar} (\epsilon^0_\alpha - \epsilon^0_\beta)\,
  P_{\alpha\beta}(t)
  + \Tr \left[ \dot\rho(t) \, c^\dagger_\beta(t) \, c_\alpha(t) \right].
\label{dotPalphabeta}
\end{equation}
For the evaluation of the second term on the right-hand side of the
last equation, we employ the standard master equation~\eqref{mastereq}
presented in Section \ref{sec:intro.master}.  Using twice the relation $\Tr
A[B,C] = \Tr [A,B]C$, which directly results from the cyclic
invariance of the trace, we obtain
\begin{equation}
  \begin{split}
  \dot P_{\alpha\beta}(t) = {} &
  - \frac{\i}{\hbar} (\epsilon^0_\alpha - \epsilon^0_\beta)\,
  P_{\alpha\beta}(t)\\
  &
  - \frac{1}{\hbar^2}
  \int_0^\infty\d\tau\,
  \left\langle
    [[c^\dagger_\beta(t) \, c_\alpha(t),
    H_\mathrm{contacts}], \widetilde H_\mathrm{contacts}(t-\tau,t)]
  \right\rangle_{t}\,.
  \end{split}
\label{dotPalphabeta'}
\end{equation}
For the further evaluation of Eq.\ \eqref{dotPalphabeta'}, we write both
$H_\mathrm{contacts}$ and $\widetilde H_\mathrm{contacts}(t-\tau,t)$ in
terms of $\tilde c_n(t-\tau,t)$ for which we insert the
approximation~\eqref{cinter}.  After some algebra, we arrive at a closed
differential equation for $P_{\alpha\beta}(t)$.  This determines the
Fourier coefficients of the asymptotic solution, $P_{\alpha\beta,k}$, which
obey the inhomogeneous set of equations
\begin{equation}
  \begin{split}
  \frac{\i}{\hbar}
  (\epsilon^0_\alpha - \epsilon^0_\beta - k\hbar\Omega) &
  P_{\alpha\beta,k}
\\
  = {}
  \frac{1}{2}
  \sum_{\ell=\mathrm{L},\mathrm{R}}
  \sum_{k'} 
  \Big\{ &
  \Gamma_\ell(\epsilon^0_{\alpha} + k' \hbar\Omega)
  \langle\phi_{\alpha,k'}|n_\ell\rangle\langle n_\ell|\phi_{\beta,k'+k}\rangle\,
  f(\epsilon^0_{\alpha}+ k' \hbar\Omega-\mu_\ell)
  \\
  + &
  \Gamma_\ell(\epsilon^0_{\beta} + k' \hbar\Omega)
  \langle\phi_{\alpha,k'-k}|n_\ell\rangle\langle n_\ell|\phi_{\beta,k'}\rangle\,
   f(\epsilon^0_{\beta}+ k' \hbar\Omega - \mu_\ell)
  \\
  - &
  \sum_{\alpha',k''}
  \Gamma_\ell(\epsilon^0_{\alpha'} + k''\hbar\Omega)
  \langle\phi_{\alpha,k'+k''-k}|n_\ell\rangle\langle n_\ell|\phi_{\alpha',k''}\rangle\,
  P_{\alpha'\beta,k'}
  \\
  - &
  \sum_{\beta',k''}
  \Gamma_\ell(\epsilon^0_{\beta'} + k''\hbar\Omega)
  \langle\phi_{\beta',k''}|n_\ell\rangle\langle n_\ell|\phi_{\beta,k+k''-k'}\rangle\,
  P_{\alpha\beta',k'}
  \Big\}\,.
  \end{split}
  \label{mastereqfloquet}
\end{equation}
Here, we have assumed that the ideal leads always stay in thermal
equilibrium and, thus, are described by the expectation
values~\eqref{ic}.  Moreover, principal value terms stemming from an
renormalization of the wire energies due to the coupling to the leads
have again been neglected.

The solution of the master equation \eqref{mastereqfloquet} together with
the current expression~\eqref{Imastereq} derived earlier, permits an
efficient numerical calculation of the dc current through the molecular
wire even for rather large systems or for energy-dependent couplings.
Furthermore, as we shall exemplify below, this
approach is still applicable in the presence of degeneracies in the
quasienergy spectrum.

\subsection{Rotating-wave approximation}

\label{sec:rwa}

The current formula~\eqref{I:RWA} valid for very weak wire-lead
coupling, which was derived in Section \ref{sec:approx.weakcoupling},
can also be obtained from the master equation approach within a
rotating-wave approximation.  Thereby, one assumes that the coherent
oscillations of all $P_{\alpha\beta}(t)$ are much faster than their
decay.  Then it is useful to factorize $P_{\alpha\beta}(t)$ into a
rapidly oscillating part that takes the coherent dynamics into account
and a slowly decaying prefactor.  For the latter, one can derive a new
master equation with oscillating coefficients.  Under the assumption
that the coherent and the dissipative time-scales are well separated,
it is possible to replace the time-dependent coefficients by their
time-average.  The remaining master equation is generally of a simpler
form than the original one.  Because we work here already with a
spectral decomposition of the master equation, we give the equivalent
line of argumentation for the Fourier coefficients
$P_{\alpha\beta,k}$.

It is clear from the Fourier representation of the master equation
\eqref{mastereqfloquet} that if
\begin{equation}
\epsilon_\alpha-\epsilon_\beta+k\hbar\Omega\gg \Gamma_{\mathrm{L}/\mathrm{R}} \,,
\label{RWA_condition}
\end{equation}
for all $\alpha,\beta,l$,
then the corresponding $P_{\alpha\beta,k}$ emerge to be small and, thus,
may be neglected.  Under the assumption that the wire-lead couplings are
weak and that the Floquet spectrum has no degeneracies, the RWA condition
\eqref{RWA_condition} is well satisfied except for
\begin{equation}
\label{RWA_condition2}
\alpha=\beta,\quad k=0,
\end{equation}
i.e.\ when the prefactor of the l.h.s.\ of Eq.~\eqref{mastereqfloquet} vanishes
exactly.  This motivates the ansatz
\begin{equation}
P_{\alpha\beta,k}=P_\alpha\,\delta_{\alpha,\beta}\,\delta_{k,0} ,
\label{RWA:ansatz}
\end{equation}
which has the physical interpretation that the stationary state consists of
an incoherent population of the Floquet modes.  The occupation
probabilities $P_\alpha$ are found by inserting the ansatz
\eqref{RWA:ansatz} into the master equation~\eqref{mastereqfloquet} and
read
\begin{equation}
\label{RWA_solution}
P_\alpha = \frac{\sum_{\ell,k} \Gamma_{\ell}(\epsilon_\alpha^0+k\hbar\Omega)
                 f(\epsilon^0_{\alpha}+k\hbar\Omega-\mu_\ell)}
                {\sum_{\ell,k} \Gamma_{\ell}(\epsilon_\alpha^0+k\hbar\Omega)} .
\end{equation}
Inserting this solution into expression \eqref{Imastereq} yields in
the wide-band limit the current formula \eqref{I:RWA}.

\subsection{Phonon damping}
\label{sec:floquet:phonon}

In order to describe the electron transport under the influence of phonon
damping, commonly a boson-like heat bath is coupled to each wire site, which
renders the on-site energies fluctuating with quantum noise
\cite{Holstein1959a, Olson1998a, Yu1999a, Brandes1999a, Emberly2000a,
Mikrajuddin2000a, Segal2000a, Ness2001a, Boese2001a, Petrov2001a,
Lehmann2002a, Petrov2002a, Segal2002a, May2002a, Aguado2004a, Lehmann2004a,
Brandes2004a}.  This can be considered as an extension of the spin-boson
model to more than two sites and the presence of leads.  For the master
equation \eqref{mastereq}, one then has in the first line in addition the
Hamiltonian of the phonon bath, while the electron-phonon coupling enters
as a further dissipative contribution to the second line.  Note that this
leaves the expression \eqref{Imastereq} for the current formally unchanged.

\subsubsection{Hartree-Fock approximation}
\label{sec:hartre-fock}

When evaluating the master equation, however, it turns out that in addition
to the terms containing the single-electron density matrix
$P_{\alpha\beta}(t)$, two-electron expectation values of the form
$\langle c^\dagger_\delta\, c^\dagger_\gamma\, c_\beta\, c_\alpha\rangle_t$
appear.  By iteration, one thus generates a hierarchy of equations up to
$N$-electron expectation values.  To obtain a description in terms of only
the single-electron expectation values, we employ the Hartree-Fock
decoupling scheme defined by the approximation
\begin{equation}
\label{hartree-fock}
  \langle c^\dagger_\delta\, c^\dagger_\gamma\, c_\beta\, c_\alpha\rangle
  \approx {}
  \langle c^\dagger_\delta\, c_\alpha\rangle\langle c^\dagger_\gamma\, c_\beta\rangle
  -\langle c^\dagger_\delta\, c_\beta\rangle\langle c^\dagger_\gamma\, c_\alpha\rangle
  =P_{\alpha\delta} P_{\beta\gamma}- P_{\beta\delta} P_{\alpha\gamma} .
\end{equation}
Clearly, such a mean-field approximation only covers certain aspects
of the full many-particle problem.  Nevertheless, it offers a feasible
and consistent description.  As a most striking consequence, the
Hartree-Fock decoupling \eqref{hartree-fock} leaves the master
equation non-linear~\cite{Lehmann2004a}.

\subsubsection{Thermal equilibrium}
\label{sec:equilibrium}

A potential problem of quantum master equations has been pointed out in
Refs.~\cite{Talkner1986a, Novotny2002a}, namely that they might not be
consistent with the second law of thermodynamics---in particular, that
they might not predict zero current even in the absence of both transport
voltage and driving.  This apparent lack of a proper equilibrium limit,
however, is not inherent to master equations of the form \eqref{mastereq}
themselves, but results from an inconsistent treatment at a later stage: It
is crucial to employ in the second line of Eq.~\eqref{mastereq} the
\textit{exact} interaction picture operators of the uncoupled subsystems.
Using any approximation indeed bears the danger of inconsistencies.
Master equations whose equilibrium limit suffer from the mentioned
problems, have, e.g.\ been derived in Ref.~\cite{Nazarov1993a} and applied
to {non}-equilibrium situations with a finite transport voltage
\cite{Stoof1996a, vanderWiel1999a} and with time-dependent fields
\cite{Brandes1999a, Hazelzet2001a} where no contradiction occurs.

Therefore, an important consistency check for quantum master equations is
an equilibrium situation, where $H_{n n'}$ is time-independent and where no
external bias is present ($\mu_\ell=\mu$ for all $\ell$).  It can be
demonstrated \cite{Lehmann2004a} that the final reduced master equation in
the absence of both driving and voltage has the solution $P_{\alpha\beta} =
\delta_{\alpha\beta}f_\alpha$, with the population $
f_\alpha=f(E_\alpha-\mu)$, determined by the Fermi distribution and the
energy $E_\alpha$ of the eigenstates $|\phi_\alpha\rangle$ which represent
the undriven limit of the Floquet states.  Consequently, the current
\eqref{Imastereq} vanishes in accordance with elementary principles of
statistical physics.

\section{Resonant current-amplification}
\label{sec:resonances}

A natural starting point for the experimental investigation of molecular
conduction under the influence of laser fields is the measurement of
fingerprints of resonant excitations of electrons in the current-voltage
characteristics.  Treating the driving as a perturbation, Keller
\etal~\cite{Keller2002a} have demonstrated that resonant electron excitations
result in peaks of the current as a function of the driving frequency.  Kohler
\etal~\cite{Kohler2002a} included within a Floquet master equation approach
the driving exactly and later derived an analytical solution
\cite{Kohler2004b} which is in good agreement with an exact numerical
solution.  In a related work \cite{Tikhonov2002b}, Tikhonov \etal~studied this
problem within a so-called independent channel approximation
\cite{Tikhonov2002a} of a Floquet transport theory.  As a central result, it
has been found that, in particular for long wires, such excitations enhance
the current significantly.  In this section, we review the analytical
treatment of Ref.~\cite{Kohler2004b} and compare to exact numerical results.

\begin{figure}[tb]
  \centerline{\includegraphics[width=.7\textwidth]{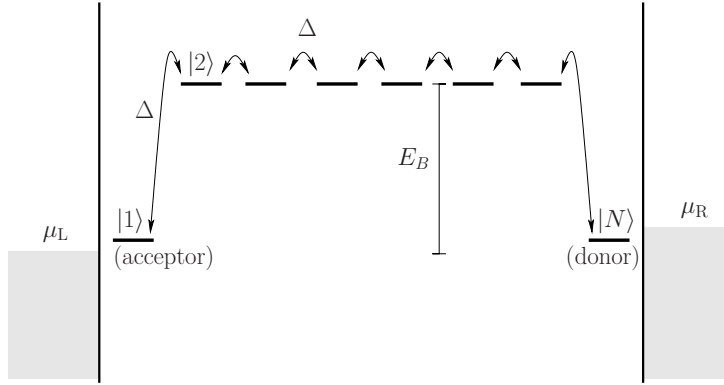}}
  \caption{Bridged molecular wire consisting of $N=8$ sites of which the
    first and the last site are coupled to leads with chemical potentials
    $\mu_\mathrm{L}$ and $\mu_\mathrm{R}=\mu_\mathrm{L}+eV$.}
  \label{fig:wire:reso}
\end{figure}

As a working model we consider a so-called bridged molecular wire
consisting of a donor and an acceptor site and $N-2$ sites in between (cf.\
Fig.~\ref{fig:wire:reso}).  Each of the $N$ sites is coupled to its nearest
neighbors by a hopping matrix element $\Delta$.  The dipole force
\eqref{Hdipole} of the laser field renders each level oscillating in time
with a position-dependent amplitude.  The energies of the acceptor and the
donor orbitals, $|1\rangle$ and $|N\rangle$, are assumed to be close to
the chemical potentials of the attached leads, $E_1=E_N \approx \mu_\mathrm{L}
\approx \mu_\mathrm{R}$.  The bridge levels $E_n$, $n=2,\dots,N-1$, lie
$E_B\gg\Delta, eV$ above the chemical potential.

\subsection{Static conductor}

Let us first discuss the static problem in the absence of the field, i.e.\
for $A=0$.  In the present case where the coupling between two neighboring
sites is much weaker than the bridge energy, $\Delta\ll E_B$, one finds two
types of eigenstates:
One group of states is located on the bridge.  It consists of $N-2$ levels
with energies in the range $[E_B-2\Delta,E_B+2\Delta]$.  In the absence of
the driving field, these bridge states mediate the super-exchange between
the donor and the acceptor.
The two remaining states form a doublet whose states are approximately
given by $(|1\rangle\pm|N\rangle)/\sqrt{2}$.  Its splitting can be
estimated in a perturbational approach \cite{Ratner1990a} and is
approximately given by $2\Delta(\Delta/E_B)^{N-2}$.  Thus, the wire can be
reduced to a two-level system with the effective tunnel matrix element
$\Delta_{DA}=\Delta\exp(-\kappa(N-2))$, where $\kappa=\ln(E_B/\Delta)$.  If
the chemical potentials of the leads are such that $\mu_\mathrm{L}>E_D$ and
$\mu_\mathrm{R}<E_A$, i.e., for a sufficiently large voltage, the current
is dominated by the total transmission and for $\Gamma\gg\Delta_{DA}$ can
be evaluated to read
\begin{equation}
I_0 = \frac{2e|\Delta|^2}{\Gamma}\mathrm{e}^{-2\kappa(N-2)} .
\label{I0a}
\end{equation}
For the explicit calculation see, e.g., Ref.~\cite{Kohler2004a}.  In
particular, one finds an exponentially decaying length dependence of the
current \cite{Mujica1994a, Mujica1994b, Nitzan2001a}.
Moreover, in this limit, it is also possible to evaluate explicitly the
zero-frequency noise to obtain the Fano factor $F=\bar S/e|\bar I|=1$.
This value has a direct physical interpretation: Because the
transmissions
of electrons across a large barrier are rare and uncorrelated events,
they obey Poisson statistics and, consequently, the mean number of
transported electrons equals its variance.  This translates to a Fano
factor $F=1$ \cite{Fano1947a}.

\subsection{Resonant excitations}

The magnitude of the current changes significantly when a driving field
with a frequency $\Omega\approx E_B/\hbar$ is switched on.  Then the
resonant bridge levels merge with the donor and the acceptor state to form
a Floquet state.  This opens a direct channel for the transport resulting
in an enhancement of the electron current.

In order to estimate the magnitude of the current through the resonantly
driven wire, we disregard all bridge levels besides the one that is in
resonance with the donor and the acceptor.  Let us assume that this
resonant bridge level $|\psi_B\rangle$ extends over the whole bridge such
that it occupies the sites $|2\rangle,\ldots,|N{-1}\rangle$ with equal
probability $1/\sqrt{N-2}$.  Accordingly, the overlap between the bridge
level and the donor/acceptor becomes
\begin{equation}
\label{overlapDB}
\langle 1| H_\mathrm{molecule} |\psi_B\rangle
= \frac{\langle 1| H_\mathrm{molecule} |2\rangle}{\sqrt{N-2}}
= \frac{\Delta}{\sqrt{N-2}}
= \langle \psi_B| H_\mathrm{molecule} |N\rangle
\end{equation}
The resonance condition defines the energy of the bridge level as
$\langle\psi_B|H_\mathrm{molecule}|\psi_B\rangle = \hbar\Omega$
(recall that we have assumed $E_D=E_A=0$).

We now apply an approximation scheme in the spirit of the one described in
Ref.~\cite{Kohler2004a} and thereby derive a \textit{static} effective
Hamiltonian that describes the \textit{time-dependent} system.  We start
out by a transformation with the unitary operator
\begin{equation}
S(t) = \exp\Big\{
-\mathrm{i}\sum_{n=2}^{N-1}|n\rangle\langle n|\Omega t
-\mathrm{i}\frac{A}{\hbar\Omega} \sum_{n=1}^N |n\rangle\langle n|
   \sin(\Omega t)
\Big\}.
\end{equation}
Note that $S(t)$ obeys the $\mathcal{T}$-periodicity of the original driven
wire Hamiltonian.  As a consequence, the transformed wire Hamiltonian
\begin{equation}
\widetilde H_\mathrm{molecule}(t)
= S^\dagger(t) H_\mathrm{molecule}(t) S(t)
  -\mathrm{i}\hbar S^\dagger(t) \dot S(t)
\end{equation}
is $\mathcal{T}$-periodic as well.
For $\hbar\Omega\gg\Delta$, we can separate time-scales and average $\widetilde
H_\mathrm{molecule}(t)$ over the driving period.  In the subspace spanned by $|1\rangle$,
$|\psi_B\rangle$, and $|N\rangle$, the time-averaged wire Hamiltonian reads
\begin{equation}
\label{Heff:reso}
H_\mathrm{molecule,eff}
= \int_0^\mathcal{T} \frac{\mathrm{d}t}{\mathcal{T}}\,\widetilde
  H_\mathrm{molecule}(t)
= b \left(\begin{array}{ccc}
   0 & 1 & 0 \\
   1 & 0 & 1 \\
   0 & 1 & 0
  \end{array}\right)
\end{equation}
with the effective tunnel matrix element
\begin{equation}
b=\frac{J_1(A/\hbar\Omega)}{\sqrt{N-2}}\Delta ,
\end{equation}
and $J_1$ the first-order Bessel function of the first kind.

The situation described by the Hamiltonian \eqref{Heff:reso} is essentially the
following: The central site $|\psi_B\rangle$ is coupled by matrix elements
$b$ to the donor and the acceptor site.  Since the latter in turn couple to
the external leads with a self energy $\Gamma/2$, their density of states
is
\begin{equation}
\rho(E) = \frac{1}{\pi}\frac{\Gamma/2}{E^2 + \Gamma^2/4} .
\end{equation}
Then, the tunneling of the electrons from and to the central site is
essentially given by the golden rule rate
\begin{equation}
\label{goldenrule}
w = \frac{2\pi}{\hbar} |b|^2 \rho(0) .
\end{equation}
Like in the static case, we assume that the chemical potential of the left
(right) lead lies above (below) the on-site energy of the acceptor (donor)
and that therefore the donor is always occupied while the acceptor is
always empty.  Then, the rate of electrons tunneling from the central site
to the acceptor is given by the golden rule rate \eqref{goldenrule} times
the occupation probability $p$ of the state $|\psi_B\rangle$.  Accordingly,
the rate of electrons from the donor to $|\psi_B\rangle$ is given by $w$
times the probability $1-p$ to find the central site empty.
Consequently, the occupation of the resonant bridge level evolves according
to
\begin{equation}
\label{master}
\dot p =  w(1-p) - wp .
\end{equation}
Equation \eqref{master} has the stationary solution $p=1/2$ and, thus,
for resonant excitations, the dc contribution of the time-dependent
current is given by
\begin{equation}
\label{I1}
\bar I_\mathrm{res} =  e\, w\, p
= e \frac{2 A^2 \Delta^2}{(N-2) \hbar^3\Omega^2 \Gamma} .
\end{equation}
Here, we have used for small arguments of the Bessel function the
approximation $J_1(x)\approx x$.  The dc current~\eqref{I1} obeys an
intriguing scaling behavior as a function of the wire length: Instead of the
exponentially decaying length dependence~\eqref{I0a} that has been found for
the static case, in the presence of resonant driving, a scaling $\bar I
\propto 1/N$ emerges.  In particular for longer wires, this means that the
external field enhances the conductance by several orders of magnitude.

\subsection{Numerical results}

\begin{figure}[tbp]
\centerline{\includegraphics[width=.6\textwidth]{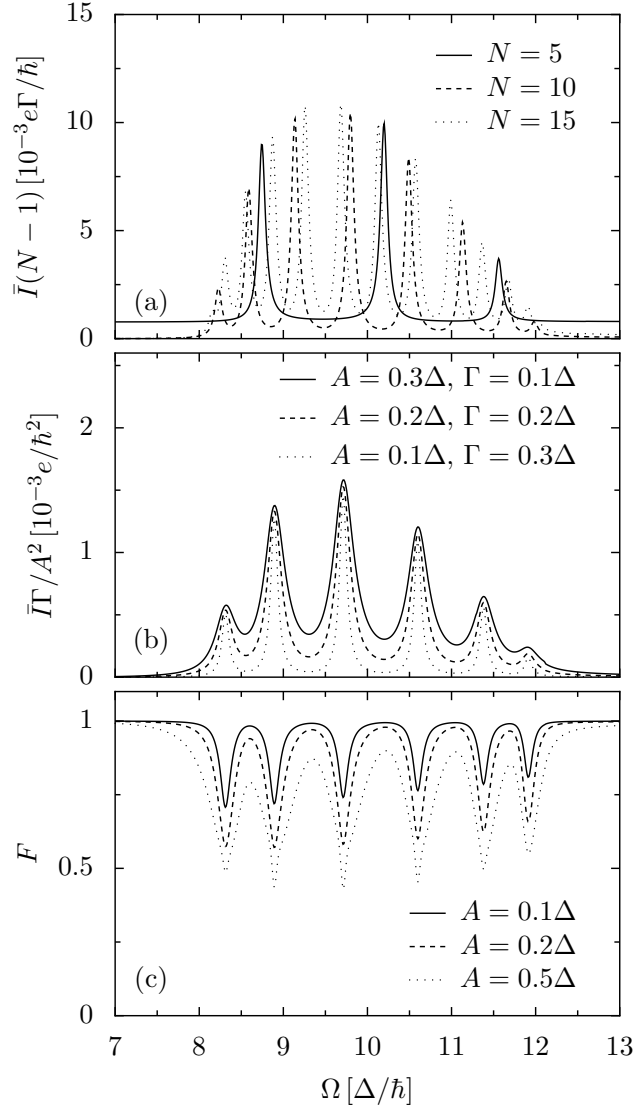}}
\caption{ \label{fig:resonances}
(a)
Average current $\bar I$ as a function of the the driving frequency
$\Omega$ for various wire lengths $N$.  The scaled amplitude is
$A=0.1\Delta$; the applied voltage $\mu_\mathrm{R}-\mu_\mathrm{L}=5\Delta/e$.  The other
parameters read $\Gamma=0.1\Delta$ and $\kBT=0$.
(b)
Average current for various driving amplitudes $A$ and coupling strengths
$\Gamma$ for a wire of length $N=8$.
(c)
Fano factor $F=\bar S/e\bar I$ for the wire length $N=8$ and the wire-lead
coupling $\Gamma=0.1\Delta$.  From Ref.~\cite{Kohler2004b}.
}
\end{figure}%
In order to corroborate the analytical estimates presented above, we treat
the transport problem for the driven wire sketched in
Fig.~\ref{fig:wire:reso} numerically by solving the corresponding Floquet
equation \eqref{Fs} and a subsequent evaluation of the expressions
\eqref{barI} and \eqref{barS} for the dc current and the zero-frequency
noise, respectively.  For a wire with $N=5$ sites, one finds peaks in the
current when the driving frequency matches the energy difference between
the donor/acceptor doublet and one of the $N-2=3$ bridge levels, cf.\ the
solid line in Fig.~\ref{fig:resonances}a.  The applied voltage is always
chosen so small that the bridge levels lie below the chemical potentials of
the leads.  In Figure~\ref{fig:resonances}a the scale of the abscissa is
chosen proportional to $(N-1)$ such that it suggests a common envelope
function.  Furthermore, we find from Fig.~\ref{fig:resonances}b that the dc
current is proportional to $A^2/\Gamma$ provided that $A$ is sufficiently
small and $\Gamma$ sufficiently large.  Thus, the numerical results
indicate that the height of the current peaks obeys
\begin{equation}
\label{scaling}
\bar I_\mathrm{peak} \propto \frac{A^2}{(N-1)\Gamma} \,,
\end{equation}
which is essentially in accordance with our analytical estimate
\eqref{I1}.  The main discrepancy comes from the fact that the overlap
between the resonant level and the donor/acceptor differs from the
estimate \eqref{overlapDB} by a numerical factor of the order one.
Moreover, Fig.~\ref{fig:resonances}c demonstrates that at the
resonances, the Fano factor assumes values considerably lower than one
as expected for the transport through a resonant single level
\cite{Blanter2000a}.

\section{Ratchets and non-adiabatic pumps}
\label{sec:pump}

A widely studied phenomenon in driven transport is the so-termed ratchet
effect: the conversion of ac forces without any net bias into directed
motion \cite{Hanggi1996a, Astumian1997a, Julicher1997a, Reimann2002b,
Reimann2002a, Astumian2002a}.  The investigation of this phenomenon has
been triggered by the question whether an asymmetric device can act as a
Maxwell demon, i.e., whether it is possible to ultimately convert noise
into work.  Feynman's famous ``ratchet and pawl'' driven by random
collisions with gas molecules, on first sight, indeed suggests that such a
Maxwell demon exists.  At thermal equilibrium, however, the whole
nano-device obeys the same thermal fluctuations as the surrounding gas
molecules.  Therefore, consistent with the second law of thermodynamics, no
directed motion occurs \cite{Feynman1963a} and one has to conclude that the
ratchet effect can be observed only in situations far from equilibrium.

A basic model, which captures the essential physics of ratchets, is an
asymmetric, periodic potential under the influence of an ac driving.  In
such a system, even in the absence of any net bias, directed transport has
been predicted for overdamped classical Brownian motion \cite{Hanggi1996a,
Reimann2002a} and also for dissipative quantum Brownian motion in the
incoherent regime \cite{Reimann1997a, Grifoni2002a}.  A related effect is
found in the overdamped limit of dissipative tunneling in driven
superlattices.  There, the spatial symmetry is typically preserved
and the directed transport is brought about by a driving field that
includes higher harmonics of the driving frequency \cite{Goychuk1998a,
Goychuk1998b, Goychuk2001a}.

In the context of mesoscopic conduction, it has been found that the cyclic
adiabatic change of the conductor parameters can induce a so-called pump
current, where the charge pumped per cycle is determined by the area of
parameter space enclosed during the cyclic evolution \cite{Thouless1983a,
Kouwenhoven1991a, Switkes1999a}.  This relates the pump current to a Berry
phase \cite{Brouwer1998a, Altshuler1999a}.
Beyond the adiabatic regime, pump effects have been investigated
theoretically \cite{Stafford1996a, Brune1997a, Levinson2000a,
Moskalets2002a, Wang2002a} and also been measured in coupled quantum dots
\cite{vanderWiel1999a, vanderWiel2003a, DiCarlo2003a}.  Since in the
non-adiabatic regime, the main contribution to the pump current comes from
electrons considerably below the Fermi surface, non-adiabatic electron
pumping is essentially temperature independent \cite{Wagner1999a}.

The studies presented in this chapter were mainly motivated by two aspects:
First, although infinitely extended ``ideal'' ratchets are convenient
theoretical models, any experimental realization will have finite length,
i.e., consist of a finite number of elementary units; cf.\
Fig.~\ref{fig:wire.ratchet}, below.  Thus, finite size effects become
relevant and it is intriguing to know the number of coupled wire units that
are needed to mimic the behavior of a practically infinite system.  Second,
prior studies of quantum ratchets focussed on incoherent tunneling
\cite{Reimann1997a, Grifoni2002a}.  By contrast, the present setup allows
one to investigate ratchet dynamics in the coherent quantum regime which has
not been explored previously.

The results of this section, have originally \cite{Lehmann2002b,
Lehmann2003b} been computed for finite temperatures within the master
equation approach of Section~\ref{sec:master}.  In the limit of zero
temperature, but otherwise equal parameters, the results from such a
perturbative treatment agree essentially perfect with the corresponding
exact solution obtained from Eq.~\eqref{barI}.

\subsection{Symmetry inhibition of ratchet currents}
\label{sec:pump:symmetry}

It is known from the study of deterministically rocked periodic potentials
\cite{Flach2000a} and of driven classical Brownian particles
\cite{Reimann2001a} that the symmetry of the equations of motion may rule
out any non-zero average current at asymptotic times.  Thus, before
starting to compute ratchet currents, let us first analyze what kind of
symmetries may prevent the effect.  We consider situations, where the
electron distributions in both leads are identical---in particular,
situations where both leads are in thermal equilibrium with a common
chemical potential, $f_\mathrm{L}(\epsilon) = f_\mathrm{R}(\epsilon) \equiv f(\epsilon)$ for
all $\epsilon$.  Then, no electromotive force acts and, consequently, in
the absence of driving, all currents must vanish.  An applied driving
field, however, violates the equilibrium condition and can generate a finite
dc current
\begin{equation}
\label{I.pump}
I_\mathrm{pump} = \frac{e}{h} \sum_k \int\d\epsilon\,
\big[T_{\mathrm{L}\mathrm{R}}^{(k)}(\epsilon)-T_{\mathrm{R}\mathrm{L}}^{(k)}(\epsilon) \big] f(\epsilon) .
\end{equation}
Obviously, the pump current vanishes if the condition
$T_{\mathrm{L}\mathrm{R}}^{(k)}(\epsilon) = T_{\mathrm{R}\mathrm{L}}^{(k)}(\epsilon)$ is fulfilled for all $k$
and $\epsilon$.  One might now ask whether this condition can be ensured by
any symmetry relation.  For the dipole driving considered here, the
relevant symmetries are those studied in the Appendix \ref{app:symmetry},
namely time-reversal symmetry, time-reversal parity, and generalized
parity.  In Section~\ref{sec:symmetry}, we have already identified the
symmetry-related channels which possess equal transmission probabilities.

Looking at the relations \eqref{transmission:T}, \eqref{transmission:TP},
and \eqref{transmission:GP}, it becomes clear that the generalized parity
$\mathcal{S}_\mathrm{GP}$ is the only symmetry that directly yields a
vanishing pump current.  This is so because it implies for the transmission
probabilities the relation \eqref{transmission:GP} and, thus, we find
$I_\mathrm{pump}=0$ \cite{Lehmann2003b}.
While time-reversal symmetry is without any consequence for the pump
current, time-reversal parity has some rather subtle effect which follows
from the fact that the transmission probabilities obey
\eqref{transmission:TP} and that in the weak-coupling limit of
Section~\ref{sec:approx.weakcoupling}, in addition, relation
\eqref{symmetry.weak} holds.  Given these two relations, we obtain
$T_{\mathrm{L}\mathrm{R}}^{(k)}(\epsilon) =
T_{\mathrm{R}\mathrm{L}}^{(k)}(\epsilon)$ and, thus, the dc current vanishes.
Since the weak-coupling approximation is correct to lowest order
in the coupling $\Gamma$, the consequence of time-reversal parity for
quantum ratchets and Brownian motors is that we no longer find the generic
behavior $I_\mathrm{pump} \propto\Gamma$, but rather $I_\mathrm{pump}
\propto\Gamma^2$.

In the following, we consider two typical cases where the generalized
parity symmetry is broken and, thus, a pump current emerges, namely
(i) an asymmetric structure under the influence of a harmonic dipole
force, the so-called rocking ratchet, and (ii) a spatially symmetric
system for which generalized parity is broken dynamically by mixing
with higher harmonics.

\subsection{Spatial symmetry-breaking: Coherent quantum ratchets}
\label{sec:pump:ratchet}

\begin{figure}[tb]
\centerline{\includegraphics[width=.7\textwidth]{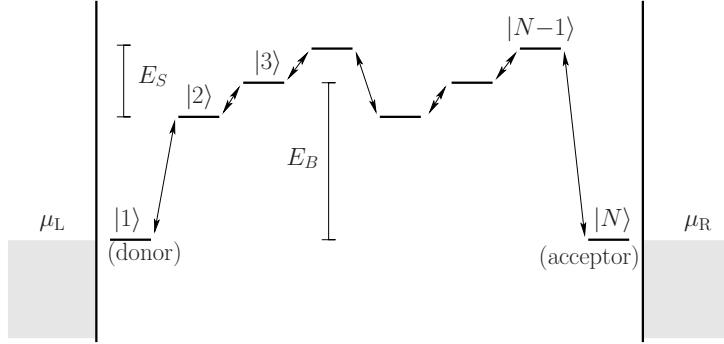}}
\caption{\label{fig:wire.ratchet}
  Level structure of the wire ratchet with $N=8$ sites, i.e., $N_g=2$
  asymmetric groups.  The bridge levels are $E_B$ above the donor and
  acceptor levels and are shifted by $\pm E_S/2$.}
\end{figure}

A straightforward way to break generalized parity, is to use a conductor
with an asymmetric level structure.  Then, already a purely harmonic dipole
driving $a(t)=A\sin(\Omega t)$ in the Hamiltonian \eqref{Hdipole} is
sufficient to generate a dc current.  As a tight-binding model of such a
structure, we have considered a wire consisting of a donor and an acceptor
site and $N_g$ asymmetric groups in the ratchet-like configuration sketched
in Fig.~\ref{fig:wire.ratchet}.  In molecular structures, such an asymmetry
can be achieved in many ways, and was explored as a source of molecular
current rectification since the early work of Aviram and Ratner
\cite{Aviram1974a}.  Later this effect has been found experimentally
\cite{Reichert2002a, Weber2002a}.  In general, an asymmetry can be created
by attaching different chemical groups to the opposite sides of an
otherwise symmetric molecular wire \cite{Chen1999a, Reichert2002a,
Weber2002a}.  In our model, the inner wire states are arranged in $N_g$
groups of three, i.e.\ $N-2=3N_g$.  In each group, the first (last) level
is lowered (raised) by an energy $E_S/2$, forming an asymmetric saw-tooth
like structure.  The energies of the donor and the acceptor orbitals are
assumed to be at the level of the chemical potentials of the attached leads
and since no voltage is applied, we thus have
$E_1=E_N=\mu_\mathrm{L}=\mu_\mathrm{R}$.  The bridge levels $E_n$ lie at $E_B$
and $E_B\pm E_S/2$, as sketched in Fig.~\ref{fig:wire.ratchet}.

\begin{figure}
\centerline{\includegraphics{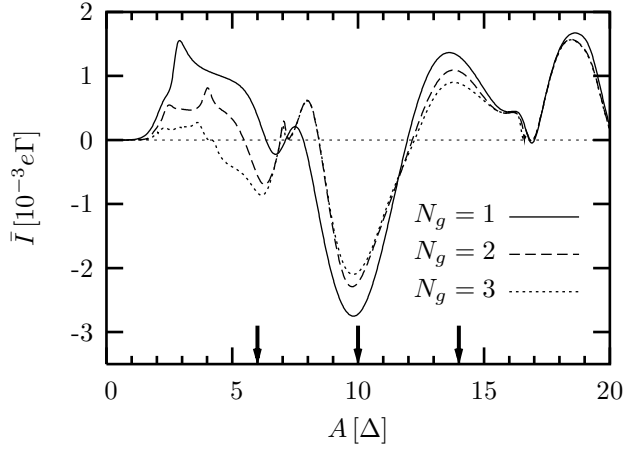}}
\caption{\label{fig:I-F}
Time-averaged current through a molecular wire ratchet that consists of $N_g$
bridge units as a function of the driving strength $A$.  The bridge
parameters are $E_B=10 \Delta$, $E_S=\Delta$, the driving frequency is
$\Omega=3\Delta/\hbar$, the coupling to the leads is chosen as
$\Gamma_\mathrm{L}=\Gamma_\mathrm{R}=0.1 \Delta/\hbar $, and the temperature
is $\kBT=0.25 \Delta$.
The arrows indicate the driving amplitudes used in Fig.~\ref{fig:I-N}.
From Ref.~\cite{Lehmann2002b}.
}
\end{figure}

\begin{figure}[tb]
  \centerline{\includegraphics{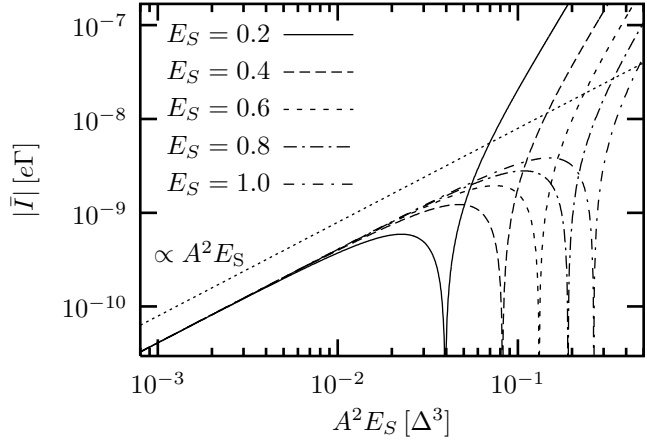}}
  \caption{\label{fig:I-EcA}
  Absolute value of the time-averaged current in a ratchet-like structure
  with $N_g=1$ as a function of $A^2 E_S$ demonstrating the proportionality
  to $A^2 E_S$ for small driving amplitudes.  All other parameters are as in
  Fig.~\ref{fig:I-F}.  At the dips on the right-hand side, the current $\bar
  I$ changes its sign.  From Ref.~\cite{Lehmann2003b}.}
\end{figure}

\begin{figure}[tb]
\centerline{\includegraphics{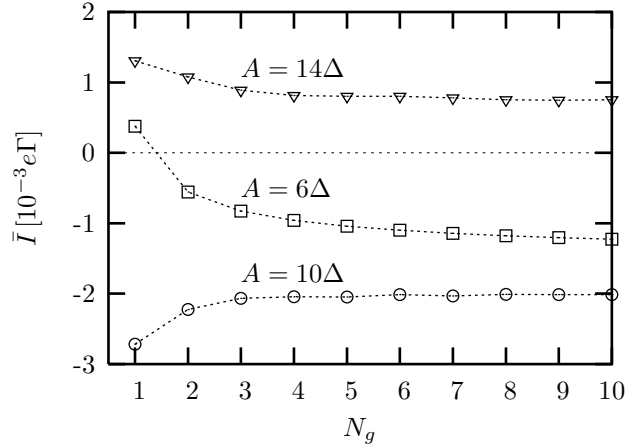}}
\caption{\label{fig:I-N}
  Time-averaged current as a function of the number of bridge units $N_g$
  for the driving amplitudes indicated in Fig.~\ref{fig:I-F}.  The other
  parameters are as in Fig.~\ref{fig:I-F}.  The connecting lines serve as a
  guide to the eye.  From Ref.~\cite{Lehmann2002b}.}
\end{figure}

Figure~\ref{fig:I-F} shows the resulting stationary time-averaged current
$\bar{I}$.  A quantitative analysis of a tight-binding model has
demonstrated that the resulting currents lie in the range of
$10^{-9}\,\mathrm{A}$ and, thus, can be measured with today's techniques
\cite{Lehmann2002b}.  In the limit of very weak driving, we find
$\bar{I}\propto E_S A^2$ (Fig.~\ref{fig:I-EcA}).  This behavior is expected
from symmetry considerations: The asymptotic current must be independent of
any initial phase of the driving field and therefore is an even function of
the field amplitude $A$.  This indicates that the ratchet effect can
only be obtained from a treatment of the field beyond Kubo theory.
For strong laser fields, Fig.~\ref{fig:I-F} also shows that $\bar{I}$ is
almost independent of the wire length.  If the driving is intermediately
strong, $\bar{I}$ depends in a short wire sensitively on the driving
amplitude $A$ and the number of asymmetric molecular groups $N_g$: even the
sign of the current may change with $N_g$, i.e.\ we find a current reversal
as a function of the wire length.  For long wires that comprise five or
more wire units, the average current becomes again length-independent, as
can be seen from Fig.~\ref{fig:I-N}.  This identifies the observed current
reversal as a finite size effect.  As practical consequence, such
relatively short wires can mimic the behavior of an (infinitely extended)
quantum ratchet.  
Moreover, the fact that $\bar{I}$ converges to a finite
value if the number of wire units is enlarged, demonstrates that the
dissipation caused by the coupling to the leads is sufficient to establish
the ratchet effect in the limit of long wires.  In this sense, no on-wire
dissipation is required.  Still, if the wire-lead model \eqref{H.wirelead}
is extended by electron-phonon coupling, the ratchet current might be
enhanced \cite{Lehmann2004a}.

\begin{figure}[tb]
\centerline{\includegraphics{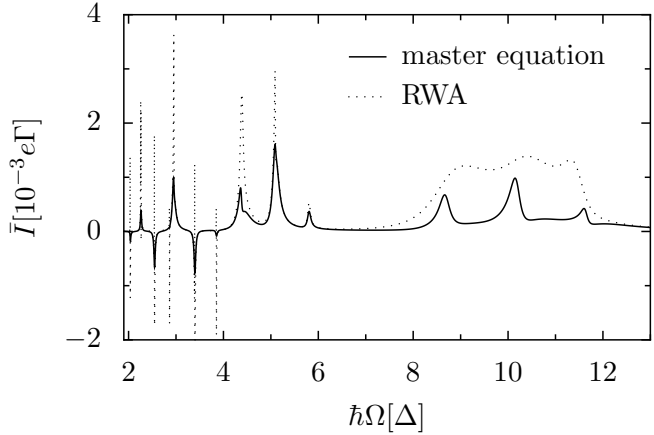}}
\caption{\label{fig:I-omega}
  Time-averaged current as a function of the driving frequency
  $\Omega$ for $A=\Delta$ and $N_g=1$ (solid line).
  All other parameters are as in Fig.~\ref{fig:I-F}.
  The dotted line depicts the solution within the
  rotating-wave approximation~\eqref{RWA_solution}.
  After Ref.~\cite{Lehmann2002b}.}
\end{figure}

\begin{figure}[tb]
\centerline{\includegraphics{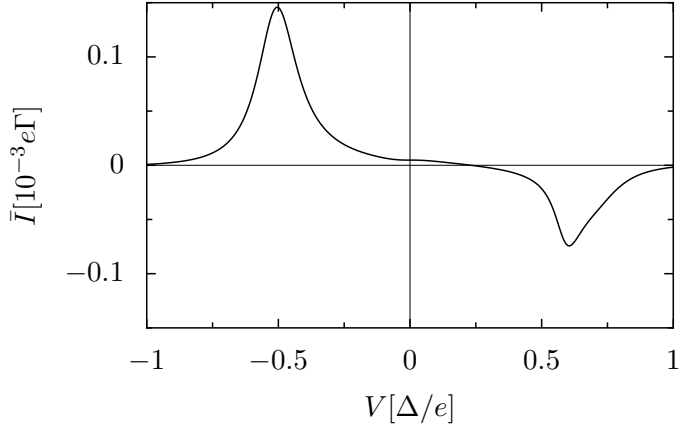}}
\caption{\label{fig:I-V}
  Time-averaged current as a function of the applied static bias
  voltage $V$, which drops solely along the
  molecule. The driving amplitude is $A=\Delta$, the driving frequency
  $\Omega=3\Delta/\hbar$, and $N_g=1$.  All other parameters are
  as in Fig.~\ref{fig:I-F}.  After Ref.~\cite{Lehmann2002b}.}
\end{figure}

Figure~\ref{fig:I-omega} depicts the average current \textit{vs.}\ the
driving frequency~$\Omega$, exhibiting resonance peaks as a striking
feature.  Comparison with the quasienergy spectrum reveals that each peak
corresponds to a non-linear resonance between the donor/acceptor and a
bridge orbital.  While the broader peaks at $\hbar\Omega\approx E_B=10
\Delta$ match the 1:1 resonance (i.e.\ the driving frequency equals the
energy difference), one can identify the sharp peaks for
$\hbar\Omega\lesssim 7 \Delta$ as multi-photon transitions.  The appearance
of these resonance peaks clearly demonstrates that the molecular bridge acts
as a \textit{coherent} quantum ratchet.
Moreover, owing to the broken spatial symmetry of the wire, one expects an
asymmetric current-voltage characteristic.  This is indeed the case as
depicted in Fig.~\ref{fig:I-V}.


\subsection{Temporal symmetry-breaking: Harmonic mixing}
\label{sec:mixing}

\begin{figure}
  \centerline{\includegraphics{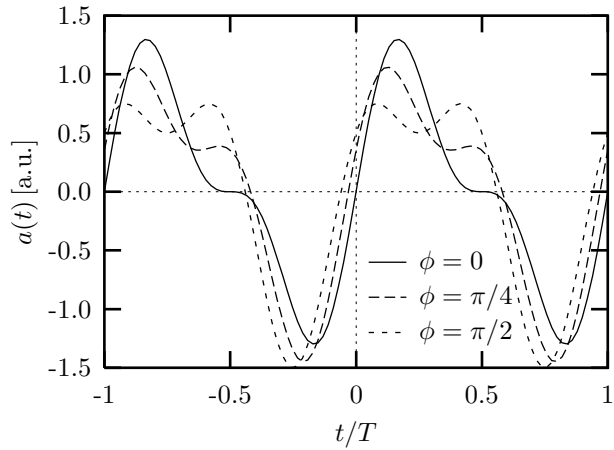}}
  \caption{\label{fig:mix_field}
  Shape of the harmonic mixing field $a(t)$ in Eq.~\eqref{mixing} for
  $A_1=2A_2$ for different phase shifts $\phi$.  For $\phi=0$, the field
  changes its sign for $t\to -t$ which amounts to the time-reversal parity
  $\mathcal{S}_\mathrm{TP}$.}
\end{figure}

The symmetry analysis in Section~\ref{sec:pump:symmetry} explains that for a
symmetric bridge without a ratchet-like structure as sketched in
Fig.~\ref{fig:wire:reso}, the pump current \eqref{I.pump} vanishes if the
driving is a purely harmonic dipole force.  This is so because then the
system is invariant under the generalized parity transformation
$\mathcal{S}_\mathrm{GP}$ and, thus, the transmission factors obey relation
\eqref{transmission:GP}.  Still, generalized parity can be broken in a
dynamical way by adding a second harmonic to the driving field, i.e., a
contribution with twice the fundamental frequency $\Omega$, such that it is
of the form
\begin{equation}
\label{mixing}
a(t) = A_1\sin(\Omega t) + A_2\sin(2\Omega t+\phi) ,
\end{equation}
as sketched in Fig.~\ref{fig:mix_field}.  While now shifting the time $t$
by a half period $\pi/\Omega$ changes the sign of the fundamental frequency
contribution, the second harmonic is left unchanged.  The generalized
parity is therefore no longer present and we expect to find a non-vanishing
average current.

\begin{figure}
  \centerline{\includegraphics{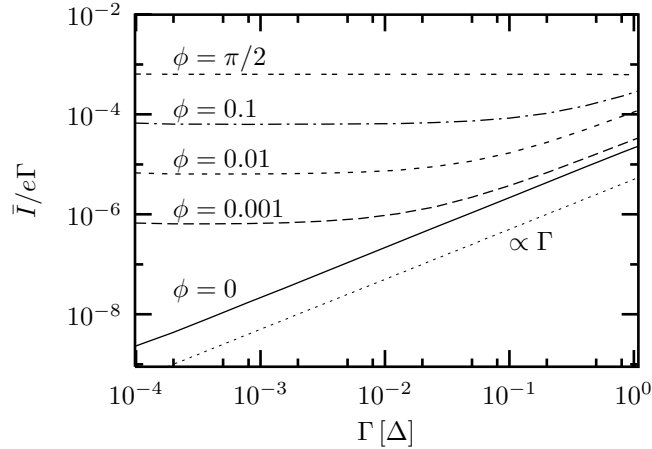}}
  \caption{\label{fig:mix_gamma}
  Average current response to the harmonic mixing signal with amplitudes
  $A_1=2A_2=\Delta$, as a function of the coupling strength for different
  phase shifts $\phi$.  The remaining parameters are $\Omega=10\Delta/\hbar$,
  $E_\mathrm{B}=5\Delta$, $\kBT=0.25\Delta$, $N=10$.  The dotted line is
  proportional to $\Gamma$; it represents a current which is proportional to
  $\Gamma^2$.  From Ref.~\cite{Lehmann2003b}.}
\end{figure}

The phase shift $\phi$ plays here a subtle role.  For $\phi=0$ (or
equivalently any multiple of $\pi$) the time-reversal parity
$\mathcal{S}_\mathrm{TP}$ is still present.  Thus, according to the symmetry
considerations in Section~\ref{sec:pump:symmetry}, the current vanishes within
the weak-coupling approximation for the transmission probability, cf.\
Eq.~\eqref{Tweak}.  Since this approximation is only correct to linear
order in $\Gamma$, the higher-order contributions typically remain finite
and, consequently, for small coupling the pump current obeys $\bar I \propto
\Gamma^2$.  Figure~\ref{fig:mix_gamma} confirms this prediction.  Yet one
observes that already a small deviation from $\phi=0$ is sufficient to
restore the usual weak coupling behavior, namely a current which is
proportional to the coupling strength $\Gamma$.  
This effect can be employed for the detection phase lags.

Other features of the harmonic mixing current resemble the ones discussed
above in the context of ratchet-like structures \cite{Lehmann2003b}.  In
particular, we again find for large driving amplitudes that the current
becomes essentially independent of the wire length.  Typically, the current
reaches convergence for a length $N\gtrsim 10$.

\subsection{Phonon damping}

Including also the coupling of the wire electrons to a phonon heat bath,
one can no longer employ the scattering formula \eqref{barI} and for the
computation of the dc current, one thus, has to resort to the master
equation approach of Section~\ref{sec:master}.  Here we only mention the
main findings and refer the reader to the original work,
Ref.~\cite{Lehmann2004a}: The presence of phonon damping, generally
increases the pump current up to one order of magnitude.  This means that
for quantum ratchets, noise plays a rather constructive role.  Moreover,
phonon damping influences the dependence of the current on the phase
lag by providing an additional shift towards a $\cos\phi$ behavior.

\section{Control setups}
\label{sec:control}

A prominent example for the control of quantum dynamics is the so-called
coherent destruction of tunneling, i.e., the suppression of the tunneling
dynamics of a particle in a double-well potential \cite{Grossmann1991a} and
in a two-level system \cite{Grossmann1991a, Grossmann1992a}.  Recently,
coherent destruction of tunneling has also been found for the dynamics of
two interacting electrons in a double quantum dot \cite{Creffield2002a,
Creffield2002b}.  A closely related phenomenon is the miniband collapse in
ac-driven superlattices which yields a suppression of quantum diffusion
\cite{Holthaus1992a, Holthaus1992b, Holthaus1993a}.
In this chapter, we address the question whether a corresponding transport
effect exists: If two leads are attached to the ends of a driven tunneling
system, is the suppression of tunneling visible in conductance properties?
Since time-dependent control schemes can be valuable in practice only if
they operate at tolerable noise levels, the question is also whether
the corresponding noise strength can be kept small or even be controlled.

\subsection{Coherent destruction of tunneling}

In order to introduce the reader to the essentials of coherent destruction
of tunneling in isolated quantum systems, we consider a single particle in
a driven two-level system described by the Hamiltonian
\begin{equation}
\mathcal{H}_\mathrm{TLS}(t)
= -\frac{\Delta}{2}\sigma_x + \frac{A}{2}\sigma_z \cos(\Omega t).
\end{equation}
If the energy of the quanta $\hbar\Omega$ of the driving field exceeds the
energy scales of the wire, one can apply the high-frequency approximation
scheme of Section~\ref{sec:approx.hf} \cite{Grossmann1992a, Kohler2004a} and
finds that the dynamics can be described approximately by the static
effective Hamiltonian \eqref{hf:Heff} which for the present case reads
\begin{equation}
\mathcal{H}_\mathrm{TLS,eff} = -\frac{\Delta_\mathrm{eff}}{2}\sigma_x ,
\end{equation}
with the tunnel matrix element renormalized according to
\begin{equation}
\label{Delta.eff}
\Delta \longrightarrow \Delta_\mathrm{eff} = J_0(A/\hbar\Omega) \Delta .
\end{equation}
Again, $J_0$ denotes the zeroth order Bessel function of the first kind.
If the ratio $A/\hbar\Omega$ equals a zero of the Bessel function $J_0$
(i.e., for the values 2.405.., 5.520.., 8.654.., \ldots), the effective
tunnel matrix vanishes and the tunneling is brought to a standstill.

This reasoning is readily generalized to other tight-binding systems: If
neighboring sites are coupled by a hopping matrix element $\Delta$ and the
difference of their on-site energies oscillates with an amplitude $A$, one
finds that the physics is determined by the renormalized matrix element
\eqref{Delta.eff}, provided that $\hbar\Omega$ is the largest energy scale.

\subsection{Current and noise suppressions}
\label{sec:ISsuppression}

\begin{figure}[t]
\centerline{\includegraphics[width=0.55\textwidth]{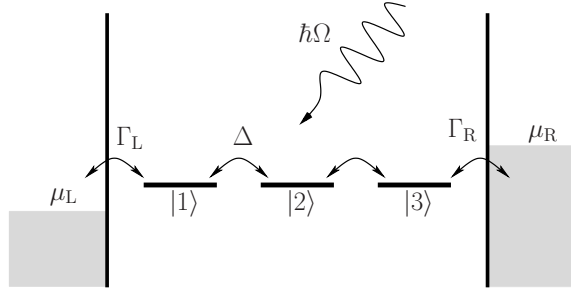}}
\caption{\label{fig:wireCamalet}
Level structure of the molecular wire with $N=3$ orbitals.  The end sites
are coupled to two leads with chemical potentials $\mu_\mathrm{L}$ and
$\mu_\mathrm{R}=\mu_\mathrm{L}-eV$.}
\end{figure}

In order to investigate coherent destruction of tunneling in the context of
transport, we consider the wire-lead setup sketched in
Fig.~\ref{fig:wireCamalet}
where the wire is described by the dipole Hamiltonian \eqref{Hdipole} with
on-site energies $E_n=0$.  The wire is assumed to couple equally to both
leads, $\Gamma_\mathrm{L} = \Gamma_\mathrm{R} = \Gamma$, and the numerical
results are computed from the exact current formula \eqref{barI}.

For a driven wire described by the Hamiltonian \eqref{Hdipole}, it has been
found \cite{Lehmann2003a, Camalet2003a, Camalet2004a} that the oscillating
dipole force suppresses the transport if the ratio
$A/\hbar\Omega$ is close to a zero of the Bessel function $J_0$.  Moreover,
in the vicinity of such suppressions, the shot noise characterized by the
Fano factor \eqref{Fano} assumes two characteristic minima.  These
suppression effects are most pronounced in the high-frequency regime, i.e.,
if the energy quanta $\hbar\Omega$ of the driving exceed the energy scales
of the wire.  Thus, before going into a detailed discussion, we start with
a qualitative description of the effect based on the static approximation
for a high-frequency driving that has been derived in
Section~\ref{sec:approx.hf}.

Let us consider first the limit of a voltage which is so large that in Eq.\
\eqref{IlO}, $f_{\mathrm{R},\mathrm{eff}} - f_{\mathrm{L},\mathrm{eff}}$ can be replaced by
unity.  Then, the average current is determined by the effective
Hamiltonian
\begin{equation}
\mathcal{H}_\mathrm{eff}
=  -\Delta_\mathrm{eff}\sum_{n=1}^{N-1}\big( |n\rangle\langle n{+}1|
    + |n{+}1\rangle\langle n| \big)
 + \sum_{n=1}^N E_n \,|n\rangle\langle n| ,
\end{equation}
which has been derived by inserting the time-dependent part of the
Hamiltonian \eqref{Hdipole} into Eqs.\ \eqref{Fn} and \eqref{hf:Heff}.
Then, obviously $\mathcal{H}_\mathrm{eff}$ is identical to the static part
of the Hamiltonian \eqref{Hdipole} but with the tunnel matrix element
renormalized according to Eq.~\eqref{Delta.eff}.  Since the Bessel function
$J_0$ assumes values between zero and one, the amplitude of the driving
field allows one to switch the absolute value of the effective hopping on the
wire, $\Delta_\mathrm{eff}$, between 0 and $\Delta$.  Since the
transmission probability of an undriven wire is proportional to
$|\Delta|^2$, the effective transmission probability
$T_\mathrm{eff}(\epsilon)$ acquires a factor $J_0^2(A/\hbar\Omega)$.  This
renormalization of the hopping then results  in a current suppression
\cite{Lehmann2003a, Camalet2003a, Camalet2004a}.

For the discussion of the shot noise, we employ the Fano factor
\eqref{Fano} as a measure.  In the limit of large applied voltages,
we have to distinguish two limits:
(i) weak wire-lead coupling $\Gamma\ll\Delta_\mathrm{eff}$ (i.e., weak
with respect to the effective hopping) and
(ii) strong wire-lead coupling $\Gamma\gg\Delta_\mathrm{eff}$.
In the first case, the tunnel contacts between the lead and the wire act as
``bottlenecks'' for the transport.  In that sense they form barriers.  Thus
qualitatively, we face a double barrier situation and, consequently,
expect the shot noise to exhibit a Fano factor $F\approx1/2$
\cite{Blanter2000a}.  In the second case, the links between the wire sites
act as $N-1$ barriers.  Correspondingly, the Fano factor assumes values
$F\approx 1$ for $N=2$ (single barrier) and $F\approx 1/2$ for $N=3$
(double barrier) \cite{DeJong1995a}.  At the crossover between the two
limits, the conductor is (almost) ``barrier free'' such that the Fano
factor assumes its minimum.

In order to be more quantitative, we evaluate the current and the
zero-frequency noise in more detail thereby considering a finite voltage.
This requires a closer look at the effective electron distribution
\eqref{hf:feff}; in particular, we have to quantify the concept of a
``practically infinite'' voltage.  In a static situation, the voltage can
be replaced by infinity, $f_\mathrm{R}(\epsilon) = 1 = 1-f_\mathrm{L}(\epsilon)$, if all
eigenenergies of the wire lie well inside the range $[\mu_\mathrm{L},\mu_\mathrm{R}]$.  In
contrast to the Fermi functions, the effective electron distribution
\eqref{hf:feff} which is decisive here, decays over a broad range in
multiple steps of size $\hbar\Omega$.  Since for our model,
$T_\mathrm{eff}(\epsilon)$ is peaked around $\epsilon=0$, we replace here
the effective electron distributions by their values for $\epsilon=0$,
\begin{equation}
\label{feff:K(V)}
f_{\ell,\mathrm{eff}}(0)
= \sum_{k<\mu_{\ell}/\hbar\Omega} J_k^2\Big(\frac{A(N-1)}{2\hbar\Omega}\Big) ,
\end{equation}
for zero temperature.  We have inserted the coefficients $a_{1,k} =
J_k(A(N-1)/2\hbar\Omega)$ and $a_{N,k} = J_{-k}(A(N-1)/2\hbar\Omega)$
which have been computed directly from their definition \eqref{a:hf};
$J_k$ denotes the $k$th order Bessel functions of the first kind.
The current, the noise, and the Fano factor are given by the static
expressions \eqref{current.static} and \eqref{noise.static} with the
transmission probability and the electron distribution replaced by the corresponding
effective quantities, $T_\mathrm{eff}$ and $f_{\mathrm{eff},\ell}$,
respectively.  Thus, we obtain
\begin{align}
\label{Ihf}
\bar I =& \lambda \bar I_{\infty} ,
\\
\label{Shf}
\bar S =& \lambda^2 \bar S_{\infty}
          + \frac{e}{2}(1-\lambda^2) \bar I_{\infty} ,
\\
F =& \lambda F_\infty + \frac{1-\lambda^2}{2\lambda},
\label{Fhf}
\end{align}
respectively, where the subscript $\infty$ denotes the corresponding
quantities in the infinite voltage limit,
\begin{align}
\bar I_\infty =& \frac{e}{h}\int \d\epsilon\, T_\mathrm{eff}(\epsilon) ,
\\
\bar S_\infty =& \frac{e^2}{h}\int \d\epsilon\, T_\mathrm{eff}(\epsilon)
            [1 - T_\mathrm{eff}(\epsilon)] ,
\end{align}
and $F_\infty=\bar S_\infty/e\bar I_\infty$.
The factor
\begin{equation}
\label{lambda}
\lambda
= f_{\mathrm{R},\mathrm{eff}}(0) - f_{\mathrm{L},\mathrm{eff}}(0)
= \sum_{|k|\leq K(V)} J_k^2 \Big(\frac{A(N-1)}{2\hbar\Omega}\Big)
\end{equation}
reflects the influence of a finite voltage; $K(V)$ denotes the
largest integer not exceeding $e|V|/2\hbar\Omega$.
Since $J_k(x)\approx 0$ for $|k|>x$ and $\sum_k J_k^2(x)\approx 1$, we find
$\lambda=1$ if $K(V)>A(N-1)/2\hbar\Omega$.  This means that for small
driving amplitudes $A<eV/(N-1)$, we can consider the voltage as practically
infinite.  With an increasing driving strength, $\lambda$ decreases and,
thus, the current becomes smaller by a factor $\lambda$ but still exhibits
suppressions.  By contrast, since $F_\infty\leq 1$ for all situations
considered here, we find
from Eq.~\eqref{Fhf} that the Fano factor will increase with smaller
$\lambda$.

Let us emphasize that unlike in the present case, the quenching of
transmission observed in Refs.~\cite{Wagner1994a, Wagner1995a} does not
result from a renormalized inter-well tunnel matrix element, but
rather originates from the appearance of the Bessel function $J_0$ in the
effective electron distribution \eqref{feff:K(V)}.  Therefore, at large
voltages, this quenching will not give rise to current suppressions.

\subsection{Numerical results}

\begin{figure}[t]
\centerline{\includegraphics{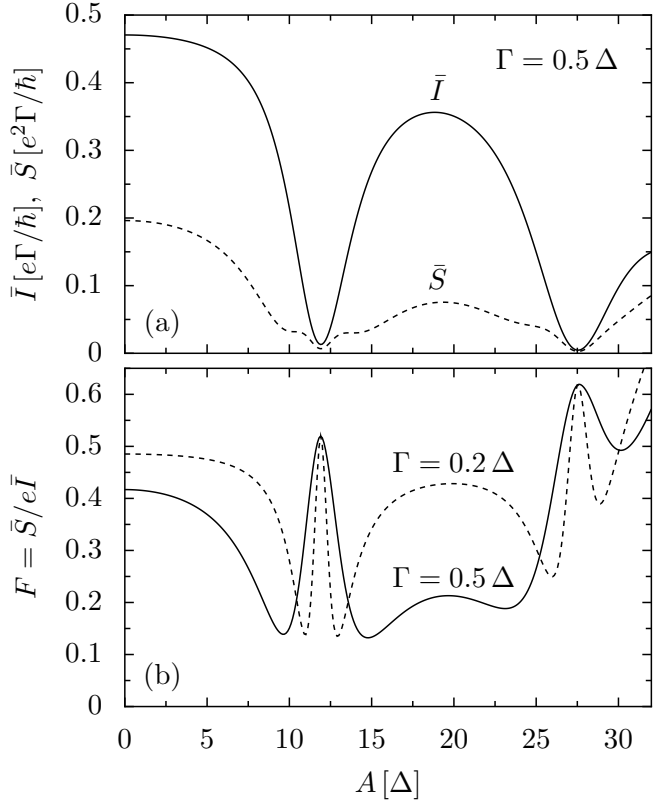}}
\caption{\label{fig:control.A}
Time-averaged current $\bar I$ and zero-frequency noise $\bar S$ (a) as a
function of the driving amplitude $A$ for a wire with $N=3$ sites with
on-site energies $E_n=0$ and chemical potentials $\mu_\mathrm{R}=-\mu_\mathrm{L}=25\Delta$.
The other parameters read $\Omega = 5 \Delta/\hbar$, $\Gamma = 0.5\Delta$,
and $\kBT=0$.  Panel (b) displays the Fano $F$ factor for these parameters
(full line) and for smaller wire-lead coupling (dash-dotted line).
From Ref.\ \cite{Camalet2003a}.}
\end{figure}

Figure \ref{fig:control.A}a depicts the dc current and the zero-frequency
noise for a wire with $N=3$ sites and a relatively large applied voltage,
$\mu_\mathrm{L}-\mu_\mathrm{R} = 50\Delta$.  As a remarkable feature, we find that for
certain values of the field amplitude $A$, the current drops to a value of
some percent of the current in the absence of the field \cite{Lehmann2003a,
Camalet2003a} with a suppression factor which is fairly independent of the
wire-lead coupling $\Gamma$ \cite{Lehmann2004a}.  The corresponding noise
strength $\bar S$ exhibits similar suppressions and, in addition, has some
small plateaus in the vicinity of the minima.  The role of the plateaus is
elucidated by the \textit{relative} noise strength characterized by the
Fano factor \eqref{Fano} which is shown in Fig.~\ref{fig:control.A}b.
Interestingly enough, we find that the Fano factor as a function of the
driving amplitude $A$ possesses both a sharp maximum at each current
suppression and two pronounced minima nearby.  For a sufficiently large
voltage, the Fano factor at the maximum assumes the value $F\approx 1/2$.
Once the driving amplitude is of the order of the applied voltage, however,
the Fano factor becomes much larger.  The relative noise minima are
distinct and provide a typical Fano factor of $F\approx0.15$.  Reducing the
coupling to the leads renders these phenomena even more pronounced since
then the suppressions occur in a smaller interval of the driving amplitude,
cf.\ Fig.\ \ref{fig:control.A}b.  The overall behavior is robust in the
sense that approximately the same values for the minima and the maximum are
also found for larger wires, different driving frequencies, different
coupling strengths, and slightly modified on-site energies, provided that
$\Delta,\Gamma,E_n \ll \hbar\Omega$ and that the applied voltage is
sufficiently large \cite{Camalet2004a}.

A comparison of these numerical results and the ones obtained in
Section~\ref{sec:ISsuppression} analytically within a high-frequency
approximation shows an excellent agreement: It quantitatively confirms both
the parameter values for which current and noise suppressions occur
and the corrections for in the large-amplitude regime $A\gtrsim eV$
\cite{Kohler2004a, Camalet2004a}.

\begin{figure}[tb]
\centerline{\includegraphics{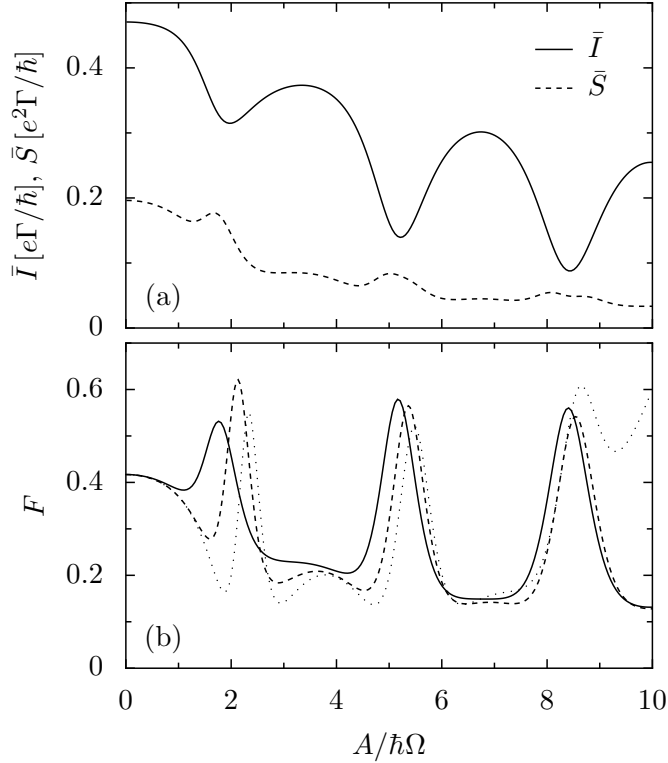}}
\caption{\label{fig:control.lf}
(a) Time-averaged current (solid line) and zero-frequency noise (dashed
line) as a function of the driving amplitude for the driving frequency
$\Omega=\Delta/\hbar$ and the transport voltage $V=48\Delta/e$.
(b) Corresponding Fano factor for the same data (solid line) and for the
driving frequencies $\Omega=1.5\Delta/\hbar$ (broken) and
$\Omega=3\Delta/\hbar$ (dash-dotted).
All other parameters are as in Fig.~\ref{fig:control.A}.
From Ref.~\cite{Camalet2004a}.
}
\end{figure}

For a much lower driving frequency of the order of the wire excitations,
$\Omega=\Delta/\hbar$, the high-frequency approximation is no longer
applicable.  Nevertheless, the average current exhibits clear minima with a
suppression factor of the order of $1/2$; see Fig.~\ref{fig:control.lf}a.
Compared to the high-frequency case, these minima are shifted towards
smaller driving amplitudes, i.e., they occur for ratios $A/\hbar\Omega$
slightly below the zeros of the Bessel function $J_0$.  At the minima of
the current, the Fano factor still assumes a maximum with a value close to
$F\approx 1/2$ (Fig.~\ref{fig:control.lf}b).  Although the sharp minima
close to the current suppressions have vanished, in-between the maxima the
Fano factor assumes remarkably low values of $F\approx 0.2$.

\begin{figure}[tb]
\centerline{\includegraphics{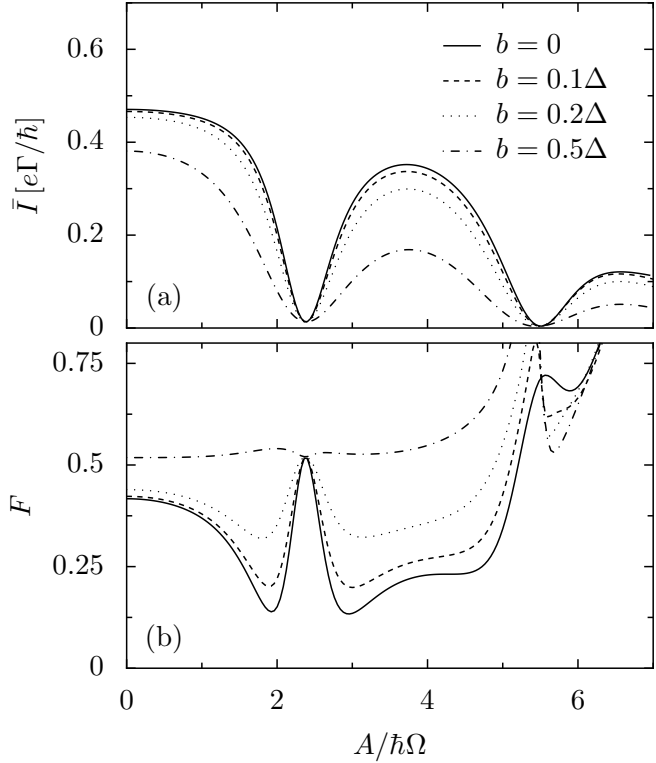}}
\caption{\label{fig:bias} Time-averaged current (a) and Fano factor (b) as
a function of the driving amplitude $A$ for a wire with $N=3$ sites in the
presence of an internal bias.  The on-site energies are $E_1=b$, $E_2=0$,
$E_3=-b$.  All other parameters are as in Fig.~\ref{fig:control.lf}.
From Ref.~\cite{Camalet2004a}.}
\end{figure}

So far, we have assumed that all on-site energies of the wire are
identical.  In an experimental setup, however, the applied transport
voltage acts also a static dipole force which rearranges the charge
distribution in the conductor and thereby causes an internal potential
profile \cite{Nitzan2002a, Pleutin2003a, Liang2004a}.  The self-consistent
treatment of such effects is, in particular in the time-dependent case,
rather ambitious and beyond the scope of this work.  Thus, here we only
derive the consequences of a static bias without determining its shape from
microscopic considerations.  We assume a position-dependent static shift of
the on-site energies by an energy $-b\,x_n$, i.e., for a wire with $N=3$
sites,
\begin{equation}
\label{Ebias}
E_1=b,\quad E_2=0,\quad E_3=-b .
\end{equation}
Figure \ref{fig:bias}a demonstrates that the behavior of the average
current is fairly stable against the bias.  In particular, we still find
pronounced current suppressions.  Note that since $b\ll\Omega$ a
high-frequency approximation is still applicable.  As a main effect of the
bias, we find reduced current maxima while the minima remain.  By
contrast, the minima of the Fano factor (Fig.~\ref{fig:bias}b) become
washed out:  Once the bias becomes of the order of the wire-lead coupling,
$b\approx\Gamma$, the structure in the Fano factor vanishes and we find
$F\approx 1/2$ for all driving amplitudes $A<eV/(N-1)$ [cf.\ the discussion
discussion after Eq.~\eqref{lambda}].  Interestingly, the value of the
Fano factor at current suppressions is bias independent.

\subsection{Current routers}

\begin{figure}[t]
  \centerline{\includegraphics{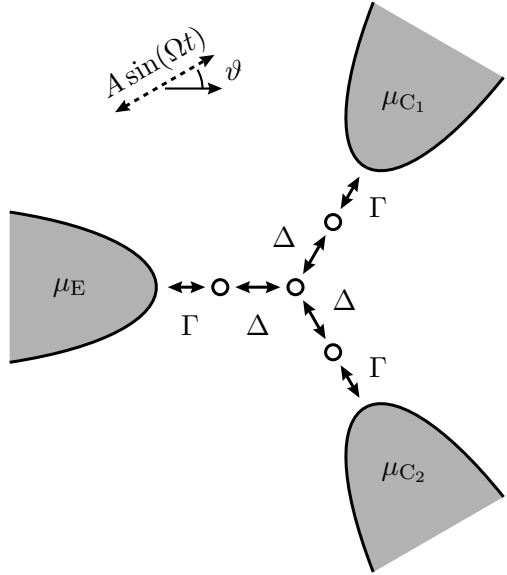}}
  \caption{Schematic top view of a setup where a molecule connected to
    three leads allows one to
  control the current flowing between the different leads (electro-chemical potentials
  $\mu_\mathrm{E}$, $\mu_{\mathrm{C}_1}$, and $\mu_{\mathrm{C}_2}$) as a function of the
  polarization angle $\vartheta$ of a linearly polarized laser field.}
  \label{fig:3contactsystem}
\end{figure}

So far, we have only considered driven transport through two-terminal
devices.  While the experimental realization of three and more
molecular contacts is rather callenging, such systems can be described
theoretically within the present formalism.  As an example, we consider a
planar three-terminal geometry with $N=4$ sites as sketched in
Fig.~\ref{fig:3contactsystem}.
We borrow from electrical engineering the designations $\mathrm{E}$,
$\mathrm{C}_1$, and $\mathrm{C}_2$.  Here, an external voltage is
always applied such that $\mathrm{C}_1$ and $\mathrm{C}_2$ have equal
electro-chemical potential, i.e.\
$\mu_{\mathrm{C}_1}=\mu_{\mathrm{C}_2}\neq\mu_\mathrm{E}$.  In a
perfectly symmetric molecule, where all on-site energies are equal,
reflection symmetry at the horizontal axis ensures that any
current which enters at $\mathrm{E}$ is equally distributed among
$\mathrm{C}_{1,2}$, thus $I_{\mathrm{C}_1} = I_{\mathrm{C}_2} =
-I_\mathrm{E}/2$.

\begin{figure}[t]
  \centerline{\includegraphics{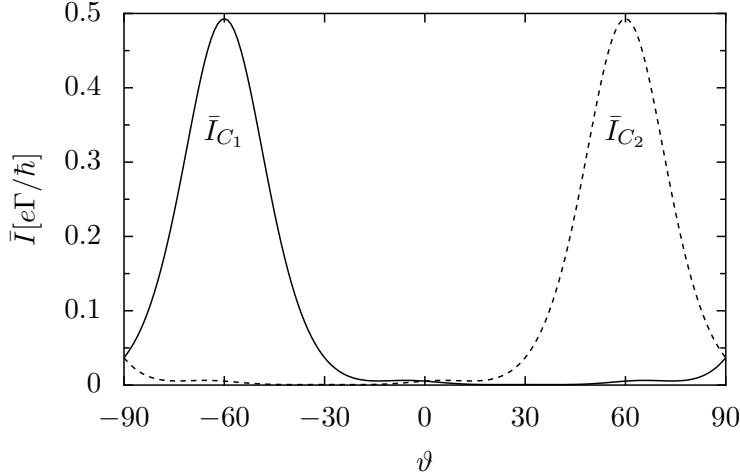}}
  \caption{Average currents (calculated within the master equation
    formalism) through contacts $\mathrm{C}_1$ (solid) and
    $\mathrm{C}_2$ (broken) as a function of the polarization angle
    $\vartheta$ for the three-terminal device depicted in the
    Fig.~\ref{fig:3contactsystem}.  The chemical potentials are
    $\mu_\mathrm{E}=-\mu_{\mathrm{C}_1} =-\mu_{\mathrm{C}_2}=50
    \Delta$; the on-site energies $E_n=0$.  The driving field is
    specified by the strength $A=25\Delta$ and the angular frequency
    $\Omega=10\Delta/\hbar$; the effective coupling is
    $\Gamma=0.1\Delta$ and the temperature $\kBT=0.25\Delta$.  From
    Ref.~\cite{Lehmann2003a}.}
  \label{fig:switch1}
\end{figure}

The fact that this structure is essentially two-dimensional, brings about a
new degree of freedom, namely the polarization of the laser field.  We
assume it to be linear with an polarization angle $\vartheta$ as sketched
in Fig.~\ref{fig:3contactsystem}.  The effective driving amplitudes of the
orbitals that are attached to the leads acquire now a geometric factor
which is only the same for both orbitals $\mathrm{C}_1$ and $\mathrm{C}_2$
when $\vartheta=0$.  For any other polarization angle, the mentioned
symmetry is broken and the outgoing currents may be different from each
other.  The difference may be huge, as exemplified in
Fig.~\ref{fig:switch1}.  There, the current ratio varies from unity for
$\vartheta=0^\circ$ up to the order of $100$ for $\vartheta=60^\circ$.
Thus, adapting the polarization angle enables one to route the current
towards the one or the other drain.

\begin{figure}[t]
  \centerline{\includegraphics{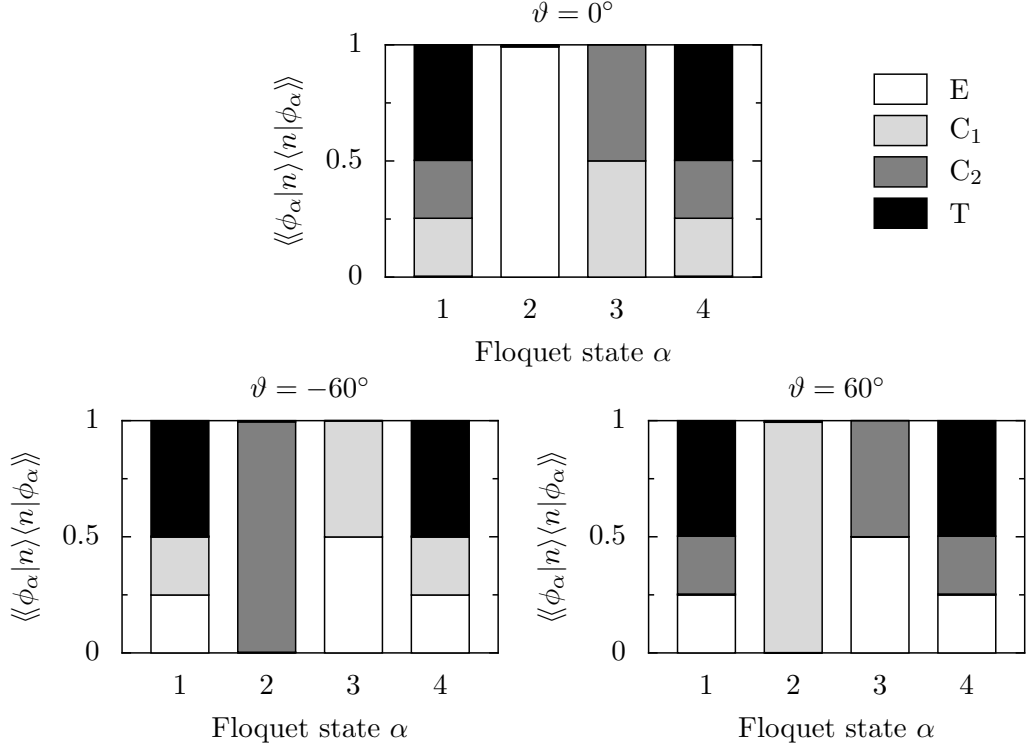}}
  \caption{Time-average $\langle\!\langle \phi_\alpha(t)|
    n\rangle \langle n|\phi_\alpha(t)\rangle\!\rangle$ of the overlaps
    $|\langle n|\phi_\alpha(t)\rangle|^2$ of the sites
    $n=\mathrm{E},\mathrm{C}_1,\mathrm{C}_2$, and $\mathrm{T}$
    (central site) to a Floquet state $|\phi_\alpha(t)\rangle$ for
    three different polarization angles $\vartheta$.  All parameters
    are as in Fig.~\ref{fig:switch1}.  Adapted from Ref.~\cite{Kohler2004c}}
  \label{fig:switch1_populations}
\end{figure}

For a qualitative explanation of the mechanism behind this effect, it
is instructive to look at the time-averages of the overlaps $|\langle
n|\phi_\alpha(t)\rangle|^2=\sum_k |\langle
n|\phi_{\alpha,k}\rangle|^2$ of the Floquet states with the terminal
sites~$n=\mathrm{E},\mathrm{C}_1,\mathrm{C}_2$, which determine the
effective tunneling rates~\eqref{GammaLalpha} and \eqref{GammaRalpha}
in the weak wire-lead coupling limit.
Figure~\ref{fig:switch1_populations} shows these overlaps for three
different polarization angles $\vartheta$.  Let us consider, for
instance, the current across contact~$\mathrm{C}_1$.  It is plausible
that only Floquet modes which have substantial overlap with both the
site~$\mathrm{C}_1$ and also the site~$\mathrm{E}$ contribute the
current through these terminals.  For a polarization
angle~$\vartheta=-60^\circ$, we can infer from
Fig.~\ref{fig:switch1_populations} that the Floquet states with
indices $\alpha=1$, $3$ and $4$ fulfill this condition and,
consequently, a current flows from lead~$\mathrm{E}$ into
lead~$\mathrm{C_1}$.  By contrast, for $\vartheta=0^\circ$ and
$\vartheta=60^\circ$ such current carrying states do not exist and the
respective current vanishes.

\subsection{Phonon damping}

A further question to be addressed is the robustness of the current
suppressions against dissipation.  In the corresponding tunneling problem,
the driving alters both the coherent and the dissipative time scale by the
same factor \cite{Fonseca2004a}.  Thus, one might speculate that a
vibrational coupling leaves the effect of the driving on the current
qualitatively unchanged.  This, however, is not the case: With increasing
dissipation strength, the characteristic current suppressions become washed
out until they finally disappear when the damping strength becomes of the
order of the tunnel coupling $\Delta$ \cite{Lehmann2004a}.  This detracting
influence underlines the importance of quantum coherence for the
observation of those current suppressions.
Moreover, for the model employed in Ref.~\cite{Lehmann2004a}, we do not
find the analogue of the effect of a stabilization of coherent destruction
of tunneling within a certain temperature range \cite{Dittrich1993a,
Dittrich1993b, Makarov1995a} or, likewise, with increasing external
noise \cite{Grossmann1993a}, as it has been reported for driven, dissipative
symmetric bistable systems.

\section{Conclusion and outlook}

In the present survey we have reviewed the role of external driving for
various transport quantities in nanosystems.  In particular, we have
focussed on the possibilities to selectively control, manipulate and
optimize transport through such systems.  In this context,
we have studied various aspects of the electron transport through
time-dependent tight-binding systems.  For the theoretical description, two
formalisms have been employed which both take advantage of the Floquet
theorem: A Floquet scattering approach provides an exact solution of the
time-averaged electrical current beyond the linear response limit and,
moreover, yields an expression for the corresponding noise power.
Interestingly, unlike in the undriven case, the noise
depends also on the phases of the transmission amplitudes.  As a drawback,
this scattering approach is limited to the case of purely coherent transport
in the absence of electron-electron interactions.  As soon as other degrees of
freedom like, e.g.\ a phonon bath, come into play, it is advantageous to
resort to other formalisms like a Floquet master equation approach which,
however, is limited to a weak wire-lead coupling.

We have investigated several driven transport phenomena such as resonant
current amplification (Section~\ref{sec:resonances}), electron pumping
(Section~\ref{sec:pump}), and coherent current control
(Section~\ref{sec:control}).
Of foremost interest in view of ongoing experiments is the enhancement
of molecular conduction by resonant excitations.  We have derived an
analytical expression for the current enhancement factor and,
moreover, have found that the relative current noise is reduced
approximately by a factor of one half.

Both molecular wires and quantum dot arrays can act as coherent quantum
ratchets and thereby operate in a regime where the quantum ratchet dynamics
has not been studied previously.  Of particular practical relevance is the
fact that already relatively short wires or arrays behave like infinite
systems.  For the investigation of such driven
nano-devices, symmetries play a crucial role: The driven nano-system
may exhibit a dynamical symmetry which includes a time transformation.
Breaking this dynamical symmetry, for instance by using a non-harmonic
driving force, can be exploited for the generation of a pump current.
Moreover, the symmetry analysis revealed that a ratchet or pump can only be
observed in the absence of the so-called generalized parity.

Coherent destruction of tunneling has a corresponding transport effect
which exhibits an even richer variety of phenomena.  For driving
parameters, where the tunneling in isolated unbiased systems is suppressed,
the dc current drops to a small residual value.  This effect is found to be
stable against a static bias.  Moreover, the investigation of the corresponding
noise level characterized by the Fano factor, has revealed that the current
suppressions are accompanied by a noise maximum and two remarkably low
minima.  This allows one to selectively control both the current
and its noise by ac fields.
Of crucial interest for potential applications are the noise properties of
non-adiabatic pumps.  For resonant excitations, these can be treated
analytically within an approximation scheme in the spirit of the one
applied in Ref.~\cite{Kohler2004a}.

An experimental realization of the phenomena discussed in this paper is
obviously not a simple problem.  The requirement for asymmetric molecular
structures is easily realized as discussed above, however difficulties
associated with the many possible effects of junction illumination have to
be surmounted \cite{Gerstner2000a}.
Firstly, there is the issue of bringing the light into the junction.  This
is a difficult problem in a break-junction setup but possible in an scanning
probe microscope configuration.
Secondly, in addition to the modulation of electronic states on the
molecular bridge as discussed in this work, other processes involving the
excitation of the metal surface may also affect electron transport.  A
complete theory of illuminated molecular junctions should consider such
possible effects.  Moreover, the junction response to an oscillating
electromagnetic field may involve displacement currents associated with the
junction capacity.
Finally, junction heating may constitute a severe problem when strong
electromagnetic fields are applied.  On the other hand, the light-induced
rectification discussed in this paper is generic in the sense that it does
not require a particular molecular electronic structure as long as an
asymmetry is present.  The prediction that with proper illumination, one
might induce a unidirectional current without any applied voltage, offers
the possibility to observe a pump effect without the background of a
voltage-induced direct current component.

An alternative experimental realization of the presented results is
possible in semiconductor hetero-structures, where, instead of a molecule,
coherently coupled quantum dots \cite{Blick1996a} form the central system.
A suitable radiation source that matches the frequency scales in this case
must operate in the microwave spectral range.  Compared to molecular
wires, these systems by are well-established.  This is evident from the fact
that in microwave-driven coupled quantum dots, electron pumping has already
been observed \cite{vanderWiel2003a}.

The authors share the belief that this survey on driven quantum transport
on the nanoscale provides the reader with a good starting point for future
own research: Many other intriguing phenomena await becoming unraveled.

\section*{Acknowledgements}

During the recent years, we enjoyed many interesting and helpful
discussions on molecular conduction and quantum dots with numerous
colleagues.  In particular, we have benefitted from discussions with
U.~Beierlein,
S.~Camalet,
G.~Cuniberti,
T.~Dittrich,
W.~Domcke,
I.~Goychuk,
M.~Grifoni,
F.~Grossmann,
G.-L.~Ingold,
J.~P.~Kotthaus,
H.~v.~L\"ohneysen,
V.~May,
A.~Nitzan,
H.~Pastawski,
E.~G.~Petrov,
G.~Platero,
M.~Ratner,
P.~Reimann,
M.~Rey,
K.~Richter,
E.~Scheer,
G.~Schmid,
F.~Sols,
M.~Strass,
P.~Talkner,
M.~Thorwart,
M.~Thoss,
H.~B.~Weber,
J.~W\"urfel,
and
S.~Yaliraki.

This work has been supported by the Volkswagen-Stiftung under Grant No.\
I/77 217, the Deutsche Forschungsgemeinschaft through SFB 486, and the
Freistaat Bayern via the quantum information initiative
``Quanteninformation l\"angs der A8''.

\appendix
\section{A primer to Floquet theory}
\label{app:floquet}

In this review, we deal with time-periodically driven quantum systems whose
dynamics is governed by the Schr\"odinger-like equation of motion
\begin{equation}
  \label{a.1}
  \i\hbar\,\frac{\d}{\d t} |\psi(t)\rangle = \big( H(t) -\i \Sigma
  \big) |\psi(t)\rangle
\end{equation}
with the $\mathcal{T}$-periodic Hamiltonian $H(t)=H(t+\T)$.  The hermitian
self-energy term $\Sigma$ results from an elimination of environmental
degrees of freedom and renders the time-evolution non-unitary.

The explicit time-dependence in the Hamiltonian rules out the standard
separation ansatz $|\psi(t)\rangle = \exp(-\i Et/\hbar)
|\varphi\rangle$, where $E$ is the (complex) eigenenergy of a state $|\varphi\rangle$,
for the solution of Eq.~\eqref{a.1}.  Yet, the time-periodicity of the
Hamiltonian allows one to apply Floquet theory, a powerful tool, which we
briefly review in this appendix.

\subsection{Floquet theorem for non-unitary time-evolution}

Floquet theory is based on the Floquet theorem which states that for a
time-periodic Hamiltonian, $H(t)=H(t+\T)$, there exists a complete set
$\{|\psi_\alpha(t)\rangle\}$ of solutions of Eq.~\eqref{a.1} which is of
the form
\begin{equation}
  \label{a.2}
  |\psi_\alpha(t)\rangle =
  \e^{-(\i\epsilon_\alpha/\hbar +\gamma_\alpha ) t} \,|u_\alpha(t)\rangle\ , \quad
  |u_\alpha(t)\rangle= |u_\alpha(t+\T)\rangle\ .
\end{equation}
The time-periodic functions $|u_\alpha(t)\rangle$ are called
Floquet modes or Floquet states and the quantities $\epsilon_\alpha$
are referred to as quasienergies with corresponding
width~$\gamma_\alpha$.  By inserting the ansatz~\eqref{a.2} into Eq.\
\eqref{a.1}, one easily verifies that the Floquet states fulfill the
eigenvalue equation
\begin{equation}
  \label{a.3}
  \Big(H(t) - \i\Sigma - \i\hbar\,\frac{\d}{\d t}\Big) |u_\alpha(t)\rangle =
  (\epsilon_\alpha -\i\hbar\gamma_\alpha)|u_\alpha(t)\rangle\,.
\end{equation}
Different methods can be used to prove the Floquet theorem. Here, we
present a constructive argument.  Upon diagonalization of the
one-period propagator $U(\T,0)$, where $U(t,t')$ is the time-evolution
operator corresponding to the dynamical equation~\eqref{a.1}, we obtain
\begin{equation}
  \label{a.4}
  U(\T,0) |u_\alpha(0)\rangle = \e^{-(\i\epsilon_\alpha/\hbar + \gamma_\alpha) \T}\,
  |u_\alpha(0)\rangle\ .
\end{equation}
Here, we have written the complex eigenvalue as exponential for some
$\epsilon_\alpha$ and $\gamma_\alpha$.  Next, we use the eigenstates
$|u_\alpha(0)\rangle$ as initial states for the time-evolution
according to Eq.~\eqref{a.1}, yielding the solutions
$|\psi_\alpha(t)\rangle = U(t,0)|u_\alpha(0)\rangle$ of
Eq.~\eqref{a.1}.  This allows us to define the Floquet modes
$|u_\alpha(t)\rangle=\exp[(\i\epsilon_\alpha/\hbar + \gamma_\alpha)
t] |\psi_\alpha(t)\rangle$, which are indeed $\T$-periodic functions:
\begin{equation}
  \begin{split}
  |u_\alpha(t+\T)\rangle
  & =
  \e^{(\i\epsilon_\alpha/\hbar+\gamma_\alpha)(t+\T)}\,
  U(t+\T,0) |u_\alpha(0)\rangle\\
  & =
  \e^{(\i\epsilon_\alpha/\hbar+\gamma_\alpha)(t+\T)}\,
  U(t,0)\, U(\T,0) |u_\alpha(0)\rangle \\
  & =
  \e^{(\i\epsilon_\alpha/\hbar+\gamma_\alpha)t}\,
  |\psi_\alpha(t)\rangle  =
  |u_\alpha(t)\rangle\ .
  \end{split}
\end{equation}
In the second line, we have used that owing to the time-periodicity of
the Hamiltonian, the relation $U(t+\T,\T)= U(t,0)$ holds true for
arbitrary time~$t$.  Finally, the completeness of the set of
solutions $\{|\psi_\alpha(t)\rangle \}$ follows, if we assume the
completeness of the eigenstates of $U(\T,0)$.

Since the one-period propagator $U(\T,0)$ is in general non-unitary,
its eigenstates $|u_\alpha(0)\rangle$ are not mutually
orthogonal. We therefore also have to consider the left eigenstates of
$U(\T,0)$, i.e., the solutions of the adjoint equation
\begin{equation}
  \label{a.5}
  \Big(H(t) + \i\Sigma^\mathrm{T} - \i\hbar\,\frac{\d}{\d t}\Big) |u^+_\alpha(t)\rangle =
  (\epsilon_\alpha + \i\hbar\gamma_\alpha)|u^+_\alpha(t)\rangle\ .
\end{equation}
Here, we have used the fact that the eigenvalues of the adjoint equation are
the complex conjugates of the eigenvalues of the original
eigenvalue equation \eqref{a.3}.  This follows from the secular equations
corresponding to the eigenvalue problems~\eqref{a.3} and \eqref{a.5}
by using the relation $\det O = \det O^\mathrm{T}$, which holds for an
arbitrary operator~$O$.  Assuming completeness of the eigenstates of
$U(\T,0)$, the Floquet modes and its adjoint modes may be chosen to form a
bi-orthonormal basis at equal times~$t$,
\begin{equation}
  \label{a.6}
  \langle u^+_\alpha(t)|u_\beta(t)\rangle =
  \delta_{\alpha\beta}
  \quad
  \text{and}
  \quad
  \sum_\alpha |u_\alpha^+(t)\rangle\langle u_\alpha(t)| =
  \mathsf{1}
  \ .
\end{equation}

The time-evolution operator $U(t,t')$ can be expressed explicitly in terms
of the Floquet modes and quasi-energies to read
\begin{equation}
  \label{a.7}
  U(t,t') = \sum_\alpha
  \e^{-\i(\epsilon_\alpha/\hbar+\gamma_\alpha)(t-t')}
  |u_\alpha(t)\rangle\langle u^+_\alpha(t')| \,.
\end{equation}
This relation is readily checked by noting that due to Eq.~\eqref{a.2}
the right-hand side solves the differential equation~\eqref{a.1}.
The initial condition $U(t,t)=\mathsf{1}$ is ensured by
the completeness~\eqref{a.6} of the Floquet modes.

It is worthwhile to remark that the conceptual importance of Floquet
theory lies in the fact that it allows one to separate the long-time
dynamics, governed by the
eigenvalues~$\epsilon_\alpha-\i\hbar\gamma_\alpha$, from the dynamics
within one driving period, determined by Floquet
modes~$|u_\alpha(t)\rangle$ [cf.\ Eq.~\eqref{a.2}].
Note also that
the quasienergies and the Floquet states in Eq.~\eqref{a.2} are not
defined uniquely.  In fact, the replacement
\begin{equation}
  \label{2.5}
  \epsilon_\alpha \to \epsilon_\alpha + k_\alpha \hbar\Omega \ ,\quad
  |u_\alpha(t)\rangle \to \e^{\i k_\alpha \Omega t} \, |u_\alpha(t)\rangle\ ,
\end{equation}
where $\Omega= 2\pi/\T$ is the angular frequency of the driving and
$\{k_\alpha\}$ is an arbitrary sequence of integers, yields a new set
of quasienergies and Floquet states corresponding to the same solutions
$\left\{|\psi_\alpha(t)\rangle\right\}$ of Eq.~\eqref{a.1}.  In other words, the
quasienergies and Floquet modes come in classes, out of which one is
allowed to select a single representative, usually with quasienergy in
a single ``Brillouin zone'' $E-\hbar\Omega/2\le \epsilon_\alpha <
E+\hbar\Omega/2$, where $E$ is an arbitrary but fixed energy.

\subsection{Extended Hilbert space formalism}

\label{sec:sambe}

According to the basic postulates of quantum mechanics, the state of a
system is described by a vector~$|\psi\rangle$ in a Hilbert
space~$\mathbb{R}$ with the inner product~$\langle \psi'| \psi
\rangle$.  Without loss of the generality, we assume that
there exists a countable and complete set~$\{|n\rangle\}$ of orthonormal
states, i.e.,
\begin{equation}
  \label{2.6}
  \langle n|n'\rangle = \delta_{nn'}
  \ ,\quad
  \sum_n |n\rangle\langle n| = \mathsf{1}\ .
\end{equation}

The Hilbert space $\mathbb{T}$ of all $\T$-periodic, complex-valued
functions possesses the inner product
\begin{equation}
  \label{2.7}
  (u,v) =  \frac{1}{\T}\int_0^\T\!\d t\, u^\ast(t) v(t)
\end{equation}
and the functions $\exp(\i k\Omega t)$ with $k=0,\pm1, \pm2, \dots$ form
the corresponding complete and orthonormal set. The decomposition of
an arbitrary $\T$-periodic, complex-valued function into this basis
yields the standard Fourier series.

As first noted by Sambe~\cite{Sambe1973a}, the time-periodicity of the
Floquet modes suggests their description in the composite Hilbert
space $\mathbb{R}\otimes\mathbb{T}$.  Its elements, for which we
adopt the notation $|u\drangle$~\cite{Sambe1973a}, are the
$\T$-periodic state vectors $|u(t)\rangle = |u(t+\T)\rangle$.
Introducing the inner product in this space in the canonical way via
\begin{equation}
  \label{2.8}
  \dlangle u'|u\drangle =
  \frac{1}{\T} \int_0^\T \!\d t \, \langle u'(t)| u(t) \rangle\ ,
\end{equation}
an orthogonal basis of $\mathbb{R}\otimes\mathbb{T}$ is given by the set of
states $\{|u^k_n\drangle\}$ defined by
\begin{equation}
  \label{2.8a}
  |u^k_n(t)\rangle = \e^{\i k\Omega t}\, |n\rangle\ .
\end{equation}
The arbitrary integer $k$ is sometimes called the sideband index. The
decomposition of a state $|u(t)\rangle$ into this basis is
equivalent to the Fourier representation
\begin{equation}
  \label{2.8b}
  \begin{split}
    |u_\alpha(t)\rangle & = \sum_k \e^{-\i k\Omega t} \,|u_{\alpha,k}\rangle\ ,\\
    |u_{\alpha,k}\rangle  & = \frac{1}{\T} \int_0^\T \!\d t\, \e^{\i k \Omega t} \, |u_\alpha(t)\rangle\ .
  \end{split}
\end{equation}
It is important, however, to always keep in mind that the states $|u_{\alpha,k}\rangle$
are not orthogonal, because the Floquet modes are only mutually orthogonal at equal times
[cf.\ Eq.~\eqref{a.6}].

By the introduction of a Hilbert space structure for the time
dependence, we have formally traced back the computation of Floquet states
to the computation of eigenstates of a time-independent Hamiltonian
with an additional degree of freedom. In particular, in the composite
Hilbert space the Floquet equation~\eqref{a.3} maps to the
time-independent eigenvalue problem
\begin{equation}
  \label{floquetsambe}
  \left(\mathcal{H}(t) - \i\Sigma \right)|u_\alpha\drangle =
  \epsilon_\alpha |u_\alpha\drangle
\end{equation}
with the so-called Floquet Hamiltonian
\begin{equation}
 \mathcal H(t) =  H(t)-\i\hbar\frac{\d}{\d t}\ .
\end{equation}

A wealth of methods for the solution of this eigenvalue problem can be
found in the literature \cite{Hanggi1998a, Grifoni1998a}.  One such
method is given by the direct numerical diagonalization of the
operator on left-hand side of Eq.~\eqref{floquetsambe}. For a harmonic
driving, the eigenvalue problem \eqref{floquetsambe} is band-diagonal
and selected eigenvalues and eigenvectors can be computed by a
matrix-continued fraction scheme \cite{Risken, Hanggi1998a}.

In cases where many Fourier coefficients (in the present context frequently
called ``sidebands'') must be taken into account for the decomposition
\eqref{2.8b}, direct diagonalization is often not very efficient and one
has to apply more elaborated schemes.  For example, in the case of a large
driving amplitude, one can treat the static part of the Hamiltonian as a
perturbation \cite{Sambe1973a, Grossmann1992a, Holthaus1992a}. The Floquet
states of the oscillating part of the Hamiltonian then form an adapted
basis set for a subsequently more efficient numerical diagonalization.

A completely different strategy to obtain the Floquet states is to
propagate the Schr\"odinger equation for a complete set of initial
conditions over one driving period to yield the one-period propagator.  Its
eigenvalues represent the Floquet states at time $t=0$, i.e.,
$|u_{\alpha}(0)\rangle$.  Fourier transformation of their time-evolution
results in the desired sidebands.  Yet another, very efficient propagation
scheme is the so-called $(t,t')$-formalism \cite{Peskin1993a}.

\subsection{Parity of a system under dipole driving}
\label{app:symmetry}

Although we focus in this work on tight-binding systems, it is more
convenient to study symmetries as a function of a continuous position and
to regard the discrete models as a limiting case.  Moreover, we consider
in this section the Hamiltonian of the entire system including the leads.
Consequently, we do not have not include any self-energy contribution.

A static Hamiltonian $H_0(x)$ is called invariant under the parity
transformation $\mathcal{P}:x\to-x$ if it is an even function of $x$.
Then, its eigenfunctions $\varphi_\alpha$ can be divided into two classes:
even and odd ones, according to the sign in $\varphi_\alpha(x) =
\pm\varphi_\alpha(-x)$.
Adding a periodically time-dependent dipole force $xa(t)$ to such a
Hamiltonian evidently breaks parity symmetry since $\mathcal{P}$ changes the
sign of the interaction with the radiation.
In a Floquet description, however, we deal with states that are functions
of both position and time---we work in the extended space
$\mathbb{R}\otimes\mathbb{T}$.
Instead of the stationary Schr\"odinger equation, we address the eigenvalue
problem
\begin{equation}
\label{floquet_app}
\mathcal{H}(x,t)\,\phi(x,t) = \epsilon\,\phi(x,t)
\end{equation}
with the Floquet Hamiltonian for zero self-energy given by
\begin{equation}
\label{H+xf}
\mathcal{H}(t)=H_0(x) + xa(t)-\i\hbar\frac{\partial}{\partial t} ,
\end{equation}
where we assume a symmetric static part, $H_0(x)=H_0(-x)$.  Our aim is now
to generalize the notion of parity to the extended space
$\mathbb{R}\otimes\mathbb{T}$ such
that the overall transformation leaves the Floquet equation
\eqref{floquet_app} invariant.  This can be achieved if the shape of the
driving $a(t)$ is such that an additional time transformation ``repairs''
the acquired minus sign.  We consider two types of transformation:
generalized parity and time-reversal parity.  Both occur for purely
harmonic driving, $a(t)=\cos(\Omega t)$.  In the following we derive their
consequences for the Fourier coefficients
\begin{equation}
\label{fourier}
\phi_k(x) = \frac{1}{T}\int_0^T \!\!\!\d t\, \e^{\i k\Omega t}\phi(x,t)
\end{equation}
of a Floquet states $\phi(x,t)$.

\subsubsection{Time-reversal symmetry}
\label{app:Tsymmetry}

Before discussing parity symmetry, let us comment on time-reversal symmetry
which is not relevant for the spectral properties but still has some
computational importance.  It is known that the energy eigenfunctions of an
non-driven Hamiltonian, which obeys time-reversal symmetry, can be chosen
real \cite{Sakurai}.  Time-reversal symmetry is typically broken by a
magnetic field (recall that a magnetic field is described by an axial
vector and, thus, changes its sign under time-reversion) or by an explicit
time-dependence of the Hamiltonian.  However, for a cosine driving,
time-reversal symmetry
\begin{equation}
\label{app:S_T}
\mathcal{S}_\mathrm{T}:\, t\to -t ,
\end{equation}
is retained and the Floquet Hamiltonian \eqref{H+xf} obeys $\mathcal{H}(t)
= [\mathcal{H}(-t)]^*$.  With the same line of reasoning as in the case of
time-reversal symmetry, but with the additional replacement $x\to -x$, we
obtain that one can choose the Floquet states with
$\phi(x,t)=\phi^*(x,-t)$.  Then, the Fourier coefficients \eqref{fourier}
are real
\begin{equation}
\label{app:phiT}
\phi_k(x)=\phi_k^*(x) ,
\end{equation}
which helps to reduce numerical effort.

\subsubsection{Time-reversal parity}
\label{app:TPsymmetry}

A further symmetry is found if $a$ is an odd function of time,
$a(t)=-a(-t)$, e.g.\ for $a(t)=\sin(\Omega t)$.  Then, time inversion
transforms the Floquet Hamiltonian \eqref{H+xf} into its complex conjugate
such that the corresponding symmetry is given by the anti-linear
transformation
\begin{equation}
\label{app:S_TP}
\mathcal{S}_\mathrm{TP}:\,(\phi,x,t)\to(\phi^*,-x,-t).
\end{equation}
This transformation represents a generalization of the parity
$\mathcal P$; we will refer to it as \textit{time-reversal parity} since
in the literature the term generalized parity is mostly used
in the context of the transformation \eqref{app:S_GP}.

Again we are interested in the Fourier decomposition~\eqref{fourier}
and obtain
\begin{equation}
\label{app:phiTP}
\phi_k(x)=\phi_k^*(-x) .
\end{equation}

The time-reversal discussed here can be generalized by an additional
time-shift to read $t\to t_0-t$.  Then, we find by the same line of
argumentation that $\phi_k(x)$ and $\phi_k^*(-x)$ differ at most by a phase
factor.  However, for convenience one may choose already from the start the
origin of the time axis such that $t_0=0$.

\subsubsection{Generalized parity}
\label{app:GPsymmetry}

It has been noted \cite{Grossmann1991a, Grossmann1991b, Peres1991a} that a
Floquet Hamiltonian of the form \eqref{H+xf} with $a(t)=\sin(\Omega t)$ may
possess degenerate quasienergies due to its symmetry under the so-called
generalized parity transformation
\begin{equation}
\label{app:S_GP}
\mathcal{S}_\mathrm{GP}:\,(x,t)\to(-x,t+\pi/\Omega) ,
\end{equation}
which consists of spatial parity plus a time shift by half a driving period.
This symmetry is present in the Floquet Hamiltonian \eqref{H+xf}, if
the driving field obeys $a(t)=-a(t+\pi/\Omega)$, since
then $\mathcal{S}_\mathrm{GP}$ leaves the Floquet equation invariant.
Owing to $\mathcal S_\mathrm{GP}^2=1$, we find that
the corresponding Floquet states are either even or odd,
$\mathcal{S}_\mathrm{GP}\phi(x,t)=\phi(-x,t+\pi/\Omega)=\pm\phi(x,t)$.
Consequently, the Fourier coefficients \eqref{fourier} obey the relation
\begin{equation}
\label{app:phiGP}
\phi_k(x)=\pm(-1)^k\phi_k(-x) .
\end{equation}




\begin{thebibliography}{100}
\expandafter\ifx\csname url\endcsname\relax
  \def\url#1{\texttt{#1}}\fi
\expandafter\ifx\csname urlprefix\endcsname\relax\def\urlprefix{URL }\fi

\bibitem{Feynman1960a}
R.~P. Feynman, There's plenty of room at the bottom, Eng. Sci. 23 (1960) 22,
  lecture given at the APS meeting 1959, see
  \texttt{http://www.its.caltech.edu/\discretionary{}{}{}\textasciitilde
  feynman/plenty.html}.

\bibitem{Binnig1984a}
G.~Binnig, H.~Rohrer, Scanning tunneling microscopy, Physica B \& C 127 (1984)
  37.

\bibitem{Aviram1974a}
A.~Aviram, M.~A. Ratner, Molecular rectifiers, Chem. Phys. Lett. 29 (1974) 277.

\bibitem{Mann1971a}
B.~Mann, H.~Kuhn, Tunneling through fatty acid salt monolayers, J. Appl. Phys.
  42 (1971) 4398.

\bibitem{Reed1997a}
M.~A. Reed, C.~Zhou, C.~J. Muller, T.~P. Burgin, J.~M. Tour, Conductance of a
  molecular junction, Science 278 (1997) 252.

\bibitem{Cui2001a}
X.~D. Cui, A.~Primak, X.~Zarate, J.~Tomfohr, O.~F. Sankey, A.~L. Moore, T.~A.
  Moore, D.~Gust, G.~Harris, S.~M. Lindsay, Reproducible measurement of
  single-molecule conductivity, Science 294 (2001) 571.

\bibitem{Reichert2002a}
J.~Reichert, R.~Ochs, D.~Beckmann, H.~B. Weber, M.~Mayor, H.~von L\"ohneysen,
  Driving current through single organic molecules, Phys. Rev. Lett. 88 (2002)
  176804.

\bibitem{Nitzan2003a}
A.~Nitzan, M.~A. Ratner, Electron transport in molecular wire junctions,
  Science 300 (2003) 1384.

\bibitem{Heath2003a}
J.~R. Heath, M.~A. Ratner, Molecular electronics, Physics Today 56~(5) (2003)
  43.

\bibitem{Hanggi2002elsevier}
{P. H\"anggi, M. Ratner, S. Yaliraki, \emph{Processes in Molecular Wires},
  Chem. Phys. 281 (2002) 111.}

\bibitem{Balzani2003a}
V.~Balzani, M.~Venturi, A.~Credi, Molecular Devices and Machines, Wiley-VCH,
  Weinheim, 2003.

\bibitem{Goser2004a}
K.~Goser, P.~Gl\"osek\"otter, J.~Dienstuhl, Nanoelectronics and Nanosystems:
  From Transisitors to Molecular and Quantum Devices, 1st Edition, Springer,
  Berlin and Heidelberg, 2004.

\bibitem{Cuniberti2005a}
G.~Cunibert, G.~Fagas, K.~Richter (Eds.), Molecular Electronics, Springer,
  Berlin, 2005.

\bibitem{DiVentra2000a}
M.~Di~Ventra, S.~T. Pantelides, N.~D. Lang, First principles calculation of
  transport properties of a molecular device, Phys. Rev. Lett. 84 (2000) 979.

\bibitem{DiVentra2002a}
M.~Di~Ventra, N.~D. Lang, Transport in nanoscale conductors from first
  principles, Phys. Rev. B 65 (2002) 045402.

\bibitem{Xue2002a}
Y.~Xue, S.~Datta, M.~A. Ratner, First-principles based matrix green's function
  approach to molecular electronic devices: general formalism, Chem. Phys. 281
  (2002) 151.

\bibitem{Damle2002a}
P.~Damle, A.~W. Ghosh, S.~Datta, First-principles analysis of molecular
  conduction using quantum chemistry software, Chem. Phys. 281 (2002) 171.

\bibitem{Heurich2002a}
J.~Heurich, J.~C. Cuevas, W.~Wenzel, G.~Sch\"on, Electrical transport through
  single-molecule junctions: From molecular orbitals to conduction channels,
  Phys. Rev. Lett. 88 (2002) 256803.

\bibitem{Evers2004a}
F.~Evers, F.~Weigend, M.~Koentopp, Conductance of molecular wires and transport
  calculations based on density-functional theory, Phys. Rev. B 69 (2004)
  235411.

\bibitem{Mujica1994a}
V.~Mujica, M.~Kemp, M.~A. Ratner, Electron conduction in molecular wires. {I.
  A} scattering formalism, J. Chem. Phys. 101 (1994) 6849.

\bibitem{Segal2000a}
D.~Segal, A.~Nitzan, W.~B. Davis, M.~R. Wasielewski, M.~A. Ratner, Electron
  transfer rates in bridged molecular systems 2: A steady-state analysis of
  coherent tunneling and thermal relaxation, J. Phys. Chem. 104 (2000) 3817.

\bibitem{Boese2001a}
D.~Boese, H.~Schoeller, Influence of nanomechanical properties on
  single-electron tunneling: A vibrating single-electron transistor, Europhys.
  Lett. 54 (2001) 668.

\bibitem{Petrov2001a}
E.~G. Petrov, P.~H\"anggi, Nonlinear electron current through a short molecular
  wire, Phys. Rev. Lett. 86 (2001) 2862.

\bibitem{Nitzan2001a}
A.~Nitzan, Electron transmission through molecules and molecular interfaces,
  Annu. Rev. Phys. Chem. 52 (2001) 681.

\bibitem{Hettler2003a}
M.~H. Hettler, W.~Wenzel, M.~R. Wegewijs, H.~Schoeller, Current collapse in
  tunneling transport through benzene, Phys. Rev. Lett. 90 (2003) 076805.

\bibitem{Olson1998a}
M.~Olson, Y.~Mao, T.~Windus, M.~Kemp, M.~A. Ratner, N.~Leon, V.~Mujica, A
  conformational study of the influence of vibrations on conduction in
  molecular wires, J. Phys. Chem. B 102 (1998) 941.

\bibitem{Yu1999a}
Z.~G. Yu, D.~L. Smith, A.~Saxena, A.~R. Bishop, Green's function approach for a
  dynamical study of transport in metal/organic/metal structures, Phys. Rev. B
  59 (1999) 16001.

\bibitem{Emberly2000a}
E.~G. Emberly, G.~Kirczenow, Landauer theory, inelastic scattering, and
  electron transport in molecular wires, Phys. Rev. B 61 (2000) 5740.

\bibitem{Mikrajuddin2000a}
Mikrajuddin, K.~Okuyama, F.~G. Shi, Mechanical effect on the electronic
  properties of molecular wires, Phys. Rev. B 61 (2000) 8224.

\bibitem{Ness2001a}
H.~Ness, S.~A. Shevlin, A.~J. Fisher, Coherent electron-phonon coupling and
  polaronlike transport in molecular wires, Phys. Rev. B 63 (2001) 125422.

\bibitem{Petrov2002a}
E.~G. Petrov, V.~May, P.~H\"anggi, Controlling electron transfer processes
  through short molecular wires, Chem. Phys. 281 (2002) 211.

\bibitem{May2002a}
V.~May, Electron transfer through a molecular wire: Consideration of
  electron-vibrational coupling within the liouville space pathway technique,
  Phys. Rev. B 66 (2002) 245411.

\bibitem{Petrov2004a}
E.~G. Petrov, V.~May, P.~H\"anggi, Spin-boson description of electron
  transmission through a molecular wire, Chem. Phys. 296 (2004) 251--266.

\bibitem{Fagas2001a}
G.~Fagas, G.~Cuniberti, K.~Richter, Electron transport in nanotube-molecular
  wire hybrids, Phys. Rev. B 63 (2001) 045416.

\bibitem{Cuniberti2002a}
G.~Cuniberti, G.~Fagas, K.~Richter, Fingerprints of mesoscopic leads in the
  conductance of a molecular wire, Chem. Phys 281 (2002) 465.

\bibitem{Gutierrez2002a}
R.~Guti\'errez, G.~Fagas, G.~Cuniberti, F.~Großmann, K.~Richter, R.~Schmidt,
  Theory of an all-carbon molecular switch, Phys. Rev. B 65 (2002) 113410.

\bibitem{Blick1996a}
R.~H. Blick, R.~J. Haug, J.~Weis, D.~Pfannkuche, K.~von Klitzing, K.~Eberl,
  Single-electron tunneling through a double quantum dot: The artificial
  molecule, Phys. Rev. B 53 (1996) 7899.

\bibitem{vanderWiel2003a}
W.~G. van~der Wiel, S.~{D}e Francesoni, J.~M. Elzerman, T.~Fujisawa,
  S.~Tarucha, L.~P. Kouwenhoven, Electron transport through double quantum
  dots, Rev. Mod. Phys. 75 (2003) 1.

\bibitem{Fujisawa1997a}
T.~Fujisawa, S.~Tarucha, Photon assisted tunnelling in single and coupled
  quantum dot systems, Superlatt. Microstruct. 21 (1997) 247.

\bibitem{Oosterkamp1998a}
T.~H. Oosterkamp, T.~Fujisawa, W.~G. van~der Wiel, K.~Ishibashi, R.~V. Hijman,
  S.~Tarucha, L.~P. Kouwenhoven, Microwave spectroscopy of a quantum-dot
  molecule, Nature 395 (1998) 873.

\bibitem{Platero2004a}
G.~Platero, R.~Aguado, Photon-assisted transport in semiconductor
  nanostructures, Phys. Rep. 395 (2004) 1.

\bibitem{Thouless1983a}
D.~J. Thouless, Quantization of particle transport, Phys. Rev. B 27 (1983)
  6083.

\bibitem{Altshuler1999a}
B.~L. Altshuler, L.~I. Glazman, Pumping electrons, Science 283 (1999) 1864.

\bibitem{Switkes1999a}
M.~Switkes, C.~M. Marcus, K.~Campman, A.~C. Gossard, An adiabatic quantum
  electron pump, Science 283 (1999) 1905.

\bibitem{Wagner1999a}
M.~Wagner, F.~Sols, Subsea electron transport: Pumping deep within the {F}ermi
  sea, Phys. Rev. Lett. 83 (1999) 4377.

\bibitem{Levinson2000a}
Y.~Levinson, O.~Entin-Wohlman, P.~W\"olfle, Acoustoelectric current and pumping
  in a ballistic point contact, Phys. Rev. Lett 85 (2000) 634.

\bibitem{Landauer1957a}
R.~Landauer, Spatial variation of currents and fields due to localized
  scatterers in metallic conduction, IBM J. Res. Dev. 1 (1957) 223.

\bibitem{Imry1986a}
Y.~Imry, Introduction to Mesoscopic Physics, Vol.~1 of Mesoscopic Physics and
  Nanotechnology, Oxford University Press, New York, 1986.

\bibitem{Datta1995a}
S.~Datta, Electronic Transport in Mesoscopic Systems, Cambridge University
  Press, Cambridge, 1995.

\bibitem{Landauer1992a}
R.~Landauer, Conductance from transmission: Common sense points, Phys. Scr. T42
  (1992) 110.

\bibitem{Imry1999a}
Y.~Imry, R.~Landauer, Conductance viewed as transmission, Rev. Mod. Phys. 71
  (1999) S306.

\bibitem{Blanter2000a}
{Y}a. M.~Blanter, M.~B\"uttiker, Shot noise in mesoscopic conductors, Phys.
  Rep. 336 (2000) 1.

\bibitem{Henseler2000a}
M.~Henseler, T.~Dittrich, K.~Richter, Signatures of chaos and tunneling in
  {AC}-driven quantum scattering, Europhys. Lett. 49 (2000) 289.

\bibitem{Henseler2001a}
M.~Henseler, T.~Dittrich, K.~Richter, Classical and quantum periodically driven
  scattering in one dimension, Phys. Rev. E 64 (2001) 046218.

\bibitem{Li1999a}
W.~Li, L.~E. Reichl, Floquet scattering through a time-periodic potential,
  Phys. Rev. B 60 (1999) 15732.

\bibitem{Datta1992a}
S.~Datta, M.~P. Anantram, Steady-state transport in mesoscopic systems
  illuminated by alternating fields, Phys. Rev. B 45 (1992) 13761.

\bibitem{Wagner2000a}
M.~Wagner, Probing {P}auli blocking factors in quantum pumps with broken
  time-reversal symmetry, Phys. Rev. Lett. 85 (2000) 174.

\bibitem{Jauho1994a}
A.-P. Jauho, N.~S. Wingreen, Y.~Meir, Time-dependent transport in interacting
  and noninteracting resonant-tunneling systems, Phys. Rev. B 50 (1994) 5528.

\bibitem{Stafford1996a}
C.~A. Stafford, N.~S. Wingreen, Resonant photon-assisted tunneling through a
  double quantum dot: An electron pump from spatial rabi oscillations, Phys.
  Rev. Lett. 76 (1996) 1916.

\bibitem{Pretre1996a}
A.~Pr\^etre, H.~Thomas, M.~B\"uttiker, Dynamic admittance of mesoscopic
  conductors: Discrete-potential model, Phys. Rev. B 54 (1996) 8130.

\bibitem{Pedersen1998a}
M.~H. Pedersen, M.~B\"uttiker, Scattering theory of photon-assisted electron
  transport, Phys. Rev. B 58 (1998) 12993.

\bibitem{Lesovik1994a}
G.~B. Lesovik, L.~S. Levitov, Noise in an ac biased junction: Nonstationary
  aharonov-bohm effect, Phys. Rev. Lett. 72 (1994) 538.

\bibitem{Camalet2003a}
S.~Camalet, J.~Lehmann, S.~Kohler, P.~H\"anggi, Current noise in ac-driven
  nanoscale conductors, Phys. Rev. Lett. 90 (2003) 210602.

\bibitem{Camalet2004a}
S.~Camalet, S.~Kohler, P.~H\"anggi, Shot-noise control in ac-driven nanoscale
  conductors, Phys. Rev. B, in press; arXiv:cond-mat/0402182.

\bibitem{Lehmann2003b}
J.~Lehmann, S.~Kohler, P.~H\"anggi, A.~Nitzan, Rectification of laser-induced
  electronic transport through molecules, J. Chem. Phys. 118 (2003) 3283.

\bibitem{Lehmann2004a}
J.~Lehmann, S.~Kohler, V.~May, P.~H\"anggi, Vibational effects in laser-driven
  molecular wires, J. Chem. Phys. 121 (2004) 2278.

\bibitem{Kohler2002a}
S.~Kohler, J.~Lehmann, S.~Camalet, P.~H\"anggi, Resonant laser excitation of
  molecular wires, Israel J. Chem. 42 (2002) 135.

\bibitem{Kohler2004b}
S.~Kohler, J.~Lehmann, M.~Strass, P.~H\"anggi, Molecular wires in
  electromagnetic fields, Adv. Solid State Phys. 44 (2004) 151.

\bibitem{Hanggi1996a}
P.~H\"anggi, R.~Bartussek, {B}rownian rectifiers: How to convert {B}rownian
  motion into directed transport, in: J.~Parisi, S.~C. M\"uller,
  W.~W.~Zimmermann (Eds.), Nonlinear Physics of Complex Systems---Current
  Status and Future Trends, Vol. 476 of Lecture Notes in Physics, Springer,
  Berlin, 1996, pp. 294--308.

\bibitem{Astumian1997a}
R.~D. Astumian, Thermodynamics and kinetics of a {B}rownian motor, Science 276
  (1997) 917.

\bibitem{Julicher1997a}
F.~J\"ulicher, A.~Adjari, J.~Prost, Modeling molecular motors, Rev. Mod. Phys.
  69 (1997) 1269.

\bibitem{Reimann2002a}
P.~Reimann, {B}rownian motors: Noisy transport far from equilibrium, Phys. Rep.
  361 (2002) 57.

\bibitem{Reimann2002b}
P.~Reimann, P.~H\"anggi, Introduction to the physics of {B}rownian motors,
  Appl. Phys. A 75 (2002) 169.

\bibitem{Astumian2002a}
R.~D. Astumian, P.~H\"anggi, {B}rownian motors, Physics Today 55~(11) (2002)
  33.

\bibitem{Reimann1997a}
P.~Reimann, M.~Grifoni, P.~H\"anggi, Quantum ratchets, Phys. Rev. Lett. 79
  (1997) 10.

\bibitem{Grifoni2002a}
M.~Grifoni, M.~S. Ferrreira, J.~Peguiron, J.~B. Majer, Qauntum ratchets with
  few bands below the barrier, Phys. Rev. Lett. 89 (2002) 146801.

\bibitem{Brouwer1998a}
P.~W. Brouwer, Scattering approach to parametric pumping, Phys. Rev. B 58
  (1998) 10135.

\bibitem{Wang2002a}
B.~Wang, J.~Wang, H.~Guo, Parametric pumping at finite temperature, Phys. Rev.
  B 65 (2002) 073306.

\bibitem{Linke1999a}
H.~Linke, T.~E. Humphrey, A.~L\"ofgren, A.~O. Shuskov, R.~Newbury, R.~P.
  Taylor, P.~Omling, Experimental tunneling ratchets, Science 286 (1999) 2314.

\bibitem{Linke2002a}
H.~Linke, T.~E. Humphrey, P.~E. Lindelof, A.~Lofgren, R.~Newbury, P.~Omling,
  A.~O. Sushkov, R.~P. Taylor, H.~Xu, Quantum ratchets and quantum heat pumps,
  Appl. Phys. A 75 (2002) 237--246.

\bibitem{Majer2003a}
J.~B. Majer, J.~Peguiron, M.~Grifoni, M.~Tusveld, J.~E. Mooij, Quantum ratchet
  effect for vortices, Phys. Rev. Lett. 90 (2003) 056802.

\bibitem{Haan2004a}
S.~de~Haan, A.~Lorke, J.~P. Kotthaus, W.~Wegscheider, M.~Bichler, Rectification
  in mesoscopic systems with broken symmetry: Quasiclassical ballistic versus
  classical transport, Phys. Rev. Lett. 92 (2004) 056806.

\bibitem{Yasutomi2004a}
S.~Yasutomi, T.~Morita, Y.~Imanishi, S.~Kimura, A molecular photodiode system
  that can switch photocurrent direction, Science 304 (2004) 1944.

\bibitem{Lehmann2002b}
J.~Lehmann, S.~Kohler, P.~H\"anggi, A.~Nitzan, Molecular wires acting as
  coherent quantum ratchets, Phys. Rev. Lett. 88 (2002) 228305.

\bibitem{Grossmann1991a}
F.~Grossmann, T.~Dittrich, P.~Jung, P.~H\"anggi, Coherent destruction of
  tunneling, Phys. Rev. Lett. 67 (1991) 516.

\bibitem{Grossmann1991b}
F.~Gro{\ss}mann, P.~Jung, T.~Dittrich, P.~H\"anggi, Tunneling in a periodically
  driven bistable system, Z. Phys. B 84 (1991) 315.

\bibitem{Grossmann1992a}
F.~Gro{\ss}mann, P.~H\"anggi, Localization in a driven two-level dynamics,
  Europhys. Lett. 18 (1992) 571.

\bibitem{Holthaus1992b}
M.~Holthaus, Collapse of minibands in far-infrared irradiated superlattices,
  Phys. Rev. Lett. 69 (1992) 351--354.

\bibitem{Creffield2002a}
C.~E. Creffield, G.~Platero, ac-driven localization in a two-electron quantum
  dot molecule, Phys. Rev. B 65 (2002) 113304.

\bibitem{Lehmann2003a}
J.~Lehmann, S.~Camalet, S.~Kohler, P.~H\"anggi, Laser controlled molecular
  switches and transistors, Chem. Phys. Lett. 368 (2003) 282.

\bibitem{Kohler2004a}
S.~Kohler, S.~Camalet, M.~Strass, J.~Lehmann, G.-L. Ingold, P.~H\"anggi, Charge
  transport through a molecule driven by a high-frequency field, Chem. Phys.
  296 (2004) 243.

\bibitem{Dayem1962a}
A.~H. Dayem, R.~J. Martin, Quantum interaction of microwave radiation with
  tunneling between superconductors, Phys. Rev. Lett. 8 (1962) 246.

\bibitem{vanderWiel1999a}
W.~G. van~der Wiel, T.~Fujisawa, T.~H. Oosterkamp, L.~P. Kouwenhoven, Microwave
  spectroscopy of a double quantum dot in the low- and high-power regime,
  Physica B 272 (1999) 31.

\bibitem{Stoof1996a}
T.~H. Stoof, {Y}u. V.~Nazarov, Time-dependent resonant tunneling via two
  discrete states, Phys. Rev. B 53 (1996) 1050.

\bibitem{Brune1997a}
P.~Brune, C.~Bruder, H.~Schoeller, Photon-assisted transport through ultrasmall
  quantum dots: Influence of intradot transitions, Phys. Rev. B 56 (1997) 4730.

\bibitem{Zrenner2002a}
A.~Zrenner, E.~Beham, S.~Stufler, F.~Findeis, M.~Bichler, G.~Abstreiter,
  Coherent properties of a two-level system based on a quantum-dot photodiode,
  Nature 418 (2002) 612.

\bibitem{Rabi1937a}
I.~I. Rabi, Space quantization in a gyrating magnetic field, Phys. Rev. 51
  (1937) 652.

\bibitem{Kergueris1999a}
C.~Kergueris, J.-P. Bourgoin, S.~Palacin, D.~Esteve, C.~Urbina, M.~Mgoga,
  C.~Joachim, Electron transport through a metal-molecule-metal junction, Phys.
  Rev. B 59 (1999) 12505.

\bibitem{Weber2002a}
H.~B. Weber, J.~Reichert, F.~Weigend, R.~Ochs, D.~Beckmann, M.~Mayor,
  R.~Ahlrichs, H.~von L\"ohneysen, Electronic transport through single
  conjugated molecules, Chem. Phys. 281 (2002) 113.

\bibitem{Datta1997a}
S.~Datta, W.~Tian, S.~Hong, R.~Reifenberger, J.~I. Henderson, C.~P. Kubiak,
  Current-voltage characteristics of self-assembled monolayers by scanning
  tunneling microscopy, Phys. Rev. Lett 79 (1997) 2530.

\bibitem{Jackel2004a}
F.~J\"ackel, M.~D. Watson, K.~M\"ullen, J.~P. Rabe, Prototypical
  single-molecule chemical-field-effect transistor with nanometer-sized gates,
  Phys.Rev. Lett. 92 (2004) 188303.

\bibitem{Wurfel}
J.~W\"urfel, H.~B. Weber, private communication.

\bibitem{Bavli1993a}
R.~Bavli, H.~Metiu, Properties of an electron in a quantum double well driven
  by a strong laser: Localization, low-frequency, and even-harmonic generation,
  Phys. Rev. A 47 (1993) 3299--3310.

\bibitem{Newns1969a}
D.~M. Newns, Self-consistent model of hydrogen chemisorption, Phys. Rev. 178
  (1969) 1123.

\bibitem{Mujica1996a}
V.~Mujica, M.~Kemp, A.~Roitberg, M.~A. Ratner, Current-voltage characteristics
  of molecular wires: {E}igenvalue staircase, {C}oulomb blockade, and
  rectificatiob, J. chem. Phys. 104 (1996) 7296.

\bibitem{Hall2000a}
L.~E. Hall, J.~R. Reimers, N.~S. Hush, K.~Silverbrook, Formalism, analytical
  model, and a priori green's-function-based calculations of the
  current-voltage characteristics of molecular wires, J. Chem. Phys. 112 (2000)
  1510.

\bibitem{Demming1998a}
F.~Demming, J.~Jersch, K.~Dickmann, P.~I. Geshev, Calculation of the field
  enhancement on laser-illuminated scanning probe tips by the boundary element
  method, Appl. Phys. B 66 (1998) 593.

\bibitem{Otto2001a}
A.~Otto, Theory of first layer and single molecule surface enhanced raman
  scattering (sers), Phys. Stat. Sol. (a) 188 (2001) 1455.

\bibitem{Fleischmann1974a}
M.~Fleischmann, P.~J. Hendra, A.~J. McQuillan, Raman spectra of pyridine
  adsorbed at a silver electrode, Chem. Phys. Lett. 26 (1974) 163.

\bibitem{Jeanmaire1977a}
D.~L. Jeanmaire, R.~P.~V. Duyne, Surface raman spectroelectrochemistry part {I.
  H}eterocyclic, aromatic, and aliphatic amines adsorbed on the anodized silver
  electrode, J. Electroanal. Chem. 84 (1977) 1.

\bibitem{Pellegrini1993a}
B.~Pellegrini, Extension of the electrokinematics theorem to the
  electromagnetic-field and quantum-mechanics, Il Nuovo Cimento 15 (1993)
  855--879.

\bibitem{Nitzan2002a}
A.~Nitzan, M.~Galperin, G.-L. Ingold, H.~Grabert, On the electrostatic
  potential profile in biased molecular wires, J. Chem. Phys. 117 (2002) 10837.

\bibitem{Pleutin2003a}
S.~Pleutin, H.~Grabert, G.-L. Ingold, A.~Nitzan, The electrostatic potential
  profile along a biased molecular wire: A model quantum-mechanical
  calculation, J. Chem. Phys. 118 (2003) 3756.

\bibitem{Liang2004a}
G.~C. Liang, A.~W. Ghosh, M.~Paulsson, S.~Datta, Electrostatic potential
  profiles of molecular conductors, Phys. Rev. B 69 (2004) 115302.

\bibitem{Tucker1979a}
J.~R. Tucker, Quantum limited detection in tunnel junction mixers, IEEE J.
  Quantum Electron. QE-15 (1979) 1234.

\bibitem{Tucker1985a}
J.~R. Tucker, M.~J. Feldman, Quantum detection at millimeter wavelength, Rev.
  Mod. Phys. 57 (1985) 1055.

\bibitem{Tien1963a}
P.~K. Tien, J.~P. Gordon, Multiphoton process observed in the interaction of
  microwave fields with the tunneling between superconductor films, Phys. Rev.
  129 (1963) 647.

\bibitem{Wingreen1990a}
N.~S. Wingreen, Rectification by resonant tunneling diodes, Appl. Phys. Lett.
  56 (1990) 253.

\bibitem{Kislov1991a}
V.~Kislov, A.~Kamenev, High-frequency properties of resonant tunneling devices,
  Appl. Phys. Lett. 59 (1991) 1500.

\bibitem{Aguado1996a}
R.~Aguado, J.~I{\~n}arrea, G.~Platero, Coherent resonant tunneling in ac
  fields, Phys. Rev. B 53 (1996) 10030.

\bibitem{Economou1981a}
E.~N. Economou, C.~M. Soukoulis, Static conductance and scaling theory of
  localization in one dimension, Phys. Rev. Lett. 46 (1981) 618.

\bibitem{Langreth1981a}
D.~C. Langreth, E.~Abrahams, Derivation of the landauer conductance formula,
  Phys. Rev. B 24 (1981) 2978.

\bibitem{Stone1988a}
A.~D. Stone, A.~Szafer, What is measured when you measure a resistence? --- the
  {L}andauer formula revisited, IBM J. Res. Develop. 32 (1988) 384.

\bibitem{Engquist1981a}
H.-L. Engquist, P.~W. Anderson, Definition and measurement of the electrical
  and thermal resistances, Phys. Rev. B 24 (1981) 1151.

\bibitem{Buttiker1986a}
M.~B\"uttiker, Four-terminal phase-coherent conductance, Phys. Rev. Lett. 57
  (1986) 1761.

\bibitem{Fisher1981a}
D.~S. Fisher, P.~A. Lee, Relation between conductivity and transmission matrix,
  Phys. Rev. B 23 (1981) 6851.

\bibitem{Sols1991a}
F.~Sols, Gauge-invariant formulation of electron linear transport, Phys. Rev.
  Lett. 67 (1991) 2874.

\bibitem{Caroli1971a}
C.~Caroli, R.~Combescot, P.~Noziere, D.~Saint-James, Direct calculation of the
  tunneling current, J. Phys. C 4 (1971) 916.

\bibitem{Meir1992a}
Y.~Meir, N.~S. Wingreen, Landauer formula for the current through an
  interacting electron region, Phys. Rev. Lett. 68 (1992) 2512.

\bibitem{Wingreen1993a}
N.~S. Wingreen, A.-P. Jauho, Y.~Meir, Time-dependent transport through a
  mesoscopic structure, Phys. Rev. B 48 (1993) 8487.

\bibitem{Fano1947a}
U.~Fano, Ionization yield of radiations. {II. T}he fluctuations of the number
  of ions, Phys. Rev. 72 (1947) 26.

\bibitem{Nakajima1958a}
S.~Nakajima, On quantum theory of transport phenomena, Prog. Theor. Phys. 20
  (1958) 948.

\bibitem{Zwanzig1960a}
R.~Zwanzig, Ensemble methods in the theory of irreversibility, J. Chem. Phys.
  33 (1960) 1338.

\bibitem{Novotny2002a}
T.~Novotn\'y, Investigation of apparent violation of the second law of
  thermodynamics in quantum transport studies, Europhys. Lett. 59 (2002) 648.

\bibitem{May2004a}
V.~May, O.~K\"uhn, Charge and Energy Transfer Dynamics in Molecular Systems,
  2nd Edition, Wiley-VCH, Weinheim, 2003.

\bibitem{Hanggi1982a}
P.~H\"anggi, H.~Thomas, Stochastic processes: Time evolution, symmetries and
  linear response, Phys. Rep. 88 (1982) 206.

\bibitem{Bruder1994a}
C.~Bruder, H.~Schoeller, Charging effects in ultrasmall quantum dots in the
  presence of time-varying fields, Phys. Rev. Lett. 72 (1994) 1076.

\bibitem{Sambe1973a}
H.~Sambe, Steady states and quasienergies of a quantum-mechanical system in an
  oscillating field, Phys. Rev. A 7 (1973) 2203.

\bibitem{Grifoni1998a}
M.~Grifoni, P.~H\"anggi, Driven quantum tunneling, Phys. Rep. 304 (1998) 229.

\bibitem{Buchleitner2002a}
A.~Buchleitner, D.~Delande, J.~Zakrzewski, Non-dispersive wave packets in
  periodically driven quantum systems, Phys. Rep. 368 (2002) 409.

\bibitem{Shirley1965a}
J.~H. Shirley, Solution of the {S}chr\"odinger equation with a {H}amiltonian
  periodic in time, Phys. Rev. 138 (1965) B979.

\bibitem{Jung1990a}
P.~Jung, P.~H\"anggi, Resonantly driven {B}rownian motion, basic concepts and
  exact results, Phys. Rev. A 41 (1990) 2977.

\bibitem{Sakurai}
J.~J. Sakurai, Modern Quantum Mechanics, 2nd Edition, Addison-Wesley, Reading,
  1995.

\bibitem{Peres1991a}
A.~Peres, Dynamical quasidegeneracies and quantum tunneling, Phys. Rev. Lett.
  67 (1991) 158.

\bibitem{Holthaus1992a}
M.~Holthaus, The quantum theory of an ideal superlattice responding to
  far-infrared laser radiation, Z. Phys. B 89 (1992) 251.

\bibitem{Keller2002a}
A.~Keller, O.~Atabek, M.~Ratner, V.~Mujica, Laser-assisted conductance of
  molecular wires, J. Phys. B 35 (2002) 4981.

\bibitem{Brandes1997a}
T.~Brandes, Truncation method for green's functions in time-dependent fields,
  Phys. Rev. B 56 (1997) 1213.

\bibitem{Hone1997a}
D.~W. Hone, R.~Ketzmerick, W.~Kohn, Time-dependent floquet theory and absence
  of an adiabatic limit, Phys. Rev. A 56 (1997) 4045.

\bibitem{Holstein1959a}
T.~Holstein, Polaron motion. {I. M}olecular-crystal model, Ann. Phys. (N.Y.) 8
  (1959) 325.

\bibitem{Brandes1999a}
T.~Brandes, B.~Kramer, Spontaneous emission of phonons by coupled quantum dots,
  Phys. Rev. Lett. 83 (1999) 3021.

\bibitem{Lehmann2002a}
J.~Lehmann, G.-L. Ingold, P.~H\"anggi, Incoherent charge transport through
  molecular wires: interplay of coulomb interaction and wire population, Chem.
  Phys. 281 (2002) 199.

\bibitem{Segal2002a}
D.~Segal, A.~Nitzan, Conduction in molecular junctions: inelastic effects,
  Chem. Phys. 281 (2002) 235.

\bibitem{Aguado2004a}
R.~Aguado, T.~Brandes, Shot noise spectrum of open dissipative quantum
  two-level systems, Phys. Rev. Lett. 92 (2004) 206601.

\bibitem{Brandes2004a}
T.~Brandes, R.~Aguado, G.~Platero, Charge transport through open driven
  two-level systems with dissipation, Phys. Rev. B 69 (2004) 205326.

\bibitem{Talkner1986a}
P.~Talkner, The failure of the quantum regression hypothesis, Ann. Phys. (N.Y.)
  167 (1986) 390.

\bibitem{Nazarov1993a}
{Y}u. V.~Nazarov, Quantum interference, tunnel junctions and resonant tunneling
  interferometer, Physica B 189 (1993) 57.

\bibitem{Hazelzet2001a}
B.~L. Hazelzet, M.~R. Wegewijs, T.~H. Stoof, {Y}u. V.~Nazarov, Coherent and
  incoherent pumping of electrons in double quantum dot, Phys. Rev. B 63 (2001)
  165313.

\bibitem{Tikhonov2002b}
A.~Tikhonov, R.~D. Coalson, Y.~Dahnovsky, Calculating electron current in a
  tight-binding model of a field-driven molecular wire: Application to
  xylyl-dithiol, J. Chem. Phys. 117 (2002) 567.

\bibitem{Tikhonov2002a}
A.~Tikhonov, R.~D. Coalson, Y.~Dahnovsky, Calculating electron current in a
  tight-binding model of a field-driven molecular wire: Floquet theory
  approach, J. Chem. Phys. 116 (2002) 10909.

\bibitem{Ratner1990a}
M.~A. Ratner, Bridge-assisted electron transfer: Effective electronic coupling,
  J. Phys. Chem. 94 (1990) 4877.

\bibitem{Mujica1994b}
V.~Mujica, M.~Kemp, M.~A. Ratner, Electron conduction in molecular wires. {II.
  A}pplication to scanning tunneling microscopy, J. Chem. Phys. 101 (1994)
  6856.

\bibitem{Feynman1963a}
R.~P. Feynman, R.~B. Leighton, M.~Sands, The Feynman Lectures on Physics,
  Vol.~1, Addison Wesley, Reading MA, 1963.

\bibitem{Goychuk1998a}
I.~Goychuk, M.~Grifoni, P.~H\"anggi, Nonadiabatic quantum {B}rownian
  rectifiers, Phys. Rev. Lett. 81 (1998) 649, erratum: ibid. \textbf{81}, 2837
  (1998).

\bibitem{Goychuk1998b}
I.~Goychuk, P.~H\"anggi, Quantum rectifiers from harmonic mixing, Europhys.
  Lett. 43 (1998) 503.

\bibitem{Goychuk2001a}
I.~Goychuk, P.~H\"anggi, Minimal quantum {B}rownian rectifiers, J. Phys. Chem.
  B 105 (2001) 6642.

\bibitem{Kouwenhoven1991a}
L.~P. Kouwenhoven, A.~T. Johnson, N.~C. van~der Vaart, C.~J. P.~M. Harmans,
  Quantized current in a quantum-dot turnstile using oscillating tunnel
  barriers, Phys. Rev. Lett. 67 (1991) 1626.

\bibitem{Moskalets2002a}
M.~Moskalets, M.~B\"uttiker, Floquet scattering theory of quantum pumps, Phys.
  Rev. B 66 (2002) 205320.

\bibitem{DiCarlo2003a}
L.~DiCarlo, C.~M. Marcus, J.~S. Harris, Jr., Photocurrent, rectification, and
  magnetic field symmetry of induced current through quantum dots, Phys. Rev.
  Lett. 91 (2003) 246804.

\bibitem{Flach2000a}
S.~Flach, O.~Yevtushenko, Y.~Zolotaryuk, Directed current due to broken
  time-space symmetry, Phys. Rev. Lett. 84 (2000) 2358.

\bibitem{Reimann2001a}
P.~Reimann, Supersymmetric ratchets, Phys. Rev. Lett. 86 (2001) 4992.

\bibitem{Chen1999a}
J.~Chen, M.~A. Reed, A.~M. Rawlett, J.~M. Tour, Large on-off ratios and
  negative differential resistance in a molecular electronic device, Science
  286 (1999) 1550.

\bibitem{Creffield2002b}
C.~E. Creffield, G.~Platero, Dynamical control of correlated states in a square
  quantum dot, Phys. Rev. B 66 (2002) 235303.

\bibitem{Holthaus1993a}
M.~Holthaus, D.~Hone, Quantum-wells and superlattices in strong time-dependent
  fields, Phys. Rev. B 47 (1993) 6499--6508.

\bibitem{DeJong1995a}
M.~J.~M. de~Jong, C.~W.~J. Beenakker, Semiclassical theory of shot-noise
  suppression, Phys. Rev. B 51 (1995) 16867.

\bibitem{Wagner1994a}
M.~Wagner, Quenching of resonant transmission through an oscillating quantum
  well, Phys. Rev. B 49 (1994) 16544.

\bibitem{Wagner1995a}
M.~Wagner, Photon-assisted transmission through an oscillating quantum well: A
  transfer-matrix approach to coherent destruction of tunneling, Phys. Rev. A
  51 (1995) 798.

\bibitem{Kohler2004c}
S.~Kohler, J.~Lehmann, P.~H\"anggi, Controlling currents through molecular
  wires, Superlatt. Microstruct. 34 (2004) 419.

\bibitem{Fonseca2004a}
K.~M. Fonseca-Romero, S.~Kohler, P.~H\"anggi, Coherence control for qubits,
  Chem. Phys. 296 (2004) 307.

\bibitem{Dittrich1993a}
T.~Dittrich, B.~Oelschl\"agel, P.~H\"anggi, Driven dissipative tunneling,
  Europhys. Lett. 22 (1993) 5.

\bibitem{Dittrich1993b}
T.~Dittrich, F.~Grossmann, P.~Jung, B.~Oelschl\"agel, P.~H\"anggi, Localization
  and tunneling in periodically driven bistable systems, Physica A 194 (1993)
  173.

\bibitem{Makarov1995a}
D.~E. Makarov, N.~Makri, Stochastic resonance and nonlinear response in
  double-quantum-well structures, Phys. Rev. E 52 (1995) R2257.

\bibitem{Grossmann1993a}
F.~Grossmann, T.~Dittrich, P.~Jung, P.~H\"anggi, Coherent transport in a
  periodically driven bistable system, J. Stat. Phys. 70 (1993) 229.

\bibitem{Gerstner2000a}
V.~Gerstner, A.~Knoll, W.~Pfeiffer, A.~Thon, G.~Gerber, Femtosecond laser
  assisted scanning tunneling microscopy, J. Appl. Phys. 88 (2000) 4851.

\bibitem{Hanggi1998a}
P.~H\"anggi, Driven quantum systems, in: Quantum Transport and Dissipation,
  Wiley-VCH, Weinheim, 1998, Ch.~5, pp. 249--286.

\bibitem{Risken}
H.~Risken, The Fokker-Planck Equation, 2nd Edition, Vol.~18 of Springer Series
  in Synergetics, Springer, Berlin, 1989.

\bibitem{Peskin1993a}
U.~Peskin, N.~Moiseyev, The solution of the time-dependent schr\"odinger
  equation by the $(t,t')$ method: Theory, computational algorithm and
  applications, J. Chem. Phys. 99 (1993) 4590.

\end{thebibliography}
\end{document}